\documentclass[preprint,12pt,authoryear]{elsarticle}
\usepackage{color}
\usepackage{graphicx}
\usepackage{upgreek}
\usepackage{subfig}
\usepackage{caption}
\usepackage{bm}
\biboptions{sort&compress}
\usepackage{tabularray,breakcites}
\usepackage{overpic}
\usepackage{booktabs}
\usepackage{colortbl}
\usepackage[hidelinks]{hyperref}
\usepackage{amssymb,amsmath}
\usepackage{geometry}
\usepackage[mathlines,pagewise]{lineno}
\usepackage{makecell}
\usepackage{xcolor}

\usepackage{caption}
	
\begin{document}

\begin{frontmatter}
		
\title{Modelling aerodynamic forces and torques of spheroid particles in compressible flows}

\author[label2,label1]{Yibin Du}
\author[label1]{Ming Yu \corref{cor1}}
\ead{yum16@tsinghua.org.cn}
\author[label2]{Chongwen Jiang}
\author[label1]{Xianxu Yuan}
\ead{yuanxianxu2023@163.com}

\address[label1]{State Key Laboratory of Aerodynamics, Mianyang, 621000, China}

\address[label2]{National Laboratory for Computational Fluid Dynamics, School of Aeronautic Science and Engineering, Beihang University, Beijing, 100191, China}

\begin{abstract}

In the present study, we conduct numerical simulations of compressible flows around spheroid
particles,
for the purpose of refining empirical formulas for drag force, lift force, and pitching torque acting on them. 
Through an analysis of approximately a thousand numerical simulation cases spanning a wide range of Mach numbers, Reynolds numbers
and particle aspect ratios, we first identify the crucial parameters that are strongly correlated with the forces and torques via Spearman correlation analysis,
based on which the empirical formulas for the drag force, lift force and pitching torque coefficients are refined.
The novel formulas developed for compressible flows exhibit consistency with their incompressible counterparts at low Mach number limits and,
moreover, yield accurate predictions with average relative errors of less than 5\%.
This underscores their robustness and reliability in predicting aerodynamic loads on spheroidal particles under various flow conditions.
\end{abstract}


\begin{keyword}
Spheroid particles \sep Drag force \sep Lift force \sep Pitching torque
\end{keyword}

\end{frontmatter}




%
\section{Introduction} \label{introduce}
Particle-laden compressible flows are frequently encountered in natural and engineering applications. 
In aerospace engineering, the impact of gas during the powered landing of a spacecraft can expel dust 
and debris, potentially causing significant damage to high-speed vehicles \citep{capecelatro_gas-particle_2023}.  
When the particle sizes are significantly smaller than the characteristic length scales of fluid motion, 
and the total volume fraction of the particle phase is low, 
the simulations of particle-laden flows can be achieved using the Eulerian-Lagrangian point-particle method. 
This method has found extensive application in various turbulence scenarios, including wall turbulence
\citep{zhao_slip_2014,zhao_four-way_2015,marchioli_relative_2016,cui_eect_nodate}, 
isotropic turbulence\citep{gustavsson_tumbling_2014,bounoua_tumbling_2018}, 
and homogeneous shear turbulence\citep{huang_rotation_2012,rosen_effect_2014}.

In solving the particle-laden flows with the Eulerian-Lagrangian point-particle method,
the accurate modeling of the forces acting on particles stands out as a critical challenge. 
Amongst the prevalent models, the drag force model tailored for spherical particles in incompressible flows 
holds a prominent position. 
Given the inherently small size of the particles, 
the particle Reynolds numbers, defined as $Re_p=\rho_f u_p d_p / \mu_f$ 
(where $\rho_f$ and $\mu_f$ represent the fluid density and dynamic viscosity, 
$u_p=|u_f-v|$ and $d_p$ denote the particle's slip velocity and diameter, $u_f$ and $v$ are the velocity of fluid and particles respectively), 
remain relatively low. 
\textcolor{black}{Therefore}, the Stokes drag force is dominant, rendering the inertial effects negligible.
For higher $Re_p$ values, \citet{clift_motion_1971} devised a formula for the drag on spherical particles 
in incompressible scenarios, drawing upon a synthesis of experimental data and insights from
\citet{stokes_mathematical_2009} and \citet{schiller_uber_1933}. 
This formula is deemed applicable for $Re_p$ values below $2\times10^5$.
  
For non-spherical particles, there will be not only the drag force, 
but also lift force and pitching torque acting on them, even in the uniform incoming flow. 
The forces and torques acting on \textcolor{black}{spheroidal} particles in such flows were initially determined by 
\citet{oberbeck_lieber_nodate}, followed by the refinement of theoretical equations for these forces 
in Stokes flows in subsequent studies \citep{brenner_oseen_1961,brenner_stokes_1964,batchelor_stress_1970}. 
Analytical expressions for the lift and drag coefficients of \textcolor{black}{spheroidal} particles under creeping flow 
conditions were provided by \citet{happel_low_1983}.
Empirical formulas for higher Reynolds numbers have been developed in recent years by many researchers 
\citep{holzer_new_2008,zastawny_derivation_2012,ouchene_new_2016,sanjeevi_drag_2018,ouchene_numerical_2020,
frohlich_correlations_2020}. 
\citet{holzer_new_2008} established a relation for the drag coefficient of non-spherical 
particles with varying sphericity, 
which is the ratio between the surface area of the volume-equivalent sphere and the surface area of 
the particle under consideration. 
\citet{zastawny_derivation_2012} established empirical correlations for drag, lift, 
pitching torque and rotational torque coefficients for \textcolor{black}{spheroidal} particles with aspect ratios of 1.25, 2.5, 
and 0.2, as well as for fibrous particles with an aspect ratio of 5. 
\citet{richter_new_2013} thoroughly examined the heat transfer, forces, and torques acting on \textcolor{black}{spheroidal} and 
cubic particles and established a comprehensive set of closure relations for heat flow and flow through 
nonspherical particles.
\citet{ouchene_new_2016} conducted direct numerical simulations of symmetric particle bypass flows and 
proposed empirical formulations for the lift, drag, and pitching torque coefficients of \textcolor{black}{spheroidal} particles 
with aspect ratios ranging from 1 to 32 at the Reynolds number ranging from 0.1 to 240. 
The fitted formulas are further corrected by \citet{arcen_prolate_2017}.
\citet{sanjeevi_drag_2018} considered \textcolor{black}{spheroidal} particles with the aspect ratios of $w=2.5,0.4$, 
and fibrous particles with $w=4$. The simulations covered a wide range of Reynolds numbers from 0.1 to 2000. 
\citet{frohlich_correlations_2020} studied the flow separation conditions of prolate spheroid particles
with aspect ratios ranging from 1 to 8 at the Reynolds number from 1 to 100, resulting in the development of 
new empirical expressions for the drag, lift, and torque of these particles. 
\citet{ouchene_numerical_2020} investigated the forces acting on oblate \textcolor{black}{spheroidal} particles with shape factors ranging from 0.2 to 1.  
The resulting expressions for lift, drag, and pitching torque coefficients are applicable within the range of $Re_p<100$.
     
In the realm of compressible flows, experimental data indicates the significance of compressibility effects \citep{nagata_investigation_2016}. 
The particle Mach number $M_p=u_p/a_f$ (where $a_f$ represents the sound speed of ambient fluid) 
serves as a crucial parameter for assessing the impacts of flow compressibility. 
\citet{loth_drag_2008} investigated the drag of spherical particles in compressible flows, 
noting a weak Mach number effect below the critical Mach number ($M_p<0.6$) and a nearly constant drag coefficient 
in the hypersonic regime ($M_p>5$). 
Subsequently, \citet{parmar_improved_2010} identified limitations in the fitting formula proposed by 
\citet{loth_drag_2008} through an analysis of data from \citet{bailey_sphere_1976}, 
offering an enhanced correlation for the drag coefficient of a sphere in compressible flows. 
\citet{loth_supersonic_2021} developed empirical expressions for a broader range of $Re_p$ 
and $M_p$ using numerical simulations and past experimental data, 
significantly enhancing their accuracy and applicability.
    
The brief literature review above highlights the extensive development and validation of forcing models 
for spherical particles in both incompressible and compressible flows,  as well as for non-spherical particles in incompressible flows. 
However, empirical formulas for the drag force, lift force, and pitching torque of non-spherical particles in 
compressible flows are currently lacking, to the best of our knowledge. 
This serves as the primary motivation for the current study. 
This paper aims to investigate the forces and torques acting on \textcolor{black}{spheroidal} particles with varying aspect ratios 
at different Mach numbers \textcolor{black}{($M_p \leq 2$)} and Reynolds numbers\textcolor{black}{($10 \leq Re_p \leq 100 $)} via numerical simulations,
\textcolor{black}{where vortex shedding will not occur.}
Based on these simulations, empirical formulas for drag, lift force, and pitching torque will be developed. 

The remainder of this paper is organized as follows. In section~\ref{sec:num} we introduce the numerical method employed in the present study as well as its validation.
In section~\ref{sec:force} we discuss at length the variation of the drag force, lift force, and pitching torque with the Reynolds and Mach numbers for particles with different aspect ratios and
the angle of attack. 
The empirical formulas of the forces and torque will be given correspondingly.
Conclusions are recapitulated in section~\ref{sec:con}.
    
\section{Numerical methods and validation} \label{sec:num}

In this study, we consider compressible Newtonian perfect gas flowing around small-sized particles in the shape of
axisymmetric ellipsoids, namely the spheroids.
The forces acting on these particles are influenced by several parameters, 
\textcolor{black}{including} the particle Mach number $M_p$, 
the particle Reynolds number $Re_p$, the aspect ratio of the particle $w$, and the angle of attack $\alpha$. 
The aspect ratio is defined as $w=a/b$, where $a$ denotes the length of the \textcolor{black}{spheroid}'s axis of rotational symmetry, 
and $b$ represents the length of the other two identical axes. 
Particles are classified as `prolate' when $w>1$ and `oblate' when $w<1$. 
The angle of attack, \textcolor{black}{denoted by} $\alpha$, is defined as the angle between the long axis of the spheroid and the direction of the incoming flow.

The coordinate system is established with its origin at the center of the particle. 
The $x-$axis aligns parallel to the direction of the incoming flow, while the $z-$axis is oriented perpendicularly to 
the plane formed by long axis orientation and the $x$ axis. 
The $y-$axis is determined according to the right-hand rule of the coordinate system. 
In the cases where the angle of attack equals zero, and the flow around the particles remains steady, 
the flow exhibits axisymmetry relative to the $x$ axis for prolate particles, rendering the $y$ and $z$ directions equivalent,
while the flow is only symmetrical relative to the \textcolor{black}{$z=0$} plane for oblate particles.

\begin{figure}[tp!]
	\begin{center}
		\includegraphics[width=0.7\textwidth]{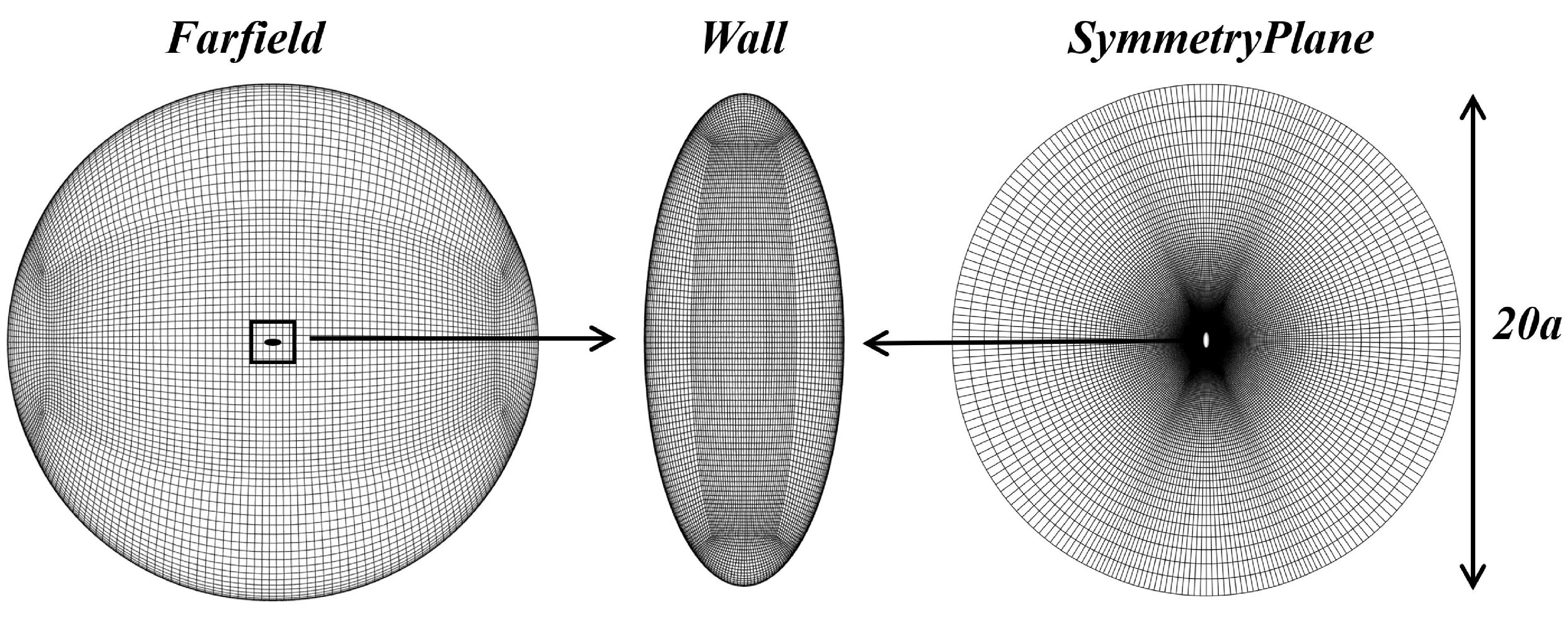}
		\caption{Sketch of computational domain and mesh grids for particles with $w=2.5$ at the far field (left) , the wall (middle), and the symmetry plane(right).}
		\label{fig1:mesh}			
	\end{center}
\end{figure}

Numerical simulations are conducted using the NNW-PHengLEI software developed by the China Aerodynamics Research and Development Center 
\citep{he2016validation}. 
The compressible Navier-Stokes equations are solved by the finite volume method. 
The convection terms are approximated using the Roe scheme, the viscous terms are computed by the central difference scheme, 
and time advancement is achieved using the implicit LU-SGS method. 
The mesh grid configuration is depicted in Figure \ref{fig1:mesh}, using the particle with
$w=2.5$ as an example. 
The computational domain is half of a sphere with a diameter of $20a$, 
which is considered to have minimal impact on particle forces \citep{richter_drag_2012}. 
The initial grid spacing is set at $0.005a$ in the first mesh layer to ensure precise computation of viscous stresses,
thereafter expanding by 1.05 times within $4a$ and by 1.2 times outward.
The total cell count varies between 0.7 and 1.2 million cells across different scenarios, 
depending on the shapes of particles.


To validate the accuracy of the numerical techniques employed in this study, 
we firstly conduct simulations for the flows around prolate spheroid particles with $w=2.5$ and 
oblate spheroid particles with $w=0.4$ at $Re_p=10$ and 100 in low Mach number $M_p=0.1$ for comparison with incompressible flows. 
The angle of attack $\alpha$ varies from $0^\circ$ to $90^\circ$. 
We primarily compare the drag force $F_D$, lift force $F_L$ and pitching torque $T$, calculated as 
\begin{equation}
	F_D=-\int_S p {\bm e}_x \cdot  {\bm n} {\rm d}S+\int_S {\bm n} \cdot {\bm \tau} \cdot {\bm e}_x {\rm d}S
\end{equation}
\begin{equation}
F_L=-\int_S p {\bm e}_y \cdot  {\bm n} {\rm d}S+\int_S {\bm n} \cdot {\bm \tau} \cdot {\bm e}_y {\rm d}S
\end{equation}
\begin{equation}
T=|(F_D {\bm e}_x +F_L {\bm e}_y) \times {\rm r}_{cp}|
\end{equation}
where $p$ is the pressure, ${\bm \tau}$ is the viscous stress tensor, ${\bm e}_x$ and ${\bm e}_y$ are the unit vector in the $x$ and $y$ directions, 
${\bm n}$ is the unit vector at the particle surface, 
and ${\rm r}_{cp}$ is location of the pressure center. 
Correspondingly, the drag force coefficient $C_D$, lift force coefficient $C_L$, and pitching torque coefficient $C_T$ are defined as
\begin{equation}
C_D=\frac{8 F_D}{\rho_\infty \pi d_p^2 U_\infty^2},~~
C_L=\frac{8 F_L}{\rho_\infty d_p^2 U_\infty^2},~~
C_T=\frac{16 T}{\rho_\infty \pi d_p^3 U_\infty^2}
\end{equation}
where $\rho_\infty$ and $U_\infty$ are the density and velocity of the incoming flow, \textcolor{black}{respectively. 
The $d_p$ denotes diameter of the volume-equivalent sphere for non-spherical particles.}

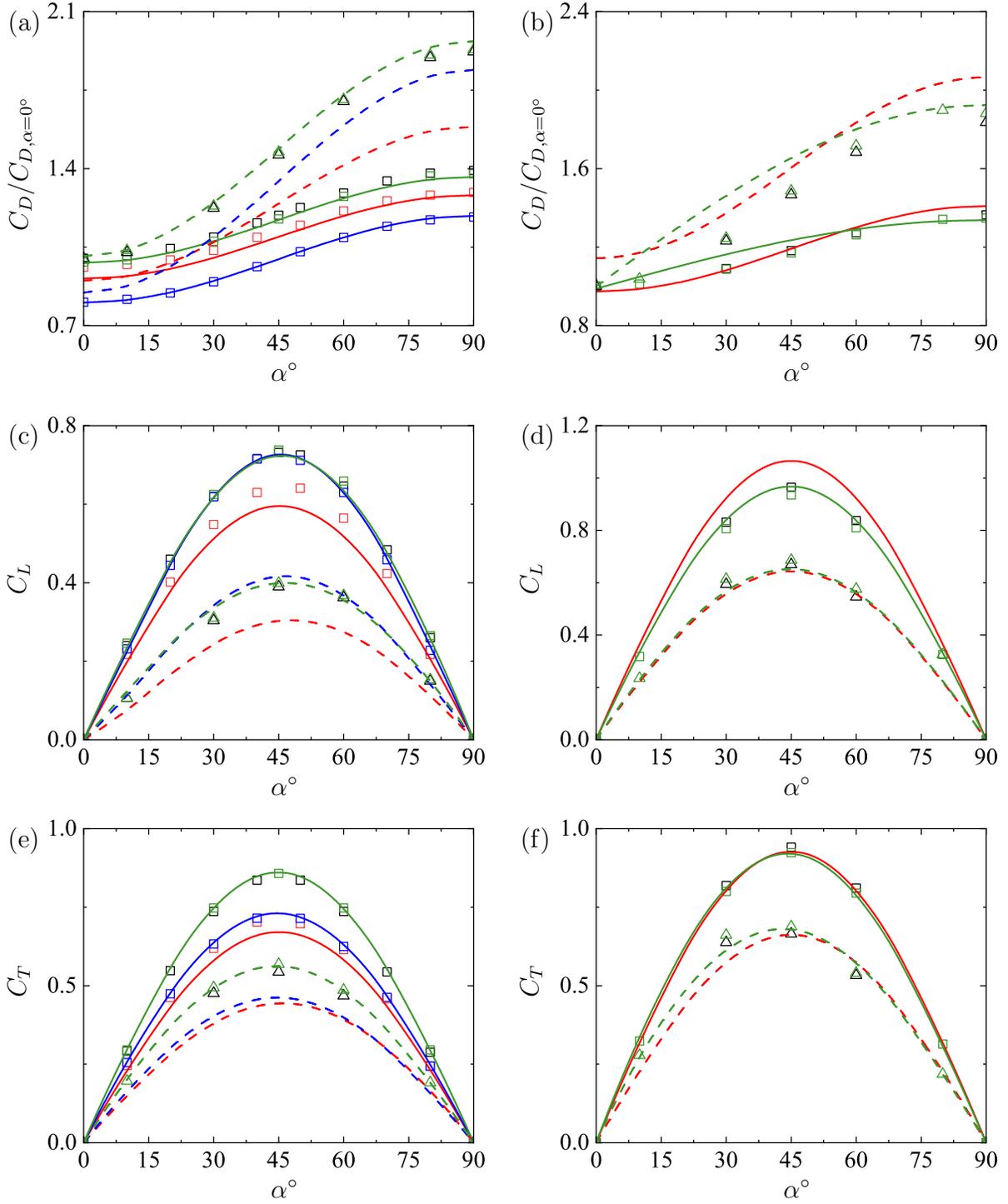
\begin{figure}[tp!]
	\centering
	\begin{overpic}[width=0.5\textwidth]{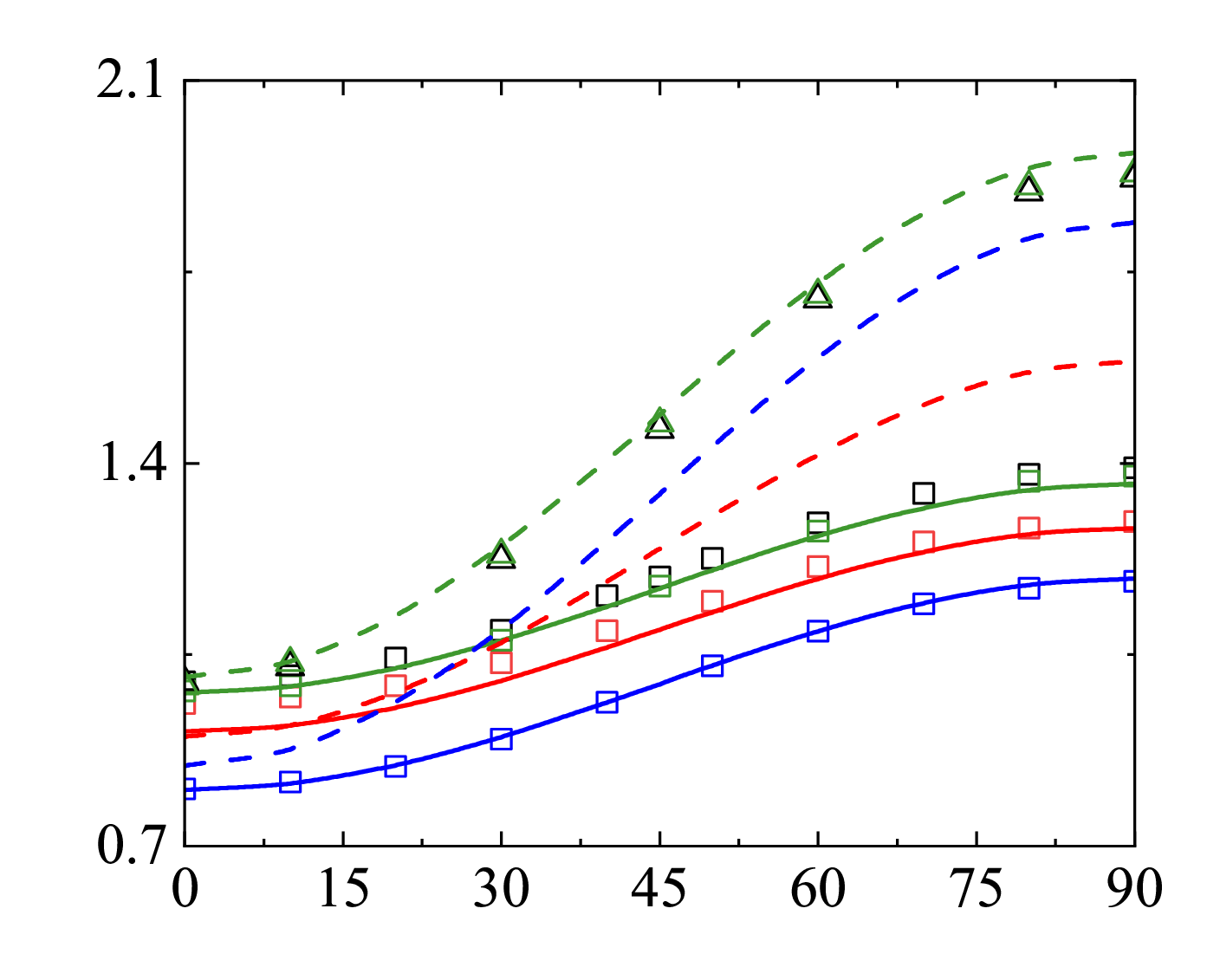}
		\put(0,70){(a)}
		\put(0,30){\rotatebox{90}{\textcolor{black}{$C_D/C_{D,\alpha=0^\circ}$}}}
		\put(52.5,0){$\alpha^\circ$}
	\end{overpic}~
	\begin{overpic}[width=0.5\textwidth]{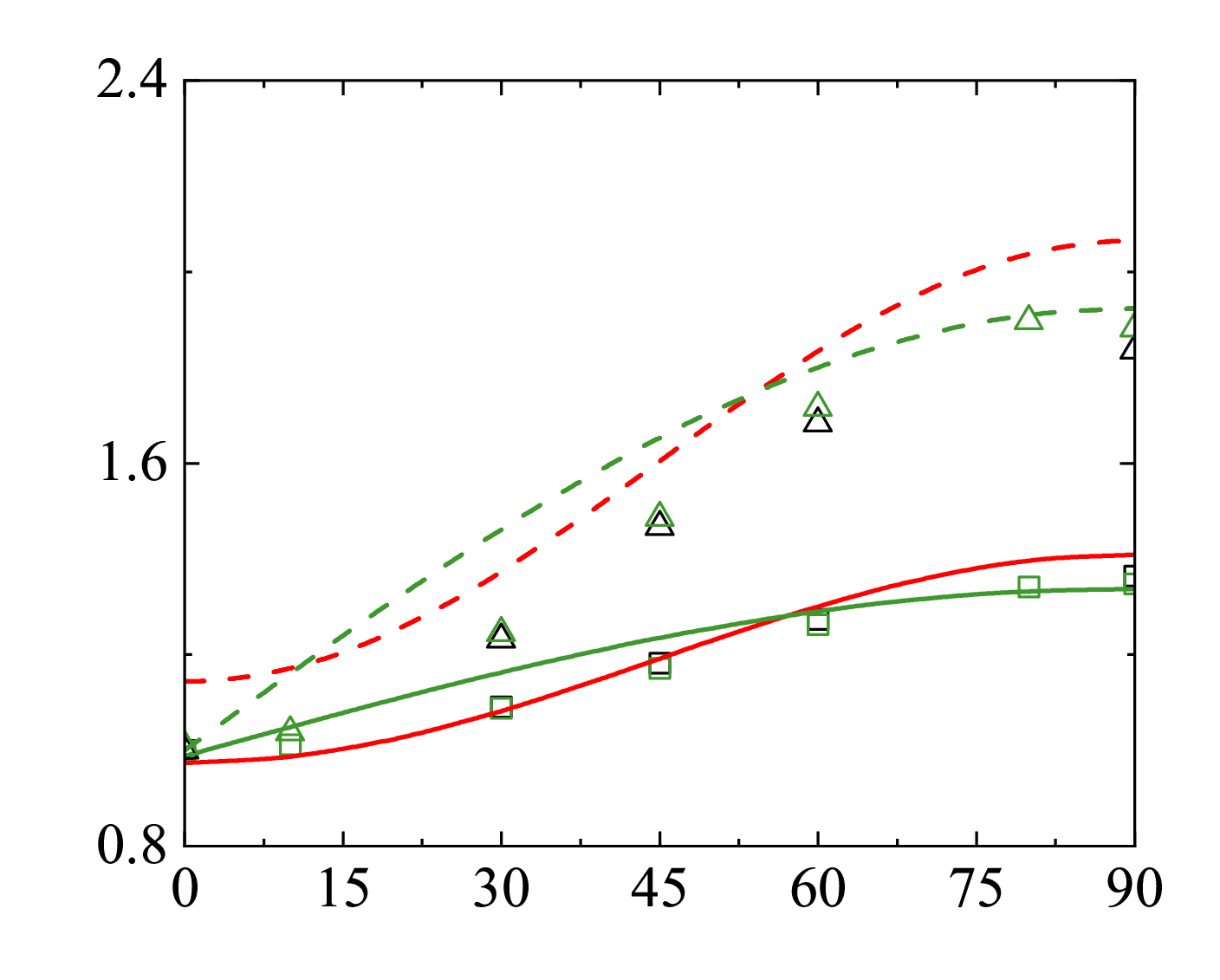}
		\put(0,70){(b)}
		\put(0,30){\rotatebox{90}{\textcolor{black}{$C_D/C_{D,\alpha=0^\circ}$}}}
		\put(52.5,0){$\alpha^\circ$}
	\end{overpic}\\[1.0ex]
	\begin{overpic}[width=0.5\textwidth]{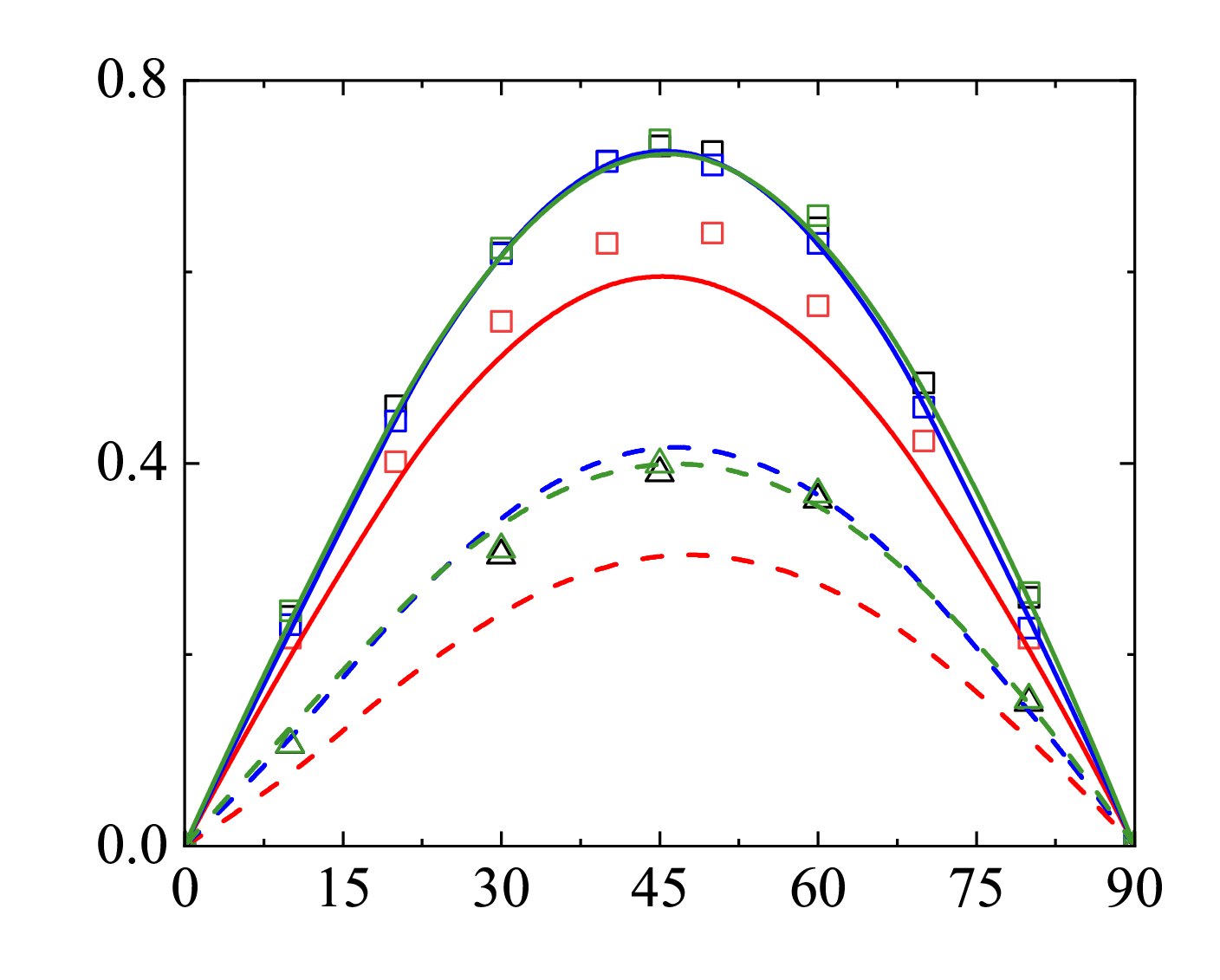}
		\put(0,70){(c)}
		\put(0,40){\rotatebox{90}{$C_L$}}
		\put(52.5,0){$\alpha^\circ$}
	\end{overpic}~
	\begin{overpic}[width=0.5\textwidth]{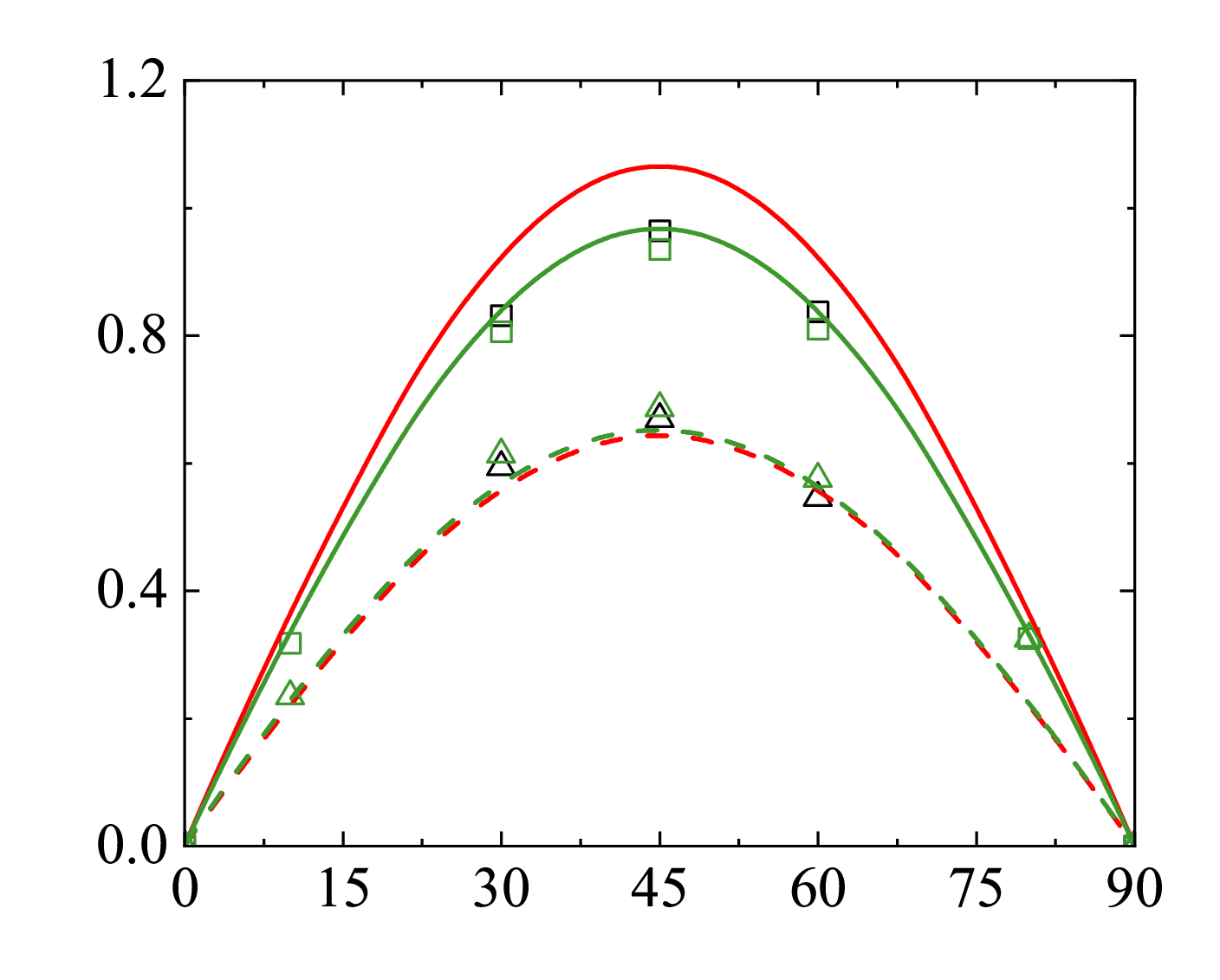}
		\put(0,70){(d)}
		\put(0,40){\rotatebox{90}{$C_L$}} 
		\put(52.5,0){$\alpha^\circ$}
	\end{overpic}\\
	\begin{overpic}[width=0.5\textwidth]{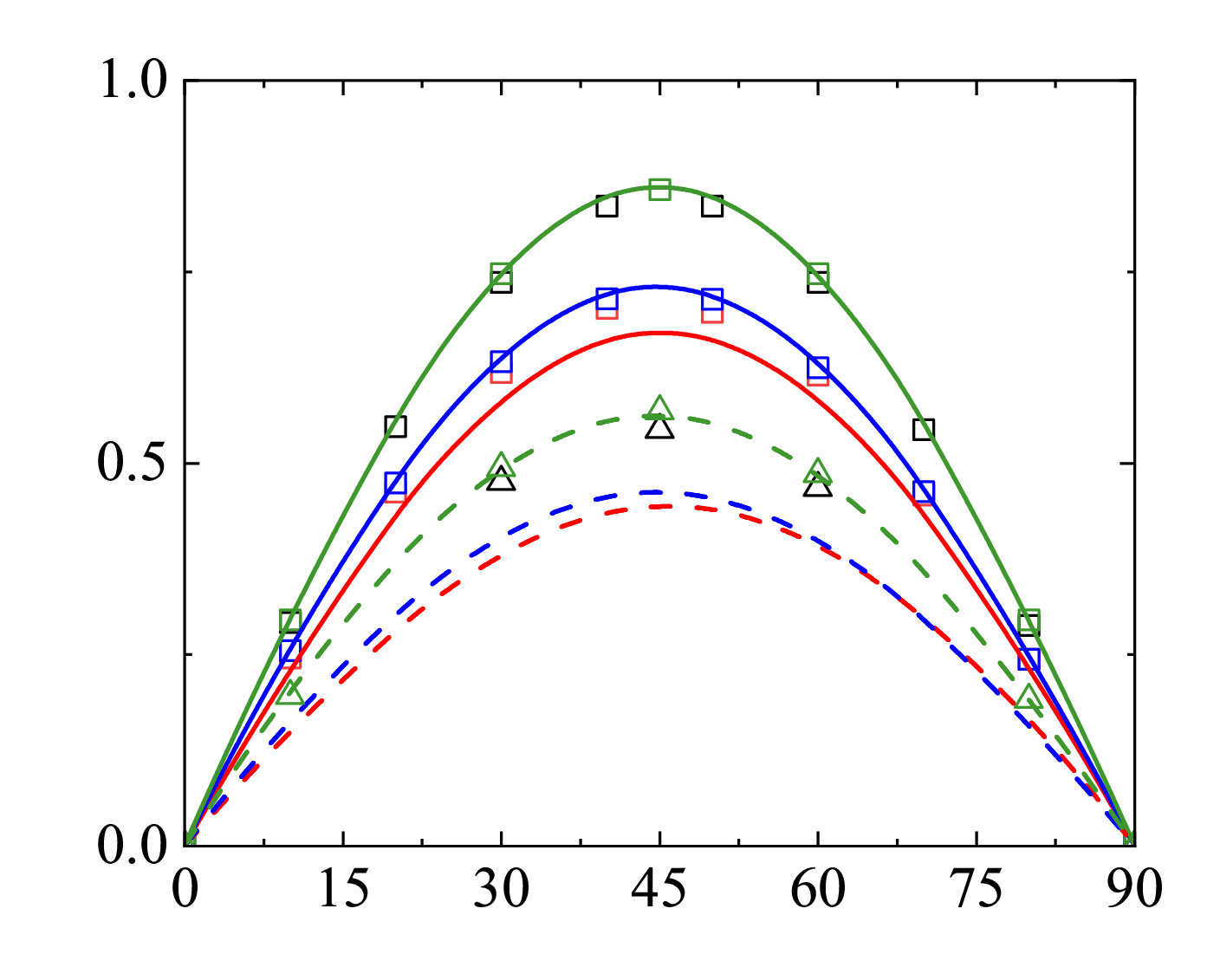}
		\put(0,70){(e)}
		\put(0,40){\rotatebox{90}{$C_T$}}
		\put(52.5,0){$\alpha^\circ$}
	\end{overpic}~
	\begin{overpic}[width=0.5\textwidth]{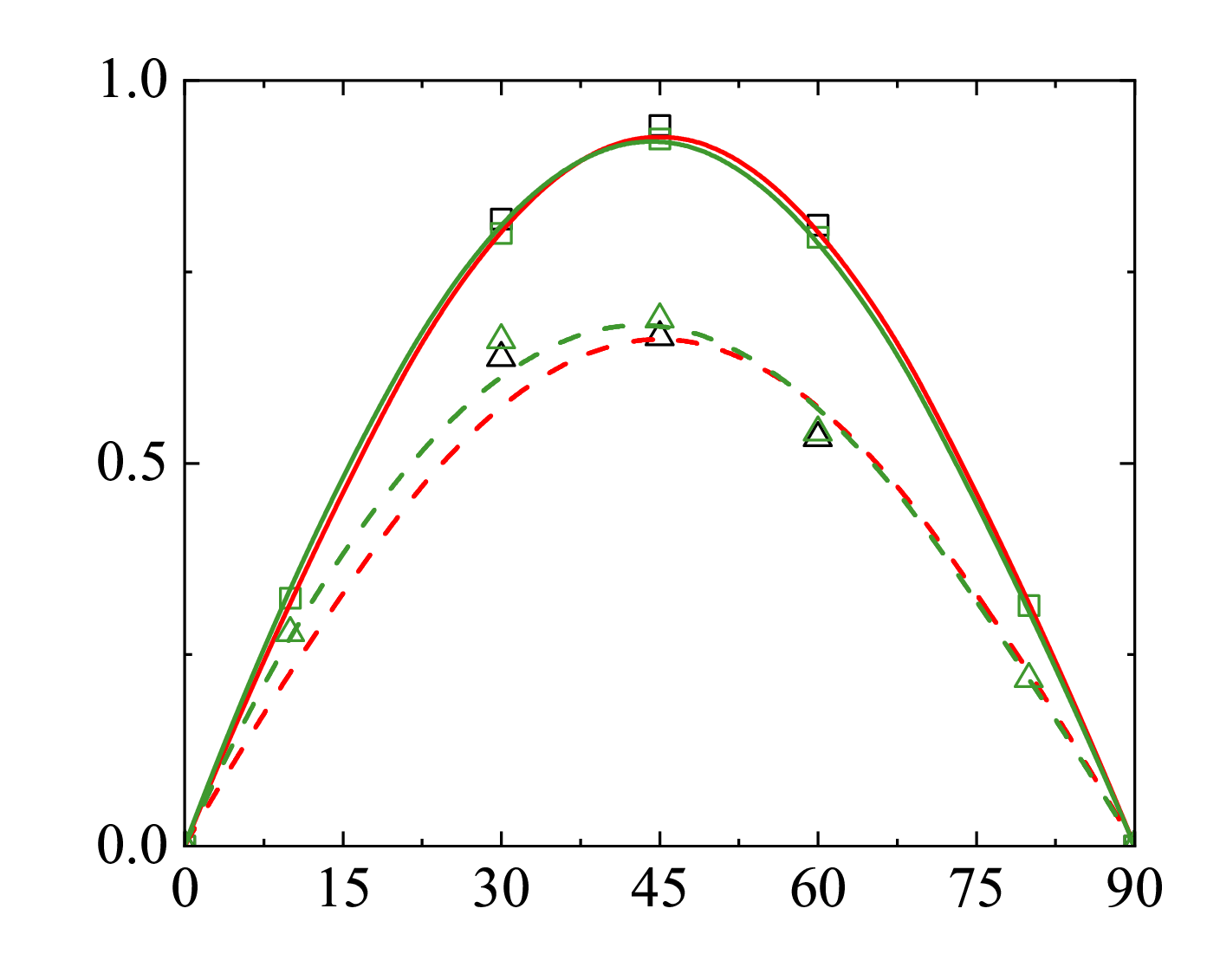}
		\put(0,70){(f)}
		\put(0,40){\rotatebox{90}{$C_T$}}
		\put(52.5,0){$\alpha^\circ$}
	\end{overpic}\\
	\caption{Variation of (a,b) \textcolor{black}{$C_D/C_{D,\alpha=0^\circ}$}, (c,d) $C_L$ and (e,f) $C_T$ against the angle of attack $\alpha$, at $M_p=0.1$.
		(a,c,e) $w=2.5$ , (b,d,f)  $w=0.4$. 
		\textcolor{black}{Solid lines: empirical formulas at $Re_p=10$,
		dashed lines: empirical formulas at $Re_p=100$.
		Squares: simulation at $Re_p=10$, 
		triangles: simulation at $Re_p=100$.
		Black lines and symbols : present results,
		red lines and symbols: (a,c,e) \citet{ouchene_new_2016}, (b,d,f) \citet{ouchene_numerical_2020}, 
		blue lines and symbols: \citet{zastawny_derivation_2012}, 
		green lines and symbols: \citet{sanjeevi_drag_2018}.}}
	\label{fig2:emample-incom}
\end{figure}

The results are shown in Figure \ref{fig2:emample-incom},  along with
\textcolor{black}{the empirical formulas and simulation data presented by}
\citet{zastawny_derivation_2012,sanjeevi_drag_2018,ouchene_new_2016} and
\citet{ouchene_numerical_2020}.
\textcolor{black}{In Figure \ref{fig2:emample-incom}(a,b), 
we divide $C_D$ by $C_{D,\alpha=0^\circ}$ to enhance their clarity in showing the trend of variation,
where $C_{D,\alpha=0^\circ}$ is also given by the present simulation, 
and a similar approach is employed in subsequent contents of this paper.}
The relative discrepancies in comparison to these formulas are detailed in
Table \ref{tab1:err-in-imcom-example}.
Overall, the current results align well with these empirical formulas
\textcolor{black}{and simulation data}.
Notably, the deviations from the results of \citet{sanjeevi_drag_2018} are minimal,
with an average relative error \textcolor{black}{generally} less than \textcolor{black}{$2.5\%$}.
Conversely, \textcolor{black}{the simulation data by
\citet{ouchene_drag_2015,ouchene_new_2016} and \citet{ouchene_numerical_2020} demonstrate
higher relative errors, notably 12\% for drag and 16\% for lift coefficients,
and empirical correlations presented by \citet{ouchene_new_2016} and \citet{ouchene_numerical_2020}
give the average relative errors of approximately 10\% 
for drag, lift and pitching torque coefficients comparing with the present results.
For the numerical results calculated by \citet{zastawny_derivation_2012},
the average relative error for drag and lift coefficients is approximately $10\%$,
while the error for torque coefficients is only $2\%$.
The correlations by \citet{zastawny_derivation_2012} give the average relative error
of approximately 14\% for drag and torque coefficients, and $3.5\%$ for lift coefficients.}
These discrepancies can probably be attributed to the numerical simulation methodologies employed
in these studies.
As a result, we consider that the numerical approach adopted is deemed valid for low Mach number flows.

\begin{table}[tbp!]
	\centering
	\small
	\caption{\textcolor{black}{Relative error between the simulation results at $M_p = 0.1$
and simulations and empirical correlations given by previous studies.}
	}
	\label{tab1:err-in-imcom-example}
	\begin{tblr}{
			cell{2}{1} = {r=6}{},
			cell{2}{2} = {r=2}{},
			cell{2}{4} = {r},
			cell{2}{5} = {r},
			cell{2}{6} = {r},
			cell{3}{4} = {r},
			cell{3}{5} = {r},
			cell{3}{6} = {r},
			cell{4}{2} = {r=2}{},
			cell{4}{4} = {r},
			cell{4}{5} = {r},
			cell{4}{6} = {r},
			cell{5}{4} = {r},
			cell{5}{5} = {r},
			cell{5}{6} = {r},
			cell{6}{2} = {r=2}{},
			cell{6}{4} = {r},
			cell{6}{5} = {r},
			cell{6}{6} = {r},
			cell{7}{4} = {r},
			cell{7}{5} = {r},
			cell{7}{6} = {r},
			cell{8}{1} = {r=6}{},
			cell{8}{2} = {r=2}{},
			cell{8}{4} = {r},
			cell{8}{5} = {r},
			cell{8}{6} = {r},
			cell{9}{4} = {r},
			cell{9}{5} = {r},
			cell{9}{6} = {r},
			cell{10}{2} = {r=2}{},
			cell{10}{4} = {r},
			cell{10}{5} = {r},
			cell{10}{6} = {r},
			cell{11}{4} = {r},
			cell{11}{5} = {r},
			cell{11}{6} = {r},
			cell{12}{2} = {r=2}{},
			cell{12}{4} = {r},
			cell{12}{5} = {r},
			cell{12}{6} = {r},
			cell{13}{4} = {r},
			cell{13}{5} = {r},
			cell{13}{6} = {r},
			hline{1-2,8,14} = {-}{},
			hline{4,6,10,12} = {2-6}{},
			colspec = {Q[c] Q[c] Q[c] Q[c] Q[c] Q[c]}
		}
		&    &      & \makecell{{\citet{ouchene_new_2016}}\\{\citet{ouchene_numerical_2020}}}
		& \citet{zastawny_derivation_2012} & \citet{sanjeevi_drag_2018} \\
		\rotatebox{90}{Simulations}  & $C_D$ & Mean(\%) & 5.69    & 11.86    & 1.06     \\
		&    & Max(\%)  & 7.14    & 16.23    & 2.44     \\
		& $C_L$ & Mean(\%) & 12.41   & 9.00        & 2.47     \\
		&    & Max(\%)  & 16.50    & 11.99    & 5.24     \\
		& $C_T$ & Mean(\%) & 15.96   & 1.93     & 2.58     \\
		&    & Max(\%)  & 16.40    & 3.78     & 4.21     \\
		\rotatebox{90}{Correlations} & $C_D$ & Mean(\%) & 9.24    & 13.45    & 1.91     \\
		&    & Max(\%)  & 17.39   & 19.74    & 4.55     \\
		& $C_L$ & Mean(\%) & 11.09   & 3.49     & 1.13     \\
		&    & Max(\%)  & 22.46   & 6.13     & 2.14     \\
		& $C_T$ & Mean(\%) & 10.40    & 14.50     & 2.20      \\
		&    & Max(\%)  & 20.96   & 15.09    & 3.10      
	\end{tblr}
\end{table}

\begin{figure}[tp!]
	\centering
	\begin{overpic}[width=0.5\textwidth]{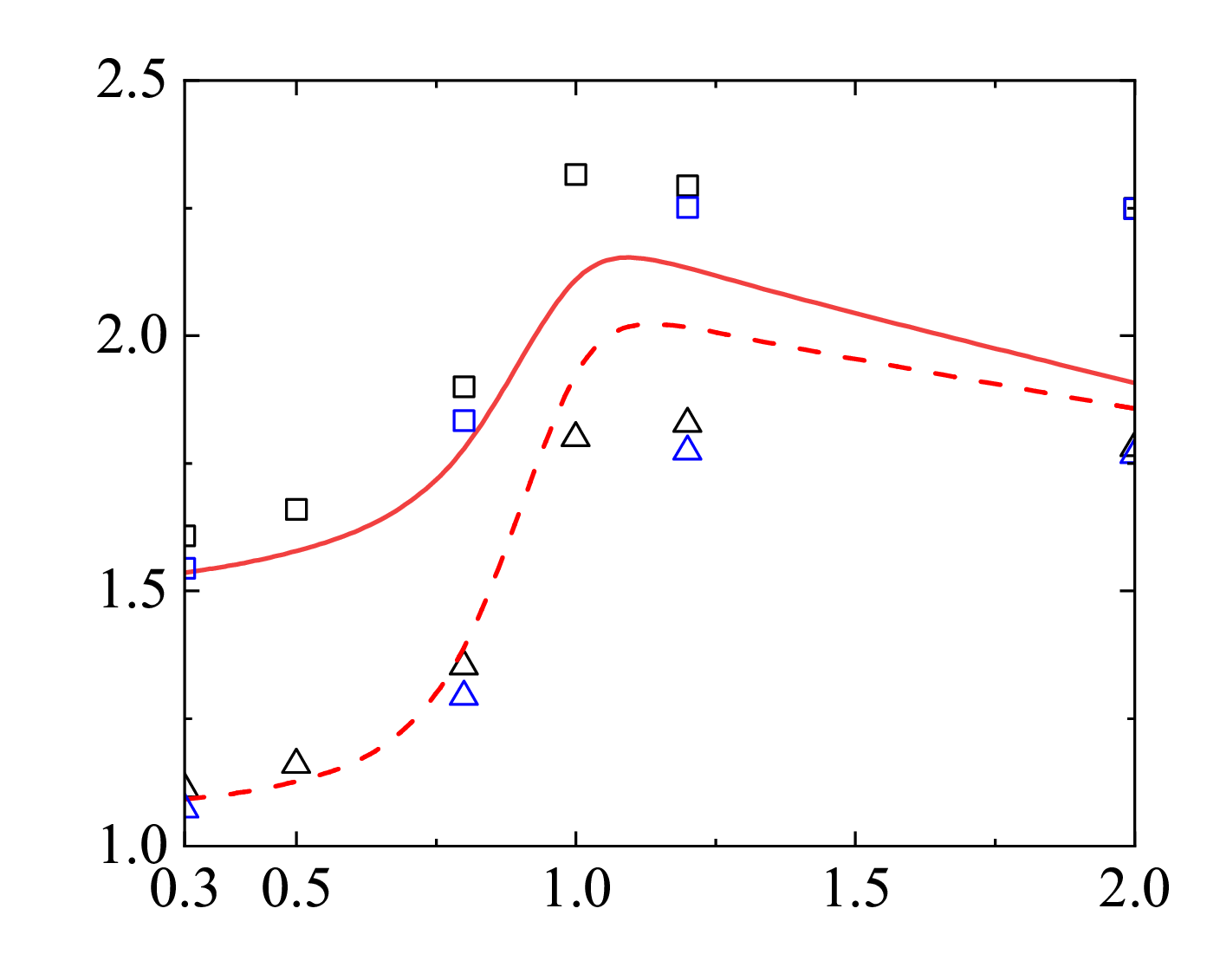}
		\put(3,40){\rotatebox{90}{$C_D$}}
		\put(52.5,0){$M_p$}
	\end{overpic}~
	\caption{Variation of $C_D$ against $M_p$, at $w=1$.
		Black symbols: present simulations,
		blue symbols: \citet{nagata_investigation_2016}. 
		Red lines: empirical formula proposed by \citet{loth_supersonic_2021}. Squares and solid lines: $Re_p=50$, triangles and dashed lines: $Re_p=100$.}
	\label{fig3:emample-com}
\end{figure}

\begin{table}[tp!]
	\centering
	\small
	\caption{Relative error between the simulation and previous results for spherical particles}
	\begin{tblr}{
			cell{1}{1} = {c=2}{},
			cell{2}{1} = {r=2}{},
			cell{4}{1} = {r=2}{},
			hline{1-2,4,6-7} = {-}{},
			colspec = {Q[c] Q[c] Q[c] Q[c]}
		} 
		~        &               & \citet{loth_supersonic_2021} & \citet{nagata_investigation_2016} \\
		Mean(\%) & $Re_p=50$  & 7.29     & 2.34       \\
		& $Re_p=100$ & 8.04    & 2.92       \\
		Max(\%)  & $Re_p=50$  & 15.17     & 3.95       \\
		& $Re_p=100$ & 15.21    & 4.36       \\
		Mean(\%) & All
		cases   & 7.66     & 2.63       
	\end{tblr}
	\label{tab2:err-com}
\end{table}

We also perform simulations for flow past spherical particles at higher Mach numbers $M_p$ ranging from 0.3$\sim$2.0. 
The drag coefficients $C_D$ are compared with the direct numerical simulation (DNS) results of \citet{nagata_investigation_2016} 
and the empirical formula proposed by \citet{loth_supersonic_2021}, as illustrated in Figure \ref{fig3:emample-com}. 
The relative errors are reported in Table \ref{tab2:err-com}. 
With the increasing $M_p$, the $C_D$ values initially rise and then decline, peaking around $M_p \approx 1$. 
While this trend aligns with the fitting formula by \citet{loth_supersonic_2021}, discrepancies exist, 
with higher values compared to the latter at $Re_p=50$ and lower at $Re_p=100$. 
The average relative error stands at approximately 7\%. 
The DNS results from \citet{nagata_investigation_2016} slightly undershoot the present findings, with an average relative error below 3\%. 
These relatively minor discrepancies underscore the validity of the numerical methodology employed in this study.

\textcolor{black}{
Lastly, we evaluate the influences of grid quality on the numerical results
using the simulation cases listed in Table~\ref{tab0:grid}.
Four grid types with different minimal grid sizes, cell numbers and the use of full/half model
were compared at $Re_p=10, 100$ and $\alpha = 45^\circ$ for prolate spheroids with $w=2.5$.
The relative errors between these cases and the numerical results of \citet{sanjeevi_drag_2018}
are lower than $4\%$,
and the relative differences amongst the present results are less than $0.45\%$,
thereby validating the appropriateness of the grid settings herein.}

\begin{table}
    \centering
    \caption{\textcolor{black}{Drag force, lift force and pitching torque coefficients of spheroid with
            $w = 2.5$ at $\alpha = 45^{\circ}$, $Re_p = 10$ and $100$ and $M_p = 0.1$.}
    }
    \label{tab0:grid}
    \begin{tblr}{
            cell{2}{1} = {r=5}{},
            cell{2}{2} = {r},
            cell{2}{5} = {r},
            cell{2}{6} = {r},
            cell{2}{7} = {r},
            cell{2}{8} = {r=4}{},
            cell{3}{2} = {r},
            cell{3}{5} = {r},
            cell{3}{6} = {r},
            cell{3}{7} = {r},
            cell{4}{2} = {r},
            cell{4}{5} = {r},
            cell{4}{6} = {r},
            cell{4}{7} = {r},
            cell{5}{2} = {r},
            cell{5}{5} = {r},
            cell{5}{6} = {r},
            cell{5}{7} = {r},
            cell{6}{2} = {c=3}{},
            cell{6}{5} = {r},
            cell{6}{6} = {r},
            cell{6}{7} = {r},
            cell{7}{1} = {r=5}{},
            cell{7}{2} = {r},
            cell{7}{5} = {r},
            cell{7}{6} = {r},
            cell{7}{7} = {r},
            cell{7}{8} = {r=4}{},
            cell{8}{2} = {r},
            cell{8}{5} = {r},
            cell{8}{6} = {r},
            cell{8}{7} = {r},
            cell{9}{2} = {r},
            cell{9}{5} = {r},
            cell{9}{6} = {r},
            cell{9}{7} = {r},
            cell{10}{2} = {r},
            cell{10}{5} = {r},
            cell{10}{6} = {r},
            cell{10}{7} = {r},
            cell{11}{2} = {c=3}{},
            cell{11}{5} = {r},
            cell{11}{6} = {r},
            cell{11}{7} = {r},
            hline{1-2,7,12} = {-}{},
            colspec = {Q[c] Q[c] Q[c] Q[c] Q[c] Q[c] Q[c] Q[c] Q[c] Q[c]}
        }
        $Re_p$~ & Total cells     & Model & \makecell{{Minimum}\\{grid interval}} & $C_D$      & $C_L$     & $C_T$       &  \makecell{{Max}\\{relative}\\{error$(\%)$}}   \\
        10  & 668334   & half  & 0.01a              & 4.5764  & 0.7335  & 0.8490  & 0.451 \\
        & 884520   & half  & 0.005a             & 4.5618  & 0.7318  & 0.8488  &       \\
        & 1997960  & half  & 0.001a             & 4.5810   & 0.7340 & 0.8489 &       \\
        & 1916398  & whole & 0.005a             & 4.5825  & 0.7340 & 0.8488 &       \\
        & \cite{sanjeevi_drag_2018} &       &                    & 4.4984 & 0.7382 & 0.8571  & 3.916 \\
        100 & 668334   & half  & 0.01a              & 1.1764 & 0.3894 & 0.5458  & 0.220  \\
        & 884520   & half  & 0.005a             & 1.1780   & 0.3900 & 0.5457  &       \\
        & 1997960  & half  & 0.001a             & 1.1759 & 0.3891  & 0.5460    &       \\
        & 1916398  & whole & 0.005a             & 1.1767  & 0.3892 & 0.5458 &       \\
        & \cite{sanjeevi_drag_2018} &       &                     & 1.1780   & 0.3987 & 0.5679  & 2.431

    \end{tblr}
\end{table}

\section{Forces and torques on spheroid particles}  \label{sec:force}

As we have stated in the previous section, \textcolor{black}{the forces and torques acting on the spheroid particles may be influenced by several factors, including} the particle Reynolds number $Re_p$, the particle Mach numbers $M_p$, 
the aspect ratio $w$ and the angle of attack $\alpha$. 
To fully understand the effects of these parameters and establish the model for the forces and torques, 
we consider six groups of \textcolor{black}{spheroidal} particles with different aspect ratios under various incoming flow conditions.
The aspect ratio $w$ ranges from 0.2 to 5.0. 
The $Re_p$ covers the range of 10 $\sim$ 100 and $M_p$ from 0.1 $\sim$ 2.0. 
The angle of attack $\alpha$ is set as $0^\circ$, $30^\circ$, $45^\circ$, $60^\circ$ and $90^\circ$, respectively. 
The database considered herein encompasses 960 groups of simulation results.
\textcolor{black}{The majority of simulations exhibit the Knudsen number below 0.1, 
indicating that the rarefaction of flow can be effectively disregarded.}

\begin{figure}[tp!]
	\centering
	\begin{overpic}[width=0.5\textwidth]{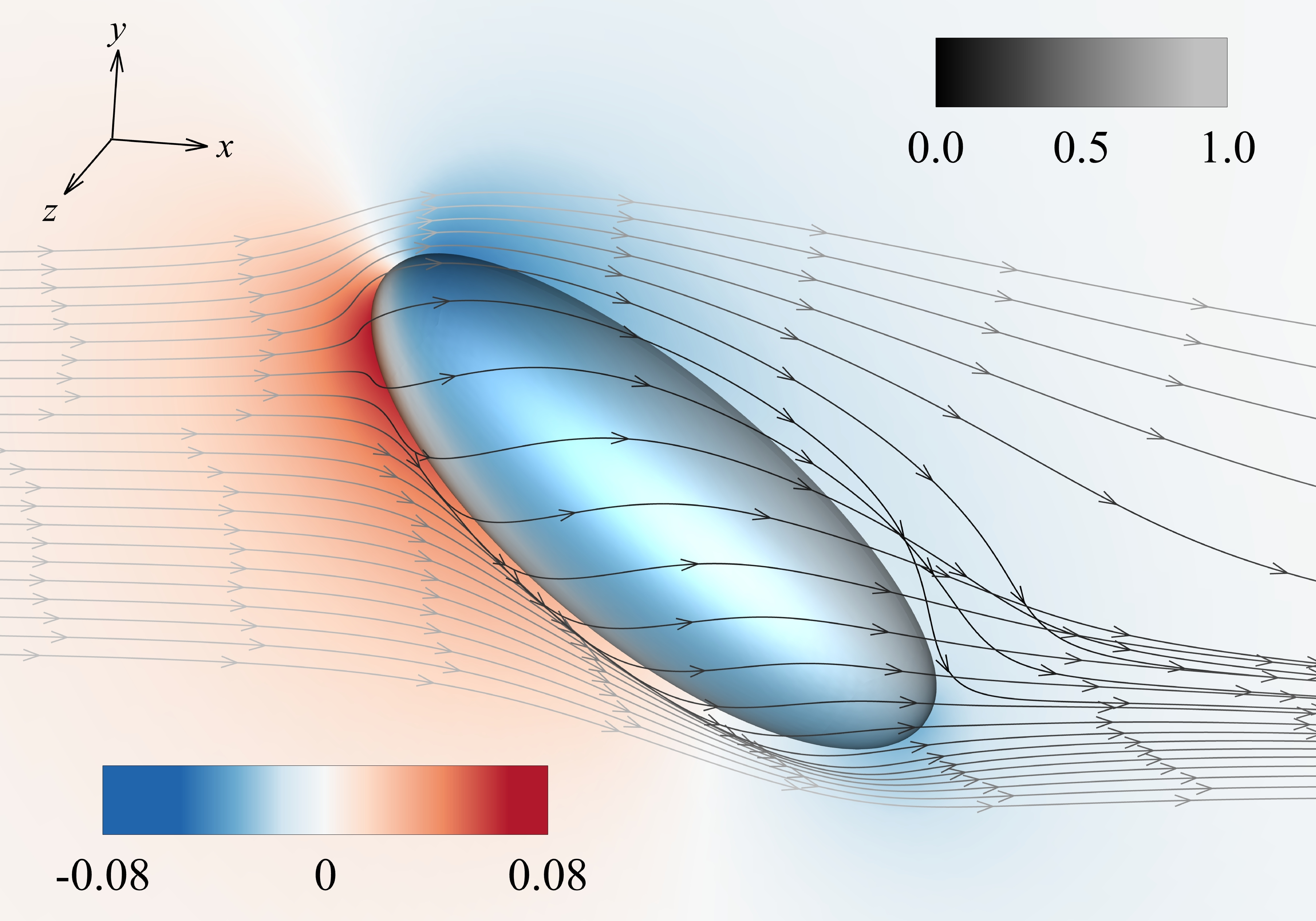}
		\put(0,65){(a)}
	\end{overpic}~
	\begin{overpic}[width=0.5\textwidth]{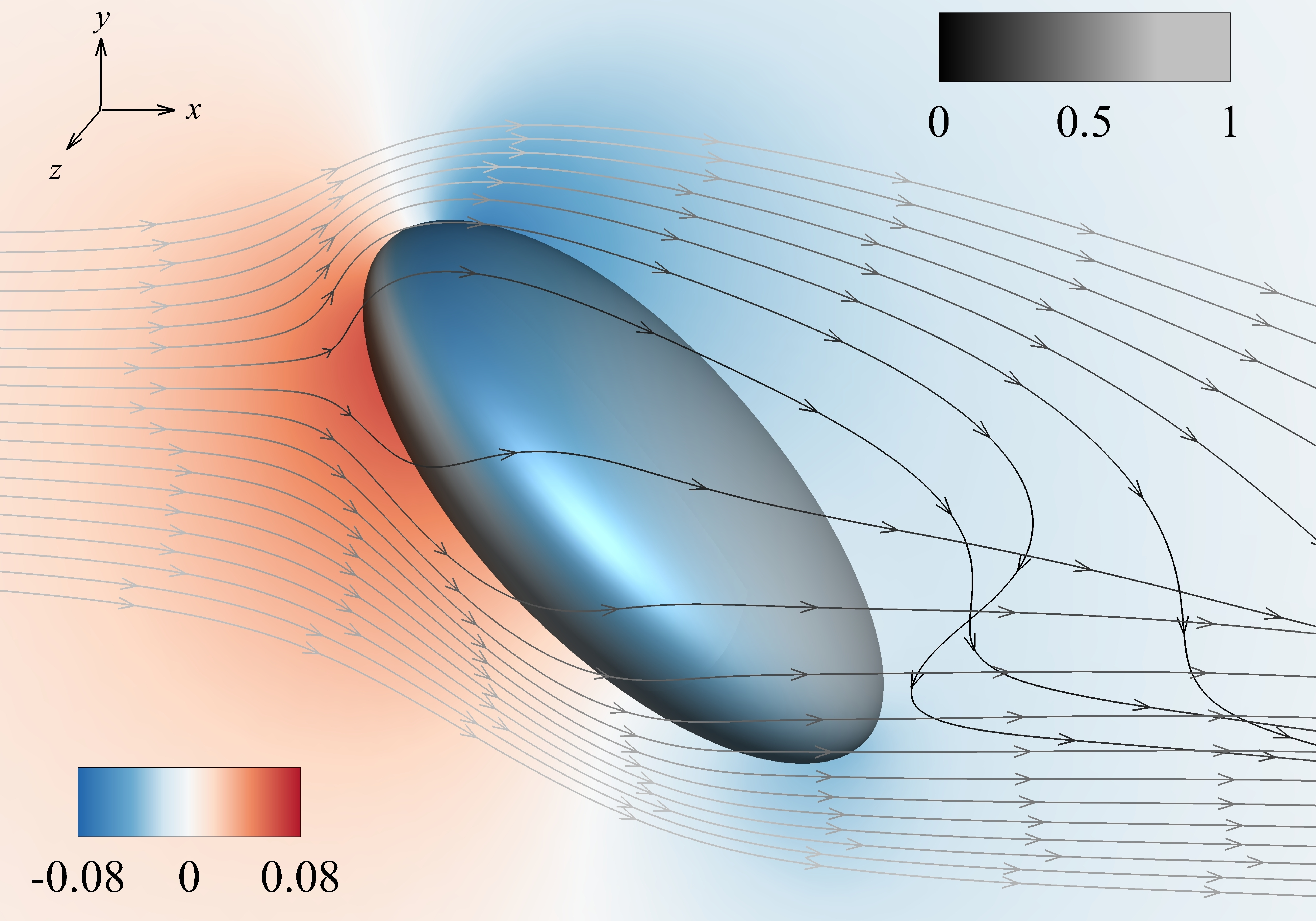}
		\put(0,65){(b)}
	\end{overpic}\\[1.0ex]
	\begin{overpic}[width=0.5\textwidth]{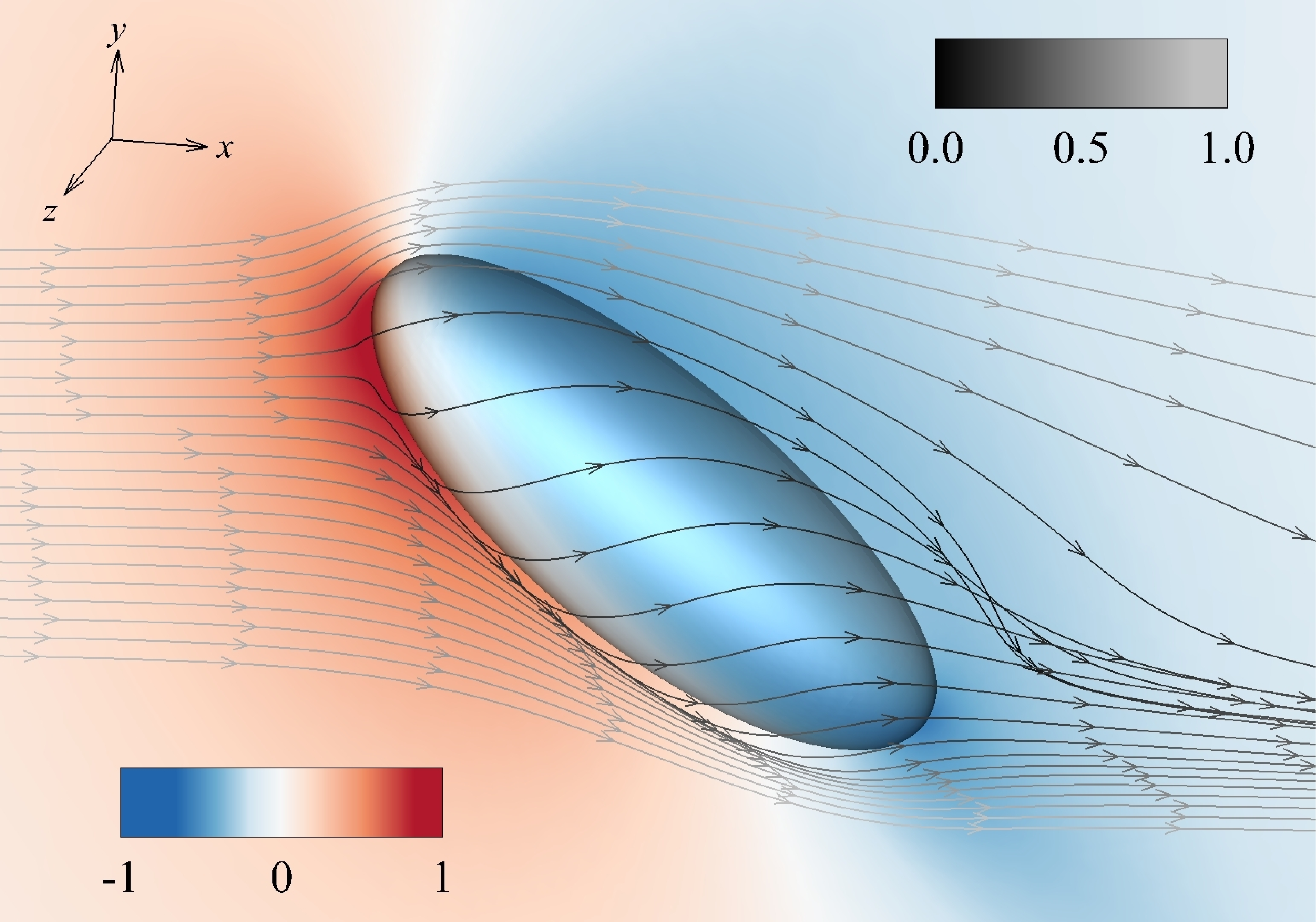}
		\put(0,65){(c)}
	\end{overpic}~
	\begin{overpic}[width=0.5\textwidth]{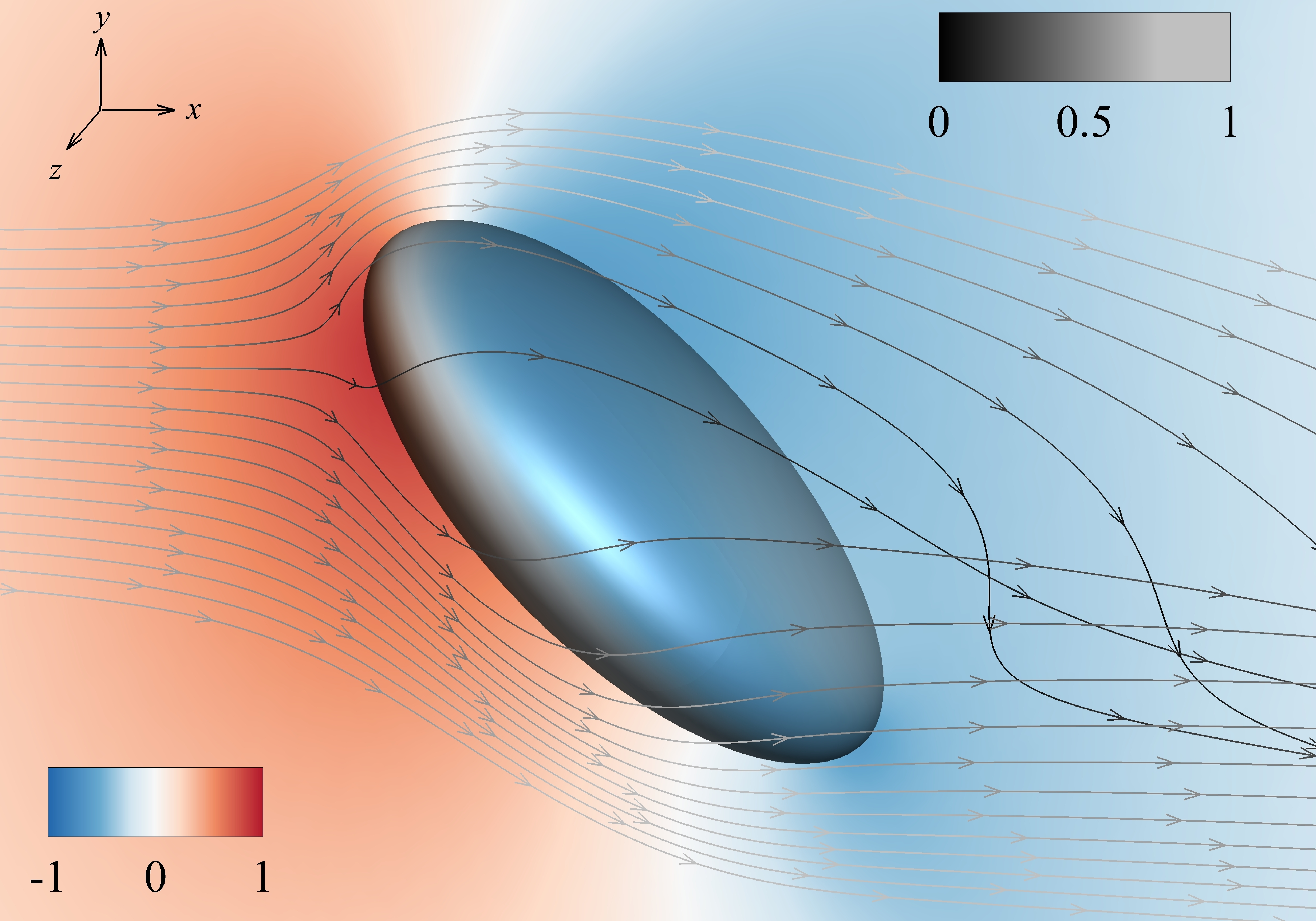}
		\put(0,65){(d)}
	\end{overpic}\\[1.0ex]
	\begin{overpic}[width=0.5\textwidth]{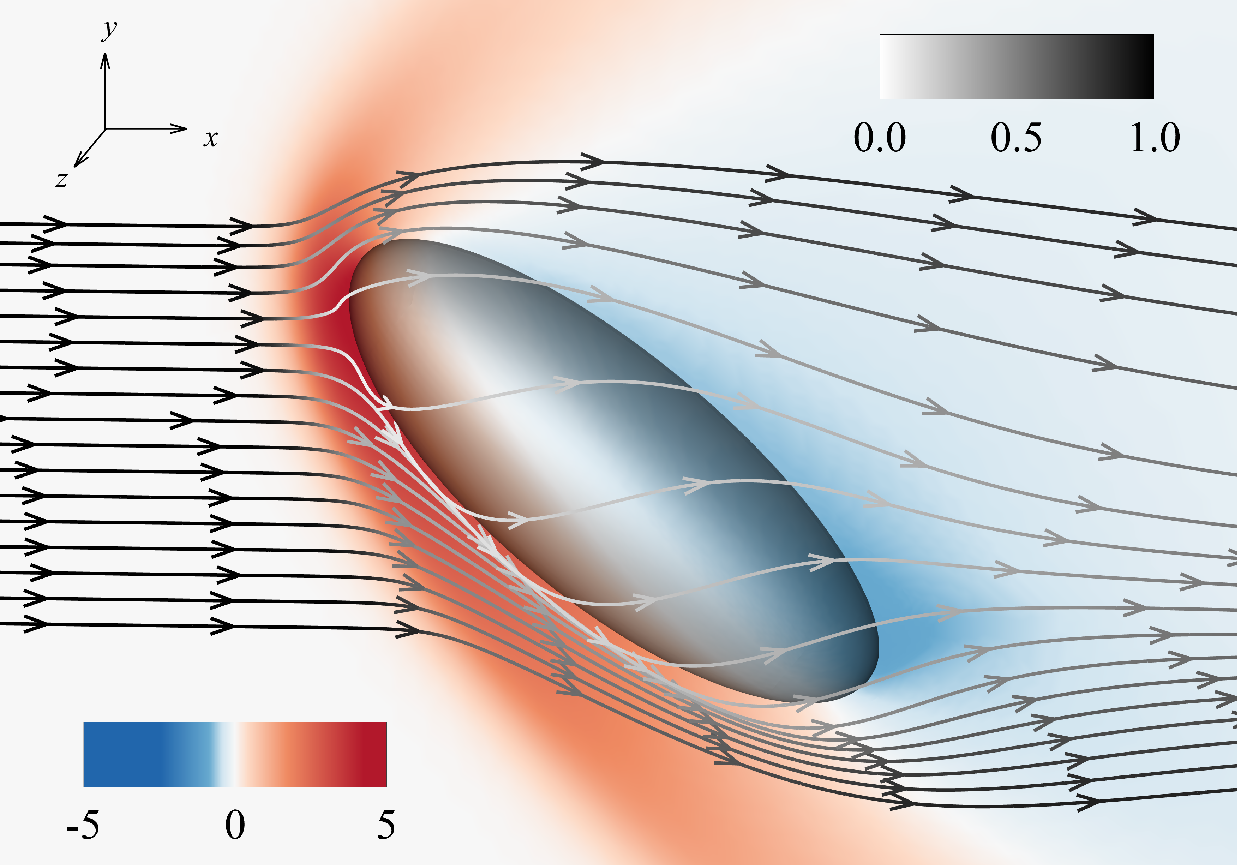}
		\put(0,65){(e)}
	\end{overpic}~
	\begin{overpic}[width=0.5\textwidth]{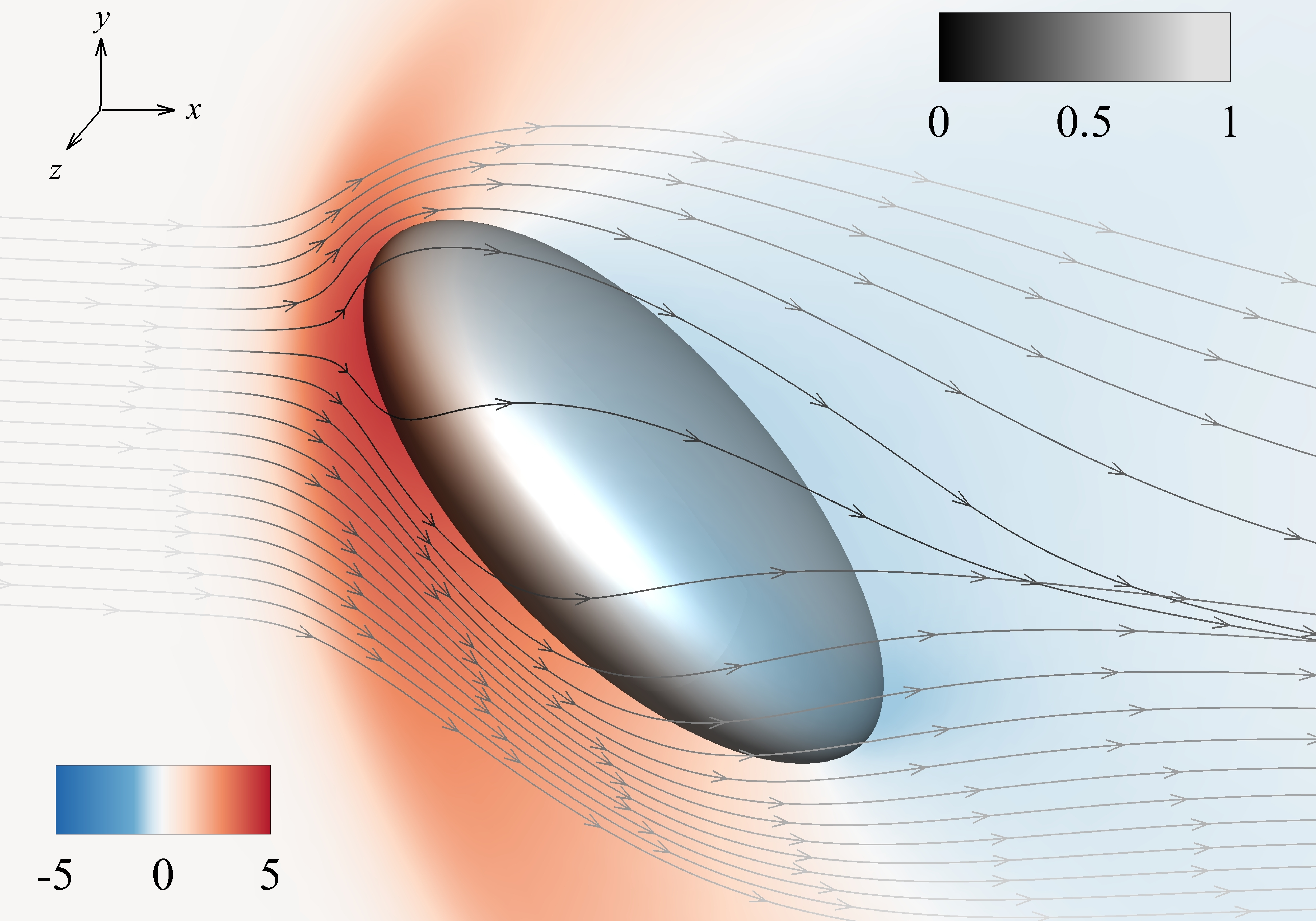}
		\put(0,65){(f)}
	\end{overpic}\\
	\caption{Distribution of relative pressure $(p-p_\infty)/p_\infty$  and the streamlines colored by streamwise velocity at $Re_p=50$ 
		and $\alpha=45^\circ$ at different Mach numbers,  (a,b) $M_p=0.3$, (c,d) $M_p=1$, (e,f) $M_p=2$,
		(a,c,e) $w=2.5$, (b,d,f) $w=0.4$.}
	\label{fig4:contour}
\end{figure}

As a first impression of the flow field, in Figure \ref{fig4:contour},
we present the distributions of the relative pressure $(p-p_\infty)/p_\infty$ and the streamlines near the spanwise symmetrical plane in the cases 
at $Re_p=50$ and $M_p$ = 0.3, 1.0 and 2.0 for particles with $w=2.5$ and 0.4 at $\alpha=45^\circ$. 
For prolate particles with $w=2.5$ at $M_p=0.3$, the streamlines originating from the upstream travel around the solid wall and 
extend to the leeward side of the particle, forming a small reverse flow region. 
The pressure is highest at the stagnation point and decreases in regions with higher shear rates, 
aligning with incompressible flow results \citep{ouchene_drag_2015,ouchene_new_2016,ouchene_numerical_2020,zastawny_derivation_2012,
sanjeevi_drag_2018, frohlich_correlations_2020}, indicating insignificant compressibility effects. 
The pressure distribution lacks symmetry about its main axis or mass center, implying the non-zero lift force and pitching torque. 
At $M_p=1.0$, a normal shock is evident due to increased flow compression, \textcolor{black}{as} indicated by a strong pressure gradient in the vicinity of the particle. 
At $M_p=2.0$, the shock wave detaches from the particle, creating a region of high pressure ahead of the particle and downstream of 
the shock wave, leading to high-pressure drag. Despite the identical Reynolds numbers $Re_p$, the reverse flow regions 
in these higher $M_p$ cases are notably smaller, suggesting that higher Mach numbers alleviate flow separation. 
Additionally, the asymmetrical pressure distribution highlights an enhanced pitching torque. 
Similar observations are made for oblate particles with $w=0.4$, except that the stronger flow separations and higher pressure behind 
the detached shock imply increased pressure drag, lift force, and pitching torque.

\subsection{Drag force}  \label{subsec:drag}

In this subsection, we investigate the variation of the drag force with the four key parameters stated above. 
It is necessary to recall first the empirical formula of the drag coefficient for incompressible flows.  
\citet{happel_low_1983} provided an analytical drag expression for \textcolor{black}{spheroidal} particles in creeping flows, 
which was \textcolor{black}{subsequently} summarized by Ouchene 
\citep{ouchene_new_2016,ouchene_numerical_2020}. 
For prolate \textcolor{black}{spheroidal} particles with $w>1$, the drag coefficient $C_D$ at the attack angle of $\alpha=0^\circ$ and 
$\alpha=90^\circ$ are \textcolor{black}{given by}
\begin{equation}
	C_{D,Stokes,\alpha	
	=0^\circ}(Re_p,w)=\frac{64}{Re_p}w^{-1/3}\left[\frac{-2w}{w^2-1}+\frac{2w^2-1}{(w^2-1)^{3/2}}ln\left(\frac{w+\sqrt{w^2-1}}{w-\sqrt{w^2-1}}\right)\right]^{-1},
	\label{3.1.1-cd0-stokes-prolate}
\end{equation}
\begin{equation}
	C_{D,Stokes,\alpha=90^\circ}(Re_p,w)=\frac{64}{Re_p}w^{-1/3}\left[\frac w{w^2-1}+\frac{2w^2-3}{(w^2-1)^{3/2}}ln(w+\sqrt{w^2-1})\right]^{-1},
	\label{3.1.2-cd90-stokes-prolate}
\end{equation}
while for oblate particles with $w<1$, the drag coefficient $C_D$ are expressed as follows,
\begin{equation}
	C_{D,Stokes,\alpha=0^\circ}(Re_p,w)=\frac{64}{Re_p}w^{-1/3}[\frac{-w}{1-w^2}-\frac{2w^2-3}{(1-w^2)^{3/2}}\arcsin(\sqrt{1-w^2})]^{-1},
	\label{3.1.3-cd0-stokes-oblate}
\end{equation}
\begin{equation}
	C_{D,Stokes,\alpha=90^\circ}(Re_p,w)=\frac{64}{Re_p}w^{-1/3}[\frac{2w}{1-w^2}+\frac{2(1-2w^2)}{(1-w^2)^{3/2}}\arctan(\frac{\sqrt{1-w^2}}w)]^{-1}.
	\label{3.1.4-cd90-stokes-oblate}
\end{equation}
At higher $Re_p$, Reynolds number effects should be considered. 
Our study primarily considers the formula proposed by \citet{frohlich_correlations_2020}, 
incorporating a correction factor $f_{d,\alpha}$ for prolate particles at $\alpha=0^\circ$ and $90^\circ$, 
\begin{equation}\label{3.1.5-cd0-prolate}
C_{D,\alpha=0^\circ} (Re_p,w) =C_{D,Stokes,\alpha=0^\circ}(Re_p,w)f_{d,\alpha=0^\circ}(Re_p,w),
\end{equation}
\begin{equation}\label{3.1.6-cd90-prolate}
C_{D,\alpha=90^\circ} (Re_p,w) =C_{D,Stokes,\alpha=90^\circ}(Re_p,w)f_{d,\alpha=90^\circ}(Re_p,w),
\end{equation}
where the correction factor $f_{d,\alpha}$ for the drag coefficient at $\alpha=0^\circ$ and $90^\circ$ of prolate particles are
\begin{equation}\label{3.1.7-fcd0}
f_{d,\alpha=0^\circ}(Re_p,w)=1+0.15Re_p^{0.687}+c_{d,1}(lnw)^{c_{d,2}Re_p^{c_{d,3}+c_{d,4}lnw}},
\end{equation}
\begin{equation}\label{3.1.8-fcd90}
f_{d,\alpha=90^\circ}(Re_p,w)=1+0.15Re_p^{0.687}+c_{d,5}(lnw)^{c_{d,6}Re_p^{c_{d,7}+c_{d,8}lnw}},
\end{equation}
and for oblate particles, $w$ should be replaced by $1/w$. 
The drag coefficient exhibits a sinusoidal function squared trend for attack angles between $0^\circ$ and $90^\circ$, 
\begin{equation}\label{3.1.9-cd-incom}
	C_{D}(Re_p,w,\alpha)=C_{D,\alpha=0^\circ}(Re_p,w)+(C_{D,\alpha=90^\circ}(Re_p,w)-C_{D,\alpha=0^\circ}(Re_p,w))\sin^2(\alpha)
\end{equation}
\textcolor{black}{
which is found to have a wide range of applicability up to $Re_p = 2000$ for 
prolate spheroid~\citep{sanjeevi_2017_orientational},
and maintain a specific degree of precision for oblate particles as
validated by the present simulations.}
As $w$ equals to 1, the formula above degenerates to the drag coefficients of spherical particles given by \citet{schiller_uber_1933}. 
The parameters in Eq.~\eqref{3.1.7-fcd0} and \eqref{3.1.8-fcd90} are detailed in Table \ref{tab3:cd-incom}, 
which are obtained by data fitting using the present numerical simulation results. 
The relative errors are given in Table \ref{tab4:err-cd-incom}.
\textcolor{black}{In comparison to the predictions provided by previous formulas, 
the empirical law presented here exhibits significantly lower relative errors than
those proposed by~\citet{ouchene_new_2016,ouchene_numerical_2020} and \cite{zastawny_derivation_2012}
for all cases considered.
The formula by \citet{sanjeevi_drag_2018} shows lower mean and maximum relative errors
for prolate particles, but higher errors for oblate particles, which should probably attributed
to the fact that the formula for former was originally proposed only for $w = 2.5$.}

\begin{table}[tp!]
	\centering
	\small
	\caption{Parameters $c_{d,i}$ in Eq.~\eqref{3.1.7-fcd0} and  \eqref{3.1.8-fcd90}.}
	\label{tab3:cd-incom}
	\begin{tabular}{lclclclclclclclcl} 
		\hline
		~ & $i=1$      & $i=2$     & $i=3$    & $i=4$     & $i=5$     & $i=6$    & $i=7$    & $i=8$      \\ 
		\hline
		$w>1$ & -0.0045 & 1.742   & 1.436 & -0.2181 & 0.05274 & 0.697  & 0.5813 & 0.0879  \\ 
		$w<1$ & 585    & -0.8002 & -4.797 & 0.6141  & 0.0536  & 0.5335 & 0.6186 & 0.1648   \\
		\hline
	\end{tabular}
\end{table}

\begin{table}[tp!]
	\centering
	\small
	\caption{Comparison of the relative errors of the empirical formulas for $C_D$ at $M_p=0.1$.}
	\label{tab4:err-cd-incom}
	\begin{tblr}{			
			cell{2}{1} = {r=2}{},
			cell{4}{1} = {r=2}{},
			hline{1-2,4,6} = {-}{},
			colspec = {Q[c] Q[c] Q[c] Q[c] Q[c] Q[c]}
		}
		\textbf{~} & ~    & Present  & \makecell{{\citet{ouchene_new_2016}}\\{\citet{ouchene_numerical_2020}}}   & \citet{zastawny_derivation_2012} & \citet{sanjeevi_drag_2018}\\
		$w>1$          & Mean(\%) & 2.23    & 12.12         & 12.37          & 1.69           \\
		& Max(\%)  & 8.49    & 29.28         & 19.75          & 2.46           \\
		$w<1$          & Mean(\%) & 1.60    & 7.31         & 13.86          & 3.27           \\
		& Max(\%)  & 15.22   & 22.41         & 29.00          & 15.72          
	\end{tblr}
\end{table}

We now start to incorporate compressibility effects into this formula. 
One approach is to introduce an additional correlation factor into the existing formulas. 
\textcolor{black}{Nevertheless}, research by \citet{loth_supersonic_2021} on compressible spherical particles reveals a coupling 
between the particle Reynolds number $Re_p$ and Mach number $M_p$. 
For the current spheroid particles under consideration, \textcolor{black}{likely}, $w$ is also linked to $M_p$. 
To clarify this relationship, we perform Spearman correlation analysis using all simulation data. 
The Spearman correlation coefficients demonstrate the consistent changes in drag coefficients with 
these flow parameters, where $\pm 1$ signifies a perfect monotonic correlation, whereas a value of 0 signifies the absence of a correlation.
The findings, presented in Table \ref{tab5:relation-cd}, suggest that both $Re_p$ and $M_p$ are significant factors 
to be included in the empirical formula for drag force coefficients, while $w$ is comparatively less influential.

\begin{table}[tp!]
	\centering
	\caption{Spearman correlation analysis for drag force coefficients.}
	\small
	\label{tab5:relation-cd}
	\begin{tblr}{
			cell{2}{1} = {r=2}{},
			cell{4}{1} = {r=2}{},
			cell{6}{1} = {r=2}{},
			cell{8}{1} = {r=2}{},
			hline{1-2,4,6,8,10} = {-}{},
			colspec={Q[c] Q[c] Q[c] Q[c] Q[c]}
		}
		~                                 & ~     & $Re_p$    & $M_p$     & $w$     \\
		{$w<1$\\$M_p\leq1$}             & $C_{D,\alpha=0}$  & 0.097  & 0.973  & 0.125  \\
		& $C_{D,\alpha=90}$ & 0.185  & 0.943  & -0.047 \\
		{$w<1$\\$M_p>1$}  & $C_{D,\alpha=0}$  & 0.854  & -0.210 & 0.079  \\
		& $C_{D,\alpha=90}$ & 0.894  & -0.144 & -0.079 \\
		{$w>1$\\$M_p\leq1$} & $C_{D,\alpha=0}$  & 0.400  & 0.633  & -0.059 \\
		& $C_{D,\alpha=90}$ & 0.537  & 0.590  & -0.042 \\
		{$w>1$\\$M_p>1$}  & $C_{D,\alpha=0}$  & -0.944 & 0.104  & -0.234 \\
		& $C_{D,\alpha=90}$ & 0.959  & -0.141 & -0.133 
	\end{tblr}
\end{table}

We analyze the influence factor mathematically by considering its relationship with various variables,
\begin{equation}\label{3.1.10-gma-cd0}
	g_{d,\alpha=0^\circ}(Re_p,M_p)=d_{0,1}Re_p^{d_{0,2}M_p^{d_{0,3}}}
\end{equation}
\begin{equation}\label{3.1.11-gma-cd90}
g_{d,\alpha=90^\circ}(Re_p,M_p)=d_{90,1}Re_p^{d_{90,2}M_p^{d_{90,3}}}
\end{equation}
The drag coefficients $C_{D,\alpha=0^\circ}$ and $C_{D,\alpha=90^\circ}$ for spheroid particles in compressible flows 
are, therefore, expressed as 
\begin{equation}\label{3.1.12-com-cd0}
	C_{D,\alpha=0^\circ}(Re_p,w,M_p)=C_{D,\alpha=0^\circ}(Re_p,w)g_{d,\alpha=0^\circ}(Re_p,M_p)
\end{equation}
\begin{equation}\label{3.1.13-com-cd90}
	C_{D,\alpha=90^\circ}(Re_p,w,M_p)=C_{D,\alpha=90^\circ}(Re_p,w)g_{d,\alpha=90^\circ}(Re_p,M_p)
\end{equation}
with $C_{D,\alpha=0^\circ}(Re_p,w)$ and $C_{D,\alpha=90^\circ} (Re_p,w)$ representing the empirical formulas for incompressible flows,
as given in Eq.~\eqref{3.1.5-cd0-prolate} and \eqref{3.1.6-cd90-prolate}. 
For other attack angles, $C_{D}$ follows a sinusoidal function squared trend as detailed 
in Eq.~\eqref{3.1.9-cd-incom}. To sum up, $C_D$ formula can be written as
\begin{equation}\label{3.1.10-cd-com}
	\begin{split}
	C_{D}(Re_p,w,M_p,\alpha) & =  C_{D,\alpha=0^\circ}(Re_p,w,M_p) \\ 
	&+(C_{D,\alpha=90^\circ}(Re_p,w,M_p)-C_{D,\alpha=0^\circ}(Re_p,w,M_p))\sin^2(\alpha)
	\end{split}
\end{equation}
As $M_p$ approaches 0, \textcolor{black}{the values of} $g_{d,\alpha=0^\circ}$ and $g_{d,\alpha=90^\circ}$ converge to 1.0, leading to a transition to 
incompressible flow scenarios. 
The formula parameters are provided in Table \ref{tab6:cd-para-com}.

\begin{table}[tp!]
	\centering
	\small
	\caption{Parameters $d_{0,i}$ and $d_{90,i}$ in Eq.~\eqref{3.1.10-gma-cd0} and Eq.~\eqref{3.1.11-gma-cd90} for drag coefficients and \textcolor{black}{average} relative errors.}
	\label{tab6:cd-para-com}
	\begin{tblr}{
			cell{2}{1} = {r=2}{},
			cell{2}{9} = {r=2}{},
			cell{4}{1} = {r=2}{},
			cell{4}{9} = {r=2}{},
			hline{1-2,4,6} = {-}{},
			colspec={Q[c] Q[c] Q[c] Q[c] Q[c] Q[c] Q[c] Q[c] Q[c]}
		}
		~  & ~  & $d_{0,1}$    & $d_{0,2}$     & $d_{0,3}$    & $d_{90,1}$   & $d_{90,2}$   & $d_{90,3}$     &  \makecell{{\textcolor{black}{Average} relative}\\{error(\%)}} \\
		$w>1$ & $M_p\leq1$ & 1      & 0.07258 & 3.674  & 1      & 0.1022 & 2.792    & 3.11               \\
		& $M_p>1$ & 0.949  & 0.08366 & 0.1941 & 0.7131 & 0.1881 & -0.06457 &                    \\
		$w<1$ & $M_p\leq1$ & 1      & 0.0788  & 3.486 & 1      & 0.1091 & 2.8466  & 2.80               \\
		& $M_p>1$ & 0.8677 & 0.1218  & -0.5433 & 0.7598 & 0.181  & -0.1093   &                    
	\end{tblr}
\end{table}

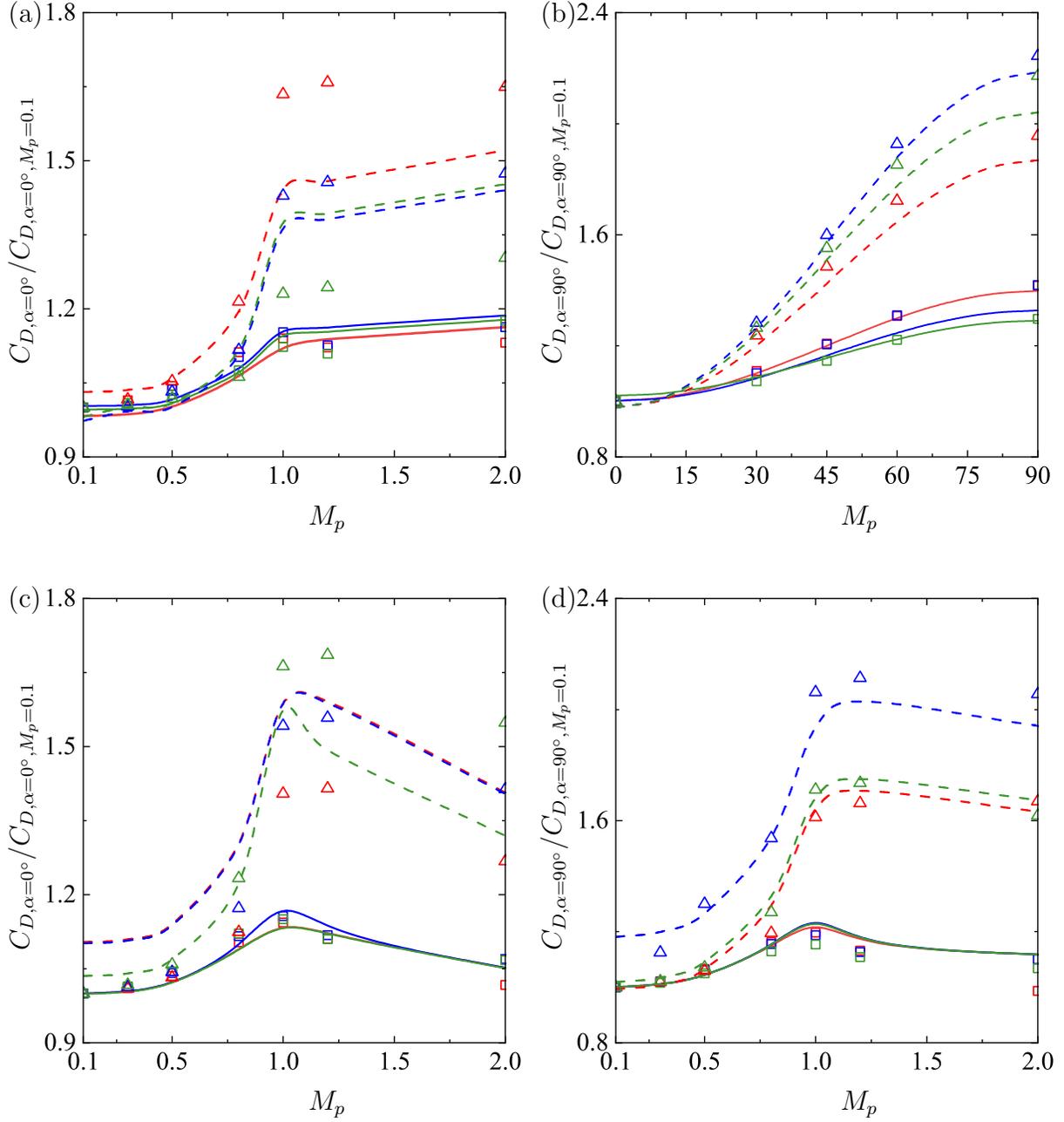
\begin{figure}[tp!]
	\centering
	\begin{overpic}[width=0.5\textwidth]{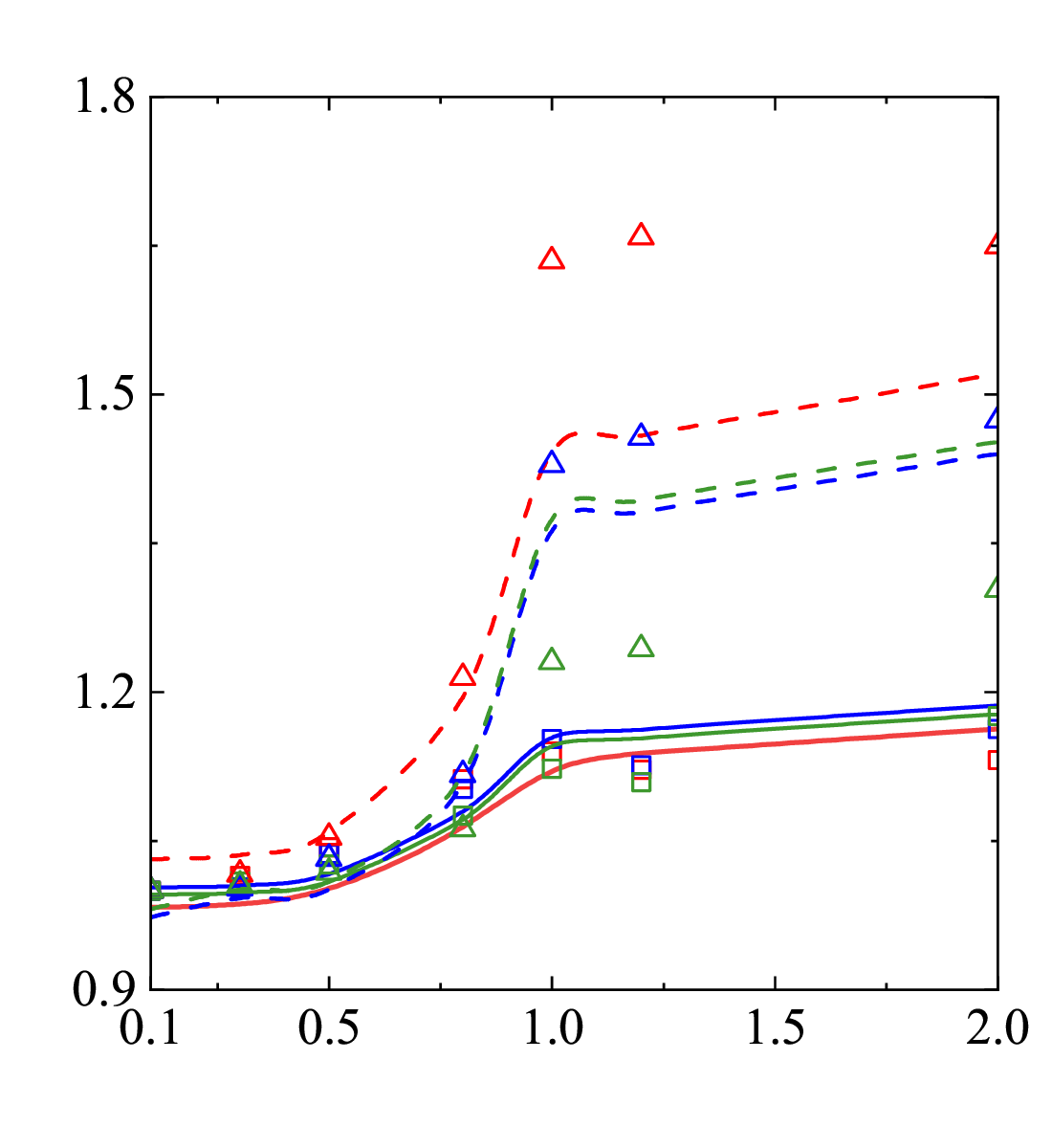}
		\put(0,90){(a)}
		\put(0,30){\rotatebox{90}{\textcolor{black}{$C_{D,\alpha=0^\circ}/C_{D,\alpha=0^{\circ},M_p=0.1}$}}}
		\put(52.5,2){$M_p$}
	\end{overpic}~
	\begin{overpic}[width=0.5\textwidth]{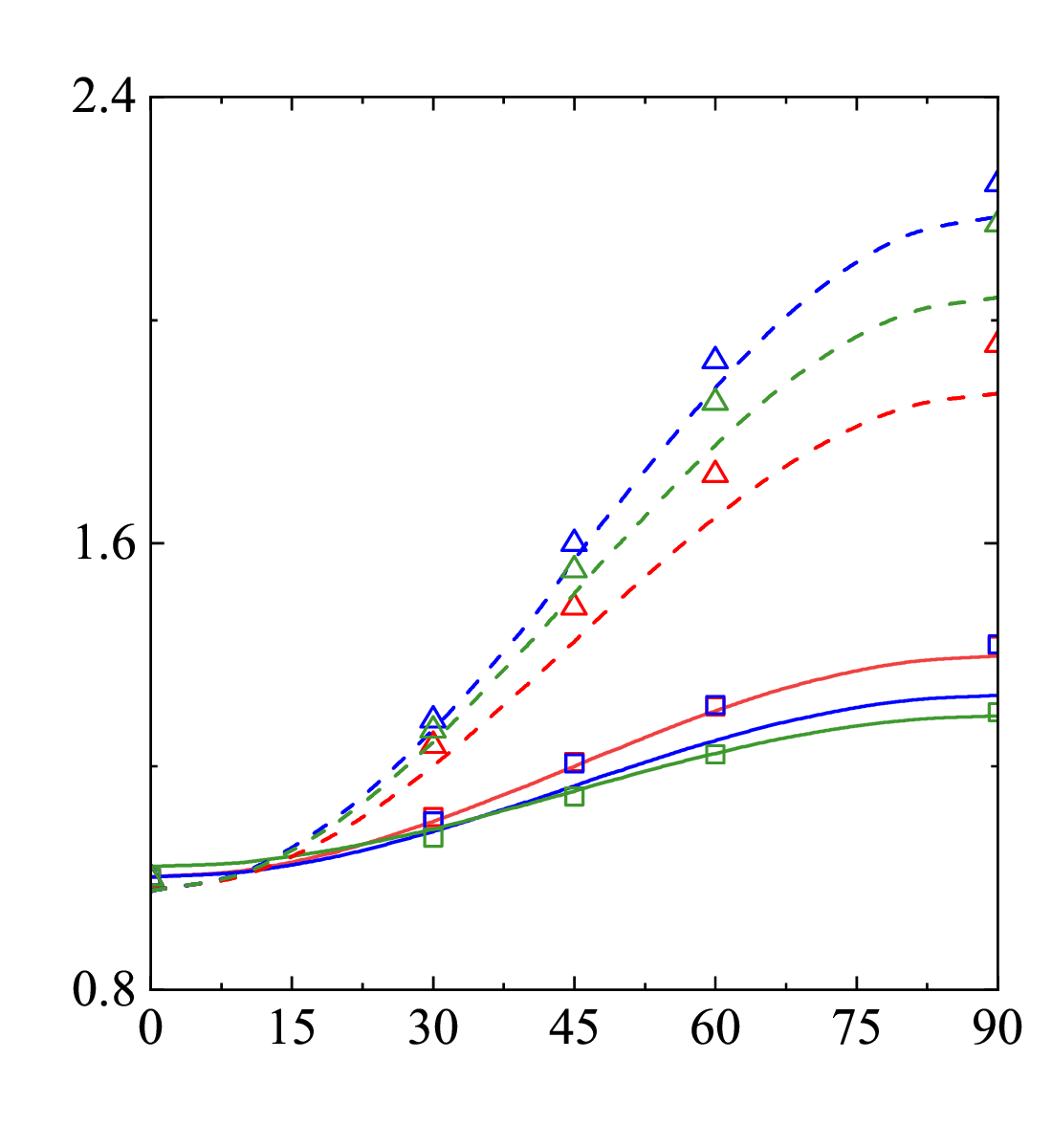}
		\put(0,90){(b)}
		\put(0,30){\rotatebox{90}{\textcolor{black}{$C_{D,\alpha=90^\circ}/C_{D,\alpha=90^{\circ},M_p=0.1}$}}}
		\put(52.5,2){$M_p$}
	\end{overpic}\\[1.0ex]
	\begin{overpic}[width=0.5\textwidth]{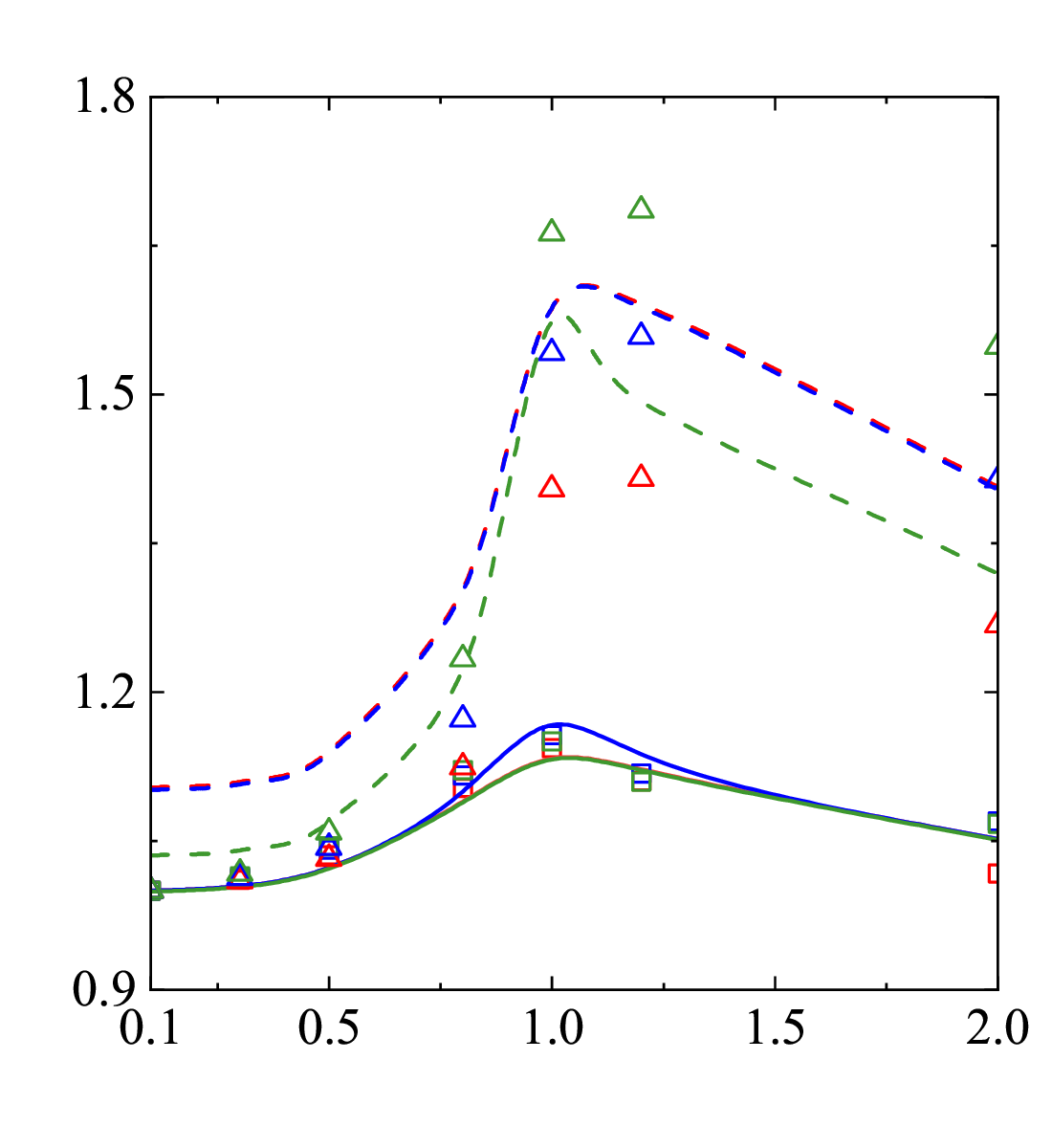}
		\put(0,90){(c)}
		\put(0,30){\rotatebox{90}{\textcolor{black}{$C_{D,\alpha=0^\circ}/C_{D,\alpha=0^{\circ},M_p=0.1}$}}}
		\put(52.5,2){$M_p$}
	\end{overpic}~
	\begin{overpic}[width=0.5\textwidth]{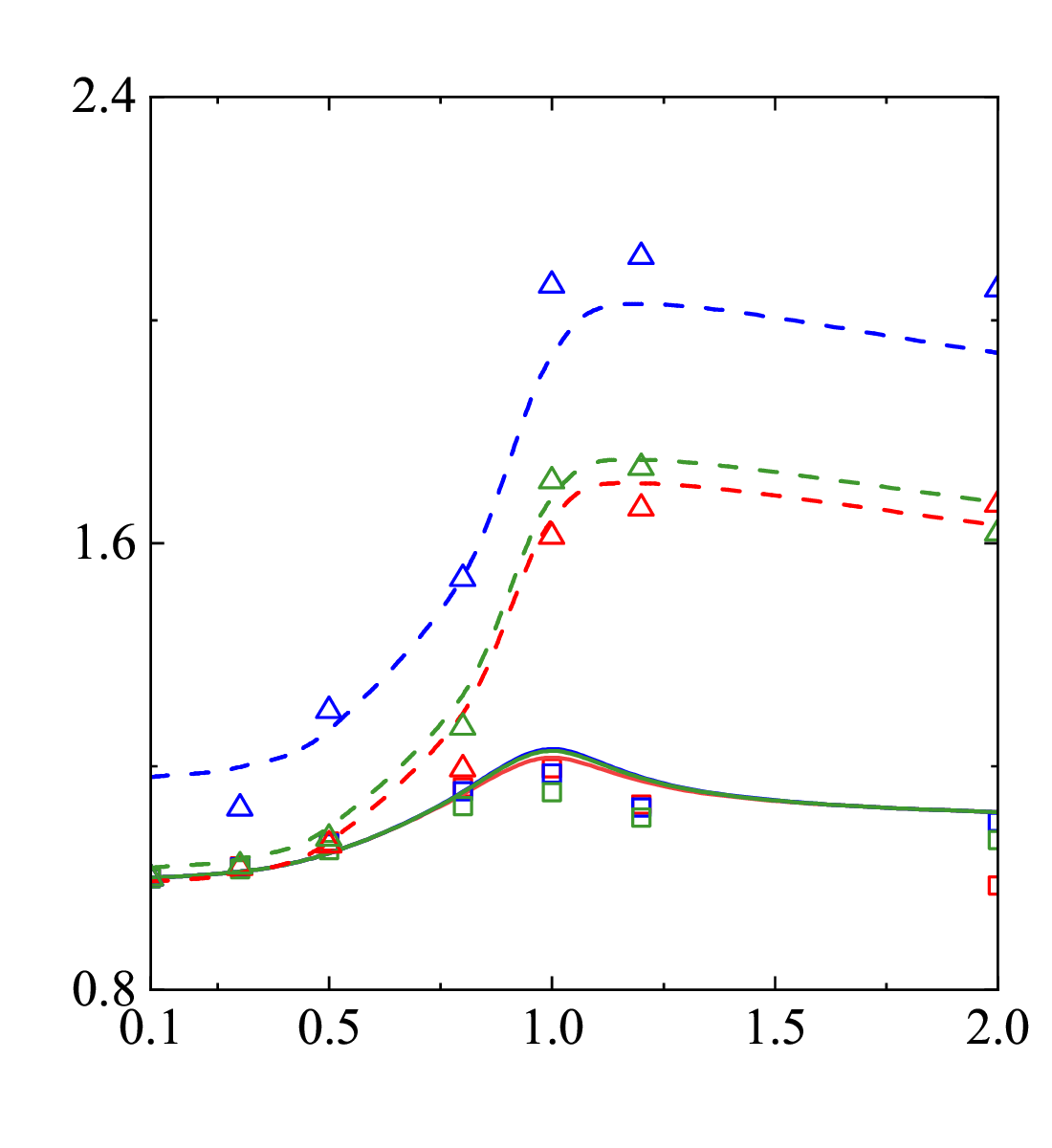}
		\put(0,90){(d)}
		\put(0,30){\rotatebox{90}{\textcolor{black}{$C_{D,\alpha=90^\circ}/C_{D,\alpha=90^{\circ},M_p=0.1}$}}}
		\put(52.5,2){$M_p$}
	\end{overpic}\\
	\caption{Variation of particle drag coefficients (a,c) \textcolor{black}{$C_{D,\alpha=0^\circ}/C_{D,\alpha=0^{\circ},M_p=0.1}$} and (b,d) \textcolor{black}{$C_{D,\alpha=90^\circ}/C_{D,\alpha=90^{\circ},M_p=0.1}$} against $M_p$, 
	(a,b) prolate particles: $w=1.25$ (red), $w=2.5$ (blue),  $w=5$ (green)
	and (c,d) oblate particles: $w=0.2$ (red), $w=0.4$ (blue),  $w=0.8$ (green). 
	Solid lines and squares: $Re_p=10$, dashed lines and triangles: $Re_p=100$. Lines: (a,c) Eq.~\eqref{3.1.12-com-cd0}, (b,d) Eq.~\eqref{3.1.13-com-cd90} , symbols: present simulations.}
	\label{fig5:cd-ma}
\end{figure}

In Figure \ref{fig5:cd-ma}, we illustrate the variation of drag coefficients with Mach numbers at $Re_p=10$ 
and 100 for particles of different aspect ratios $w$. 
Across all shapes, drag coefficients exhibit a monotonic increase at $M_p<1$ but experience a sudden decline 
at the critical point $M_p=1$. 
Beyond $M_p=1$, drag coefficients at $\alpha=0^\circ$ for prolate particles display a non-monotonic trend with $M_p$, 
initially decreasing and then increasing, while other drag coefficients decrease monotonically at $Re_p=10$. 
A similar trend of variation is observed for cases at $Re_p$ = 100.
The proposed empirical formulas accurately capture these trends, with an average relative error of approximately 3\%, 
indicating their capability to predict drag coefficients for particles of varying aspect ratios $w$ in terms of $M_p$.

\begin{figure}[tp!]
	\centering
	\begin{overpic}[width=0.5\textwidth]{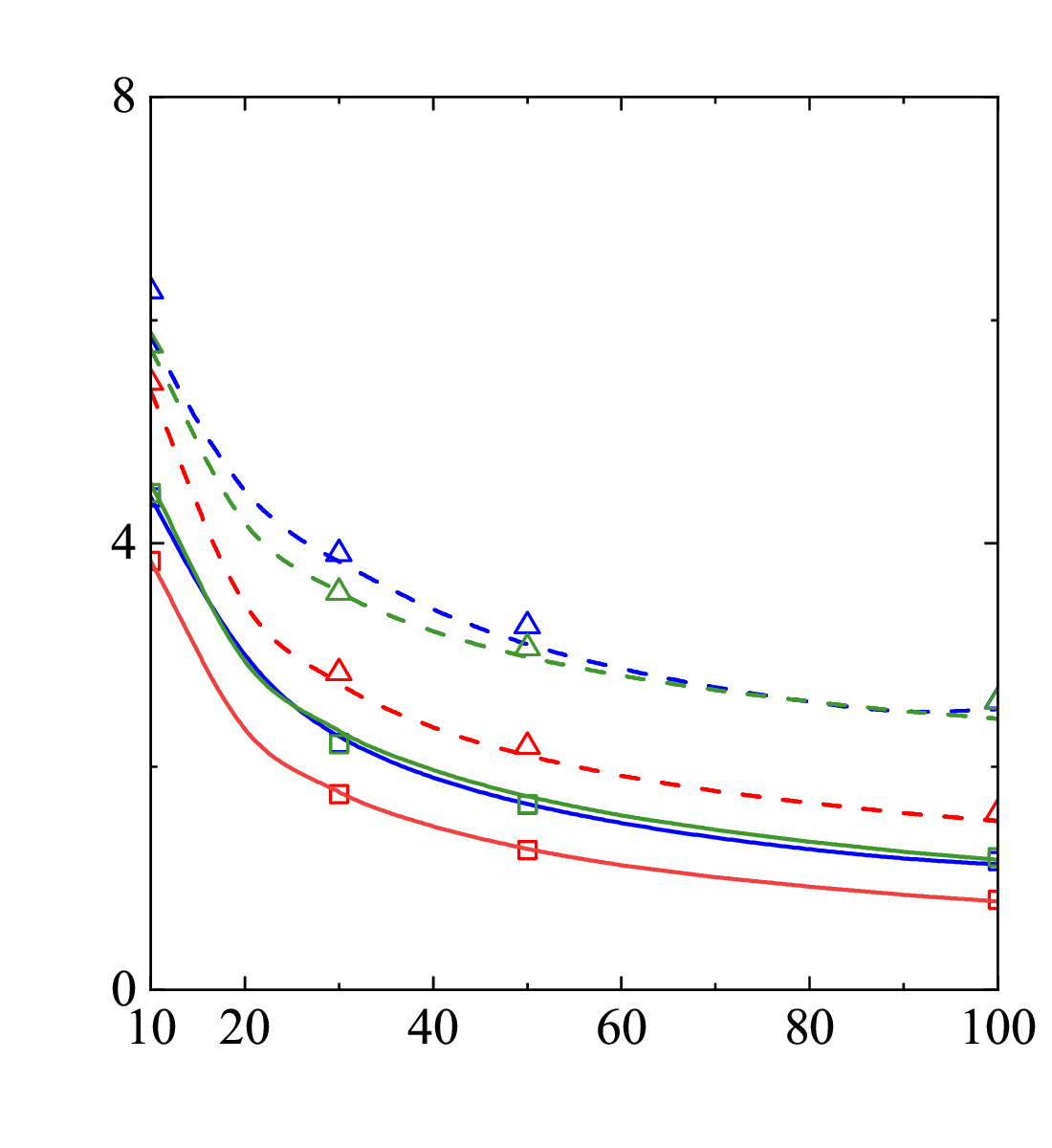}
		\put(0,90){(a)}
		\put(2,50){\rotatebox{90}{$C_D$}}
		\put(50,2){$Re_p$}
	\end{overpic}~
	\begin{overpic}[width=0.5\textwidth]{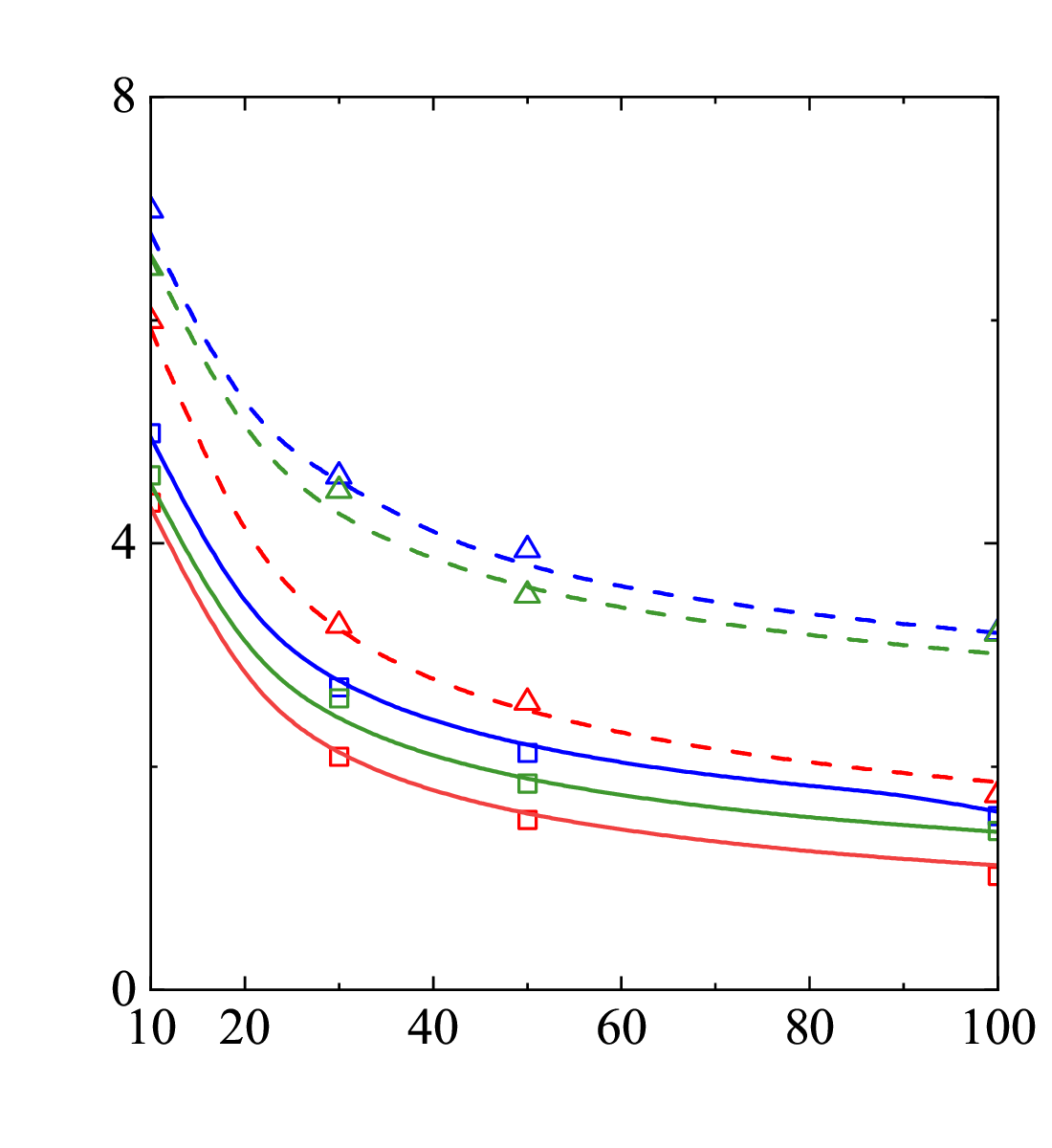}
		\put(0,90){(b)}
		\put(2,50){\rotatebox{90}{$C_D$}}
		\put(50,2){$Re_p$}
	\end{overpic}\\
	\caption{Variation of particle drag coefficients $C_D$ against $Re_p$, (a) $w=2.5$ and (b) $w=0.4$. 
	Solid lines and squares: $C_{D,\alpha=0^\circ}$, dashed lines and triangles: $C_{D,\alpha=90^\circ}$ for
	$M_p=0.3$ (red), $M_p=1.0$ (blue),  $M_p=2.0$ (green).  
	Solid lines: Eq.~\eqref{3.1.12-com-cd0} , dashed lines: Eq.~\eqref{3.1.13-com-cd90}, symbols: present simulations.
	}

	\label{fig6:cd-re}
\end{figure}

Figure \ref{fig6:cd-re} \textcolor{black}{illustrates} the variation of the drag coefficient with the particle Reynolds number $Re_p$ 
for prolate and oblate particles with aspect ratios $w=2.5$ and 0.4, respectively, grouped by three $M_p$ values. 
Across all particle aspect ratios $w$ and $M_p$ considered, 
the drag coefficients decrease monotonically with $Re_p$, 
which is well captured by the proposed formula.
This behavior holds true for particles with different aspect ratios, affirming the validity of the proposed
formula concerning $Re_p$.

\begin{figure}[tp!]
	\centering
	\begin{overpic}[width=0.5\textwidth]{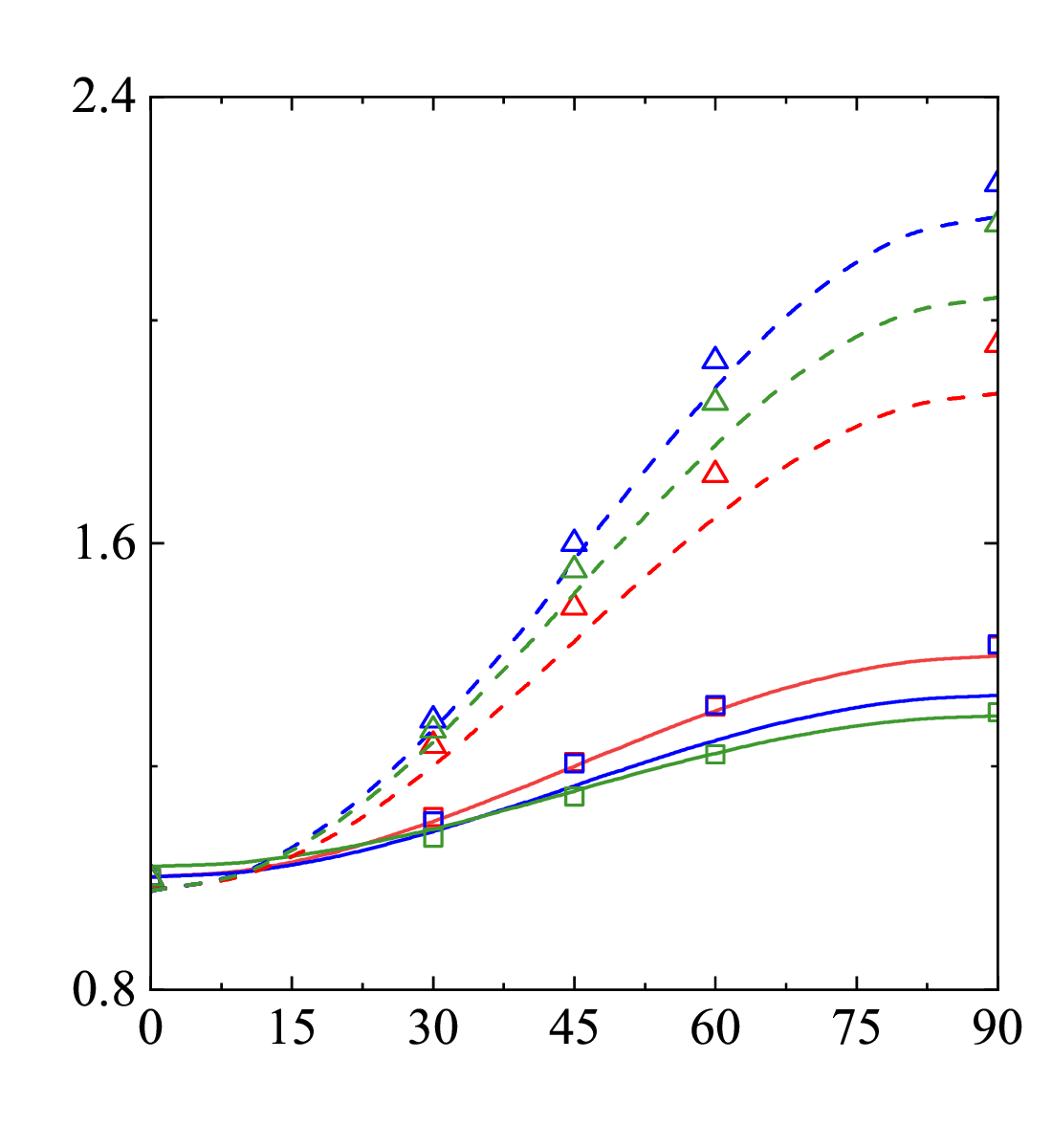}
		\put(0,90){(a)}
		\put(0,45){\rotatebox{90}{\textcolor{black}{$C_D/C_{D,\alpha=0^\circ}$}}
		\put(45,-40){$\alpha^\circ$}}
	\end{overpic}~
	\begin{overpic}[width=0.5\textwidth]{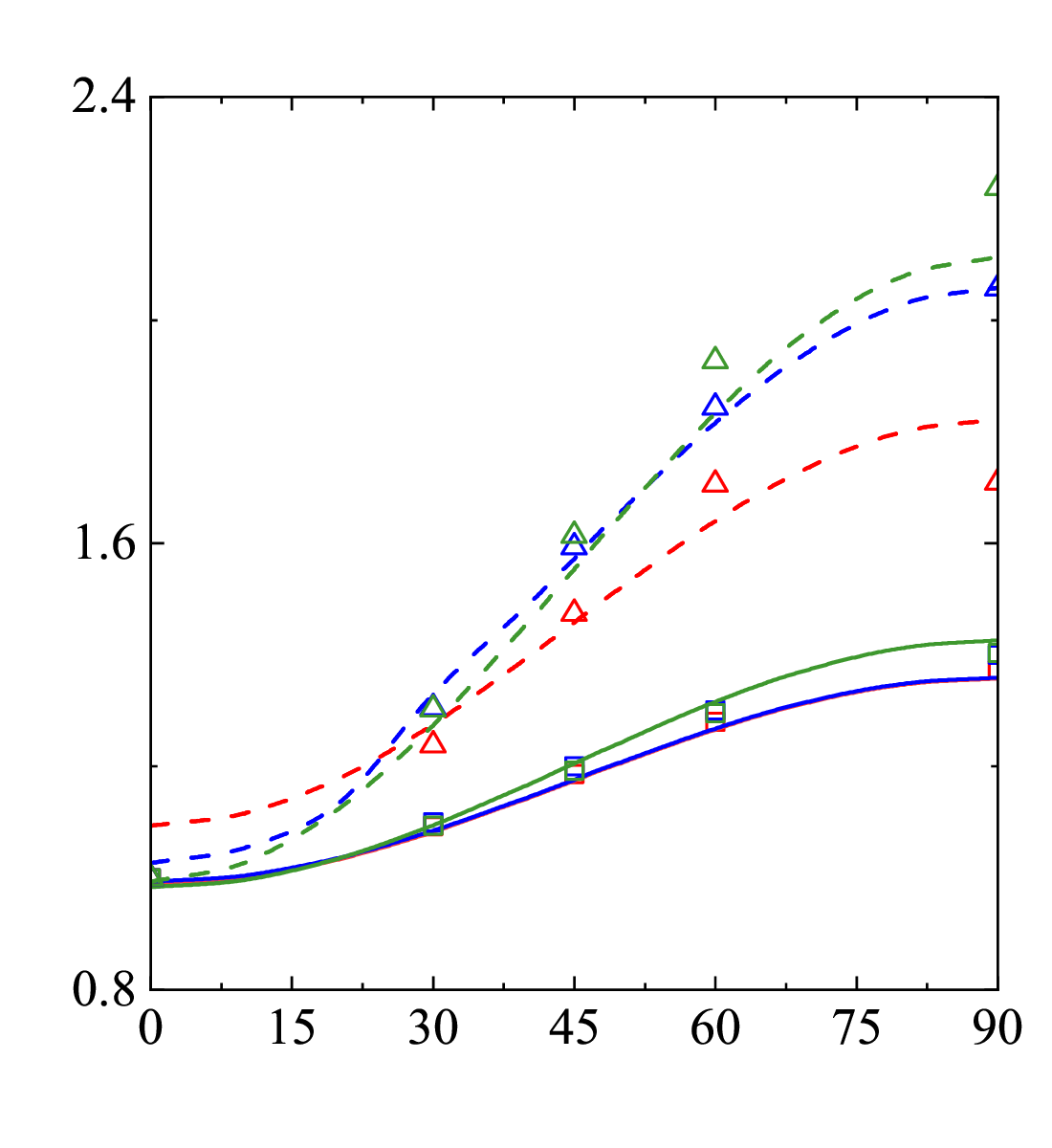}
		\put(0,90){(b)}
		\put(0,45){\rotatebox{90}{\textcolor{black}{$C_D/C_{D,\alpha=0^\circ}$}}
		\put(45,-40){$\alpha^\circ$}}
	\end{overpic}\\
	\caption{Variation of particle drag coefficients \textcolor{black}{$C_D/C_{D,\alpha=0^\circ}$} against $\alpha$, 
		(a) $w=2.5$ and 
		(b) $w=0.4$ for
		 $M_p=0.3$ (red), $M_p=1.0$ (blue),  $M_p=2.0$ (green).  
		Solid lines and squares: $Re_p=10$, 
		dashed lines and triangles: $Re_p=100$.
		Lines: Eq.~\eqref{3.1.10-cd-com}, symbols: present simulations.}
	\label{fig7:cd-a}
\end{figure}

Lastly, we examine the impact of the angle of attack on the drag coefficient at various Mach numbers, 
shown in Figure \ref{fig7:cd-a}. 
The trend in drag coefficient variation is consistent for both compressible and incompressible flows, 
suggesting that the attack angle $\alpha$ plays a minor role in drag coefficient modification. 
Overall, the predicted $C_D$ using the proposed formula aligns well with simulation results, 
validating the use of the sinusoidal function squared to derive $C_D$ at any arbitrary attack angle $\alpha$ \textcolor{black}{from 0$^\circ$ to 90$^\circ$}. 

In this subsection, we focus on adjusting drag coefficients for spheroid particles in compressible flows. 
Through an examination of drag coefficient variations with $Re_p$, $M_p$, $w$, and $\alpha$, 
we demonstrate that the proposed empirical formulas closely match simulation results, with an average relative error of 
less than 3\%.

\subsection{Lift force} \label{subsec:lift}

We further consider the lift force in this subsection.  
Theoretical derivations and empirical formulas for the lift force coefficients are usually 
associated with the modeling of their drag forces \citep{ouchene_new_2016,ouchene_numerical_2020,frohlich_correlations_2020}, 
\citet{happel_low_1983} derived lift coefficients for \textcolor{black}{spheroidal} particles in stokes flow as follows
\begin{equation}
	\label{3.2.1-cl-stokes}
	C_{L,Stokes}=(C_{D,Stokes,\alpha=90^\circ}-C_{D,Stokes,\alpha=0^\circ})\sin(\alpha)\cos(\alpha)
\end{equation}
coupling the drag coefficients \textcolor{black}{$C_{D,Stokes,\alpha=0^\circ}$ and $C_{D,Stokes,\alpha=90^\circ}$} 
with trigonometric function $\sin(\alpha)\cos(\alpha)$ to determine the lift coefficients in creeping flows. 
To incorporate high Reynolds number effects, the maximum value of the lift coefficient reached at $\alpha=45^\circ$ should be modified by multiplying the correction factor \citep{frohlich_correlations_2020} as
\begin{equation}
	\label{3.2.2-clmax-incom}
	C_{L,\alpha=45^\circ}(Re_p,w)=(C_{D,Stokes,\alpha=90^\circ}-C_{D,Stokes,\alpha=0^\circ})f_{l,\alpha=45^\circ}(Re_p,w)
\end{equation}
with the correction factor
\begin{equation}
	\label{3.2.3-fclmax-incom}
	f_{l,\alpha=45^\circ}(Re_p,w)=1+c_{l,3}Re_p^{c_{l,4}+c_{l,5}\ln w}w^{c_{l,6}+c_{l,7}Re_p}
\end{equation}
In creeping flows, the maximum lift coefficients are \textcolor{black}{observed at $\alpha=45^\circ$ and exhibit a symmetrical distribution as the angle} approaches 
$0^\circ$ and $90^\circ$.  However, such a symmetrical variation is violated at higher Reynolds numbers. 
To account for the unsymmetrical shift, we propose to use the following expression,
\begin{equation}
	\label{3.2.4-cl-incom}
	C_L(Re_p,w,\alpha)=2(\sin \alpha)^{c_{l,1}^{Re_pw}}(\cos \alpha)^{c_{l,2}^{Re_pw}} C_{L,\alpha=45^\circ}(Re_p,w)
\end{equation}
which is based on the results of our simulation and previous studies \textcolor{black}{in the field}. Note that these formulas are valid for prolate particles. 
Those for the oblate particles can be obtained by replacing the $w$ as $1/w$, 
same as the procedure employed for drag force coefficients. 
The correlation for lift coefficient proposed above has the same form as the theoretical solution in Eq.~\eqref{3.2.1-cl-stokes} when the particle Reynolds number $Re_p$ approaches zero. 
The parameters in the formulas are listed in Table \ref{tab7:cl-incom} obtained by the best fitting using the results at $M_p=0.1$.

\begin{table}[tp!]
	\centering
	\caption{Parameters $c_{l,i}$ in Eq.~\eqref{3.2.3-fclmax-incom} and Eq.~\eqref{3.2.4-cl-incom}.}
	\label{tab7:cl-incom}
	\begin{tabular}{cccccccc} 
		\hline
		~ & $i=1$      & $i=2$     & $i=3$    & $i=4$     & $i=5$     & $i=6$    & $i=7$          \\ 
		\hline
		$w>1$ & 1.0005 & 0.9995   & 0.3402 & 0.8787 & -0.0493 & 0.0 & 0.0   \\ 
		$w<1$ & 0.9998    & 1.0002 & 0.0600 & 1.1333  & 0.0905  & 0.1908 & -0.0037    \\
		\hline
	\end{tabular}
\end{table}

\begin{figure}[tp!]
	\centering
	\begin{overpic}[width=0.5\textwidth]{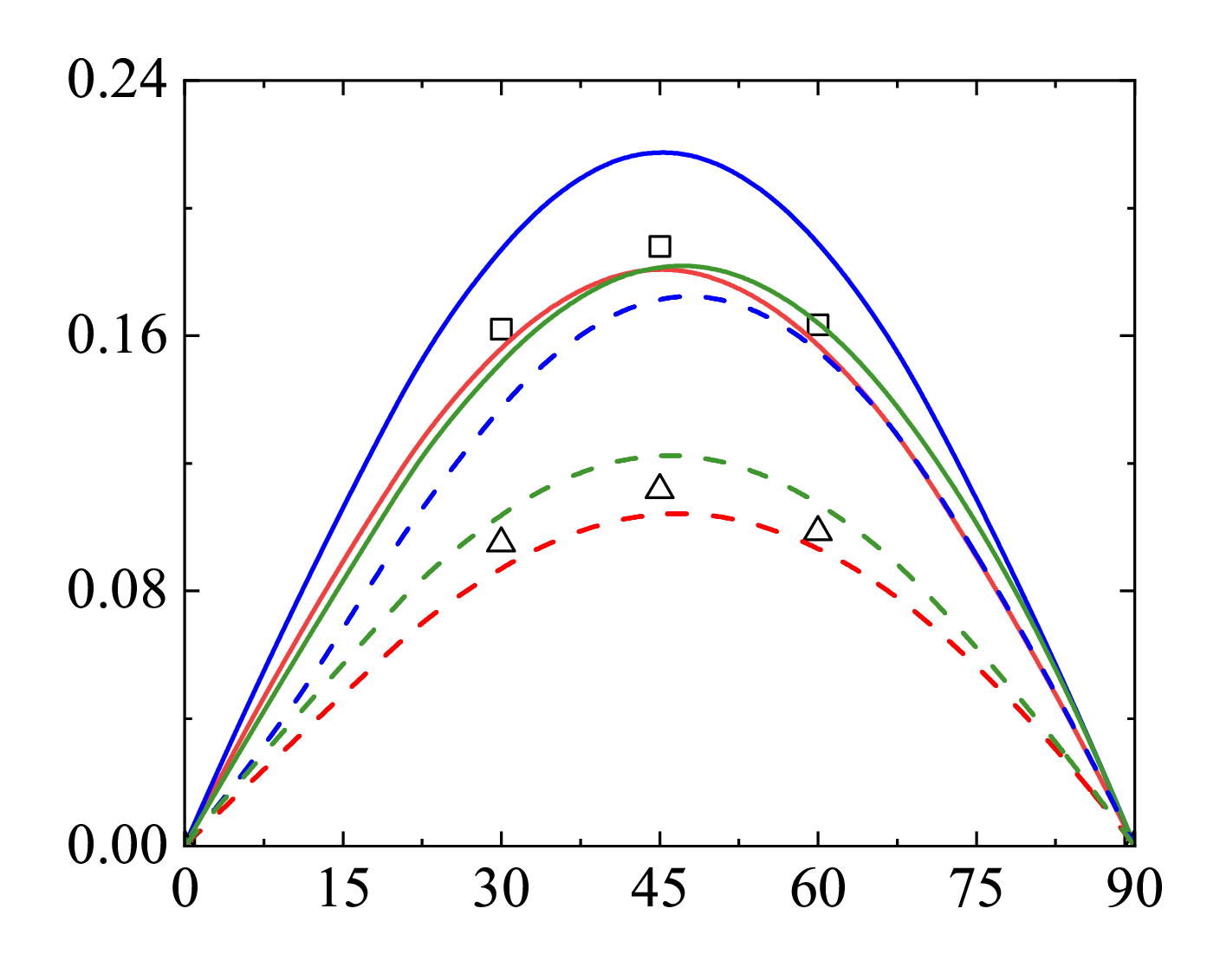}
		\put(-1,70){(a)}
		\put(2,40){\rotatebox{90}{$C_L$}}
		\put(52.5,0){$\alpha^\circ$}		
	\end{overpic}~
	\begin{overpic}[width=0.5\textwidth]{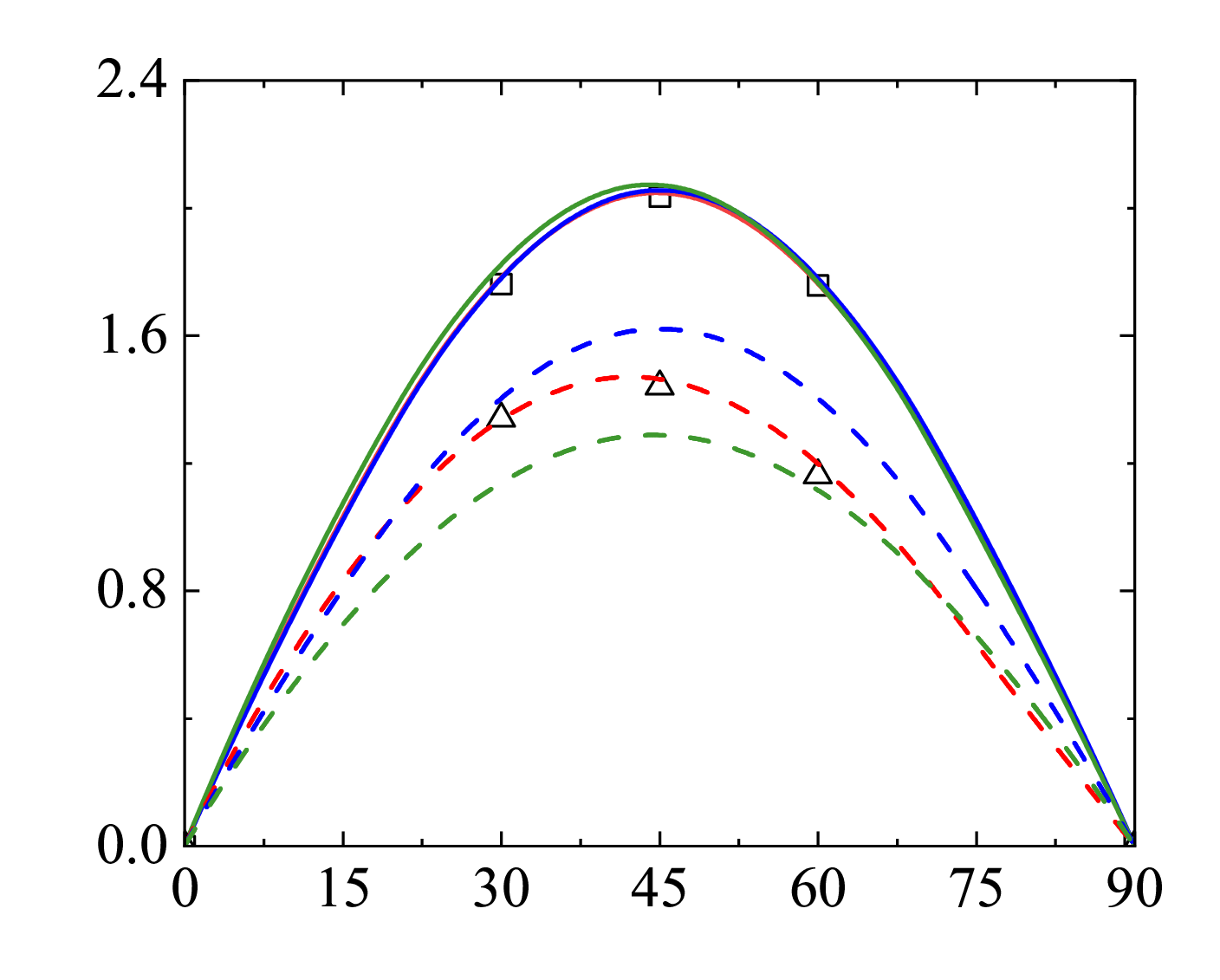}
		\put(-1,70){(b)}
		\put(3,40){\rotatebox{90}{$C_L$}}
		\put(52.5,0){$\alpha^\circ$}
	\end{overpic}\\[1.0ex]
	\begin{overpic}[width=0.5\textwidth]{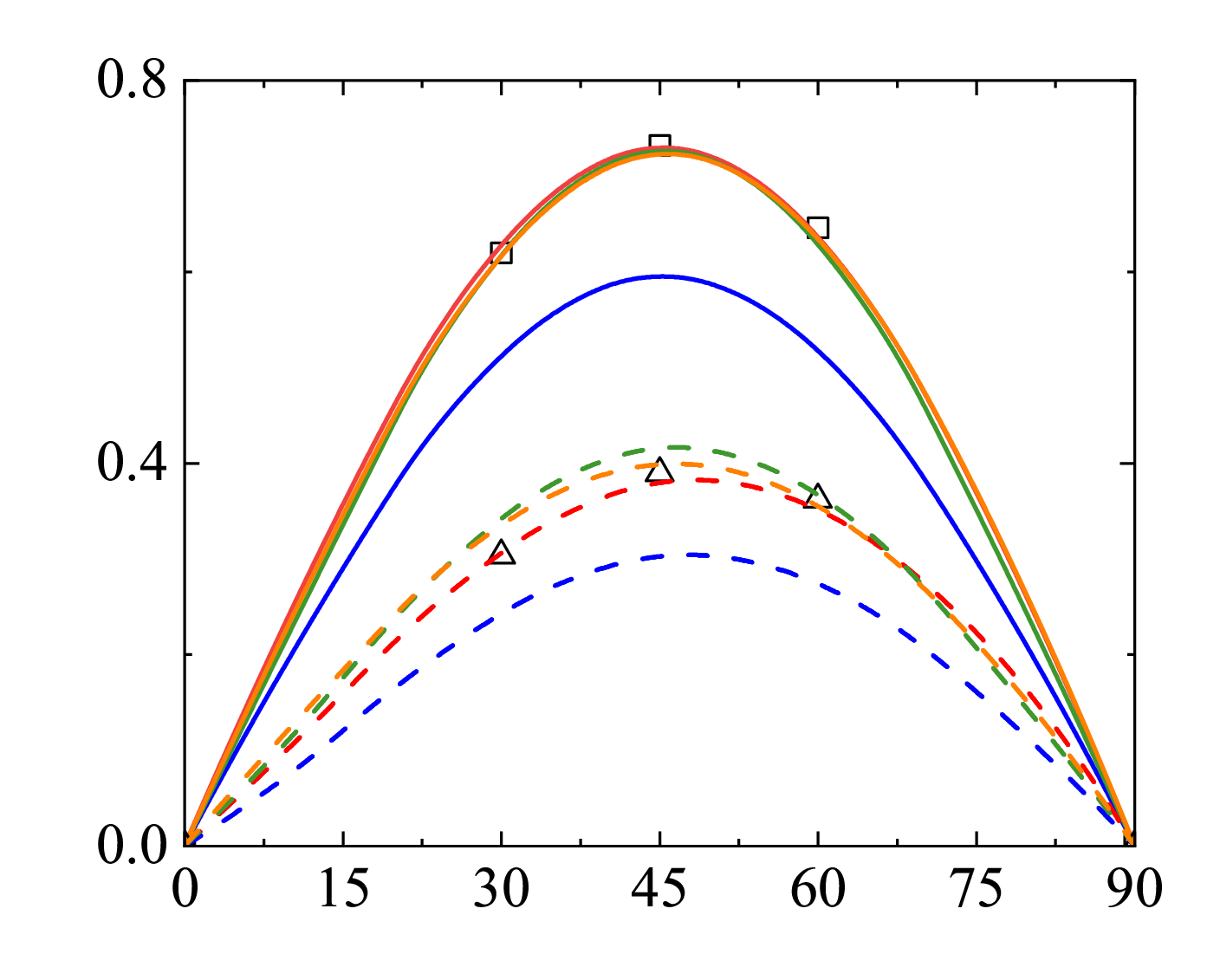}
		\put(-1,70){(c)}
		\put(3,40){\rotatebox{90}{$C_L$}}
		\put(52.5,0){$\alpha^\circ$}
	\end{overpic}~
	\begin{overpic}[width=0.5\textwidth]{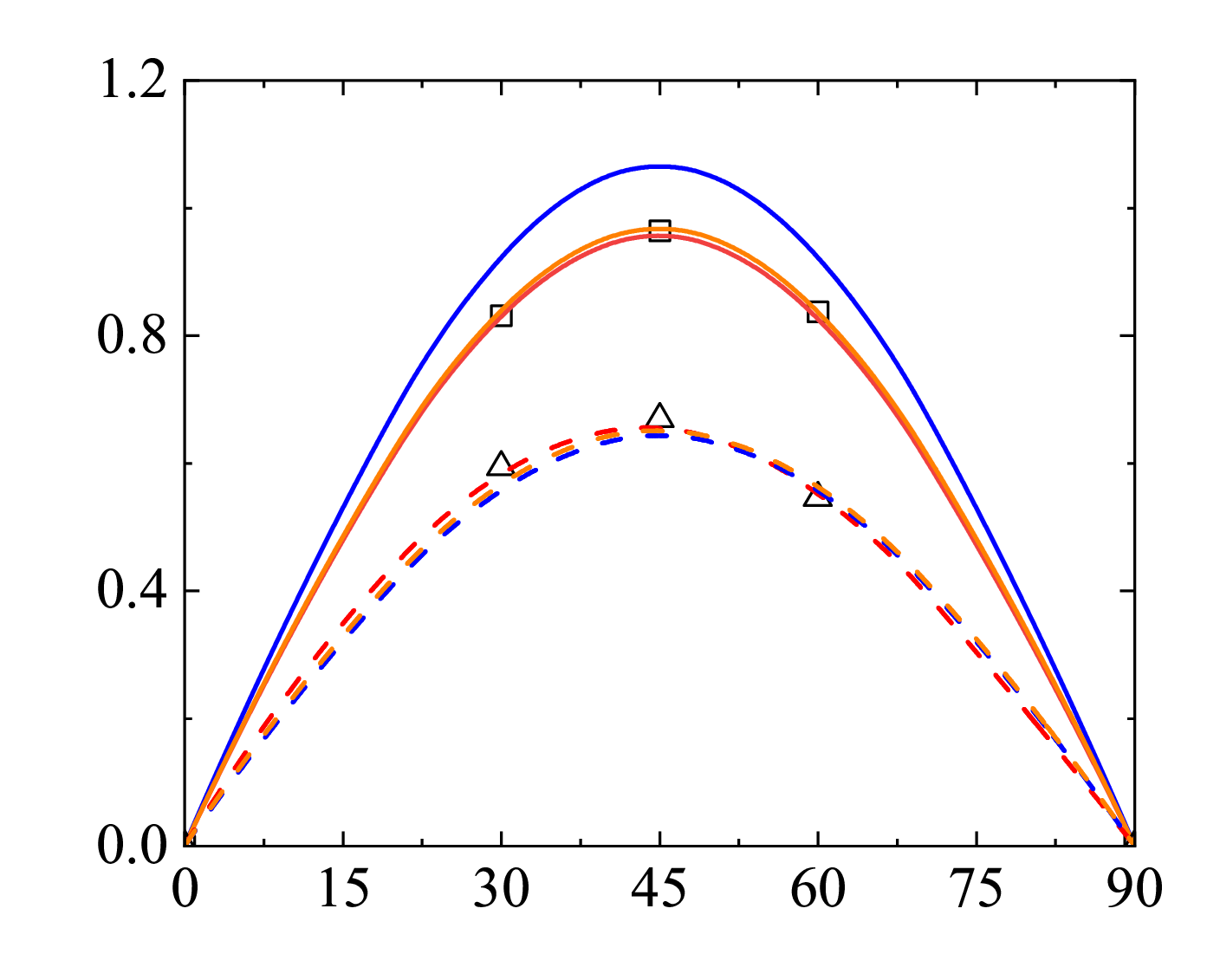}
		\put(-1,70){(d)}
		\put(3,40){\rotatebox{90}{$C_L$}}
		\put(52.5,0){$\alpha^\circ$}
	\end{overpic}\\[1.0ex]
	\begin{overpic}[width=0.5\textwidth]{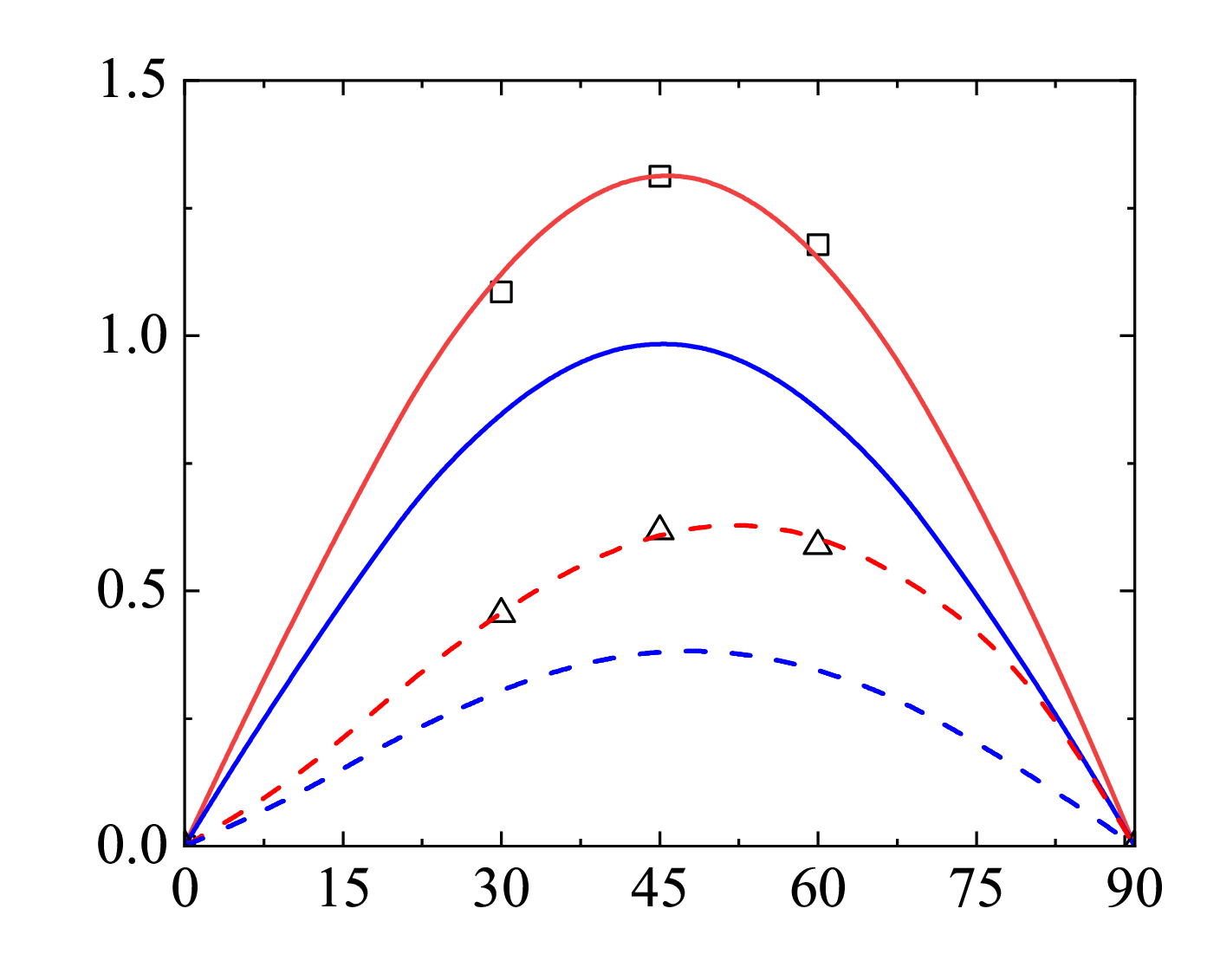}
		\put(-1,70){(e)}
		\put(3,40){\rotatebox{90}{$C_L$}}
		\put(52.5,0){$\alpha^\circ$}
	\end{overpic}~
	\begin{overpic}[width=0.5\textwidth]{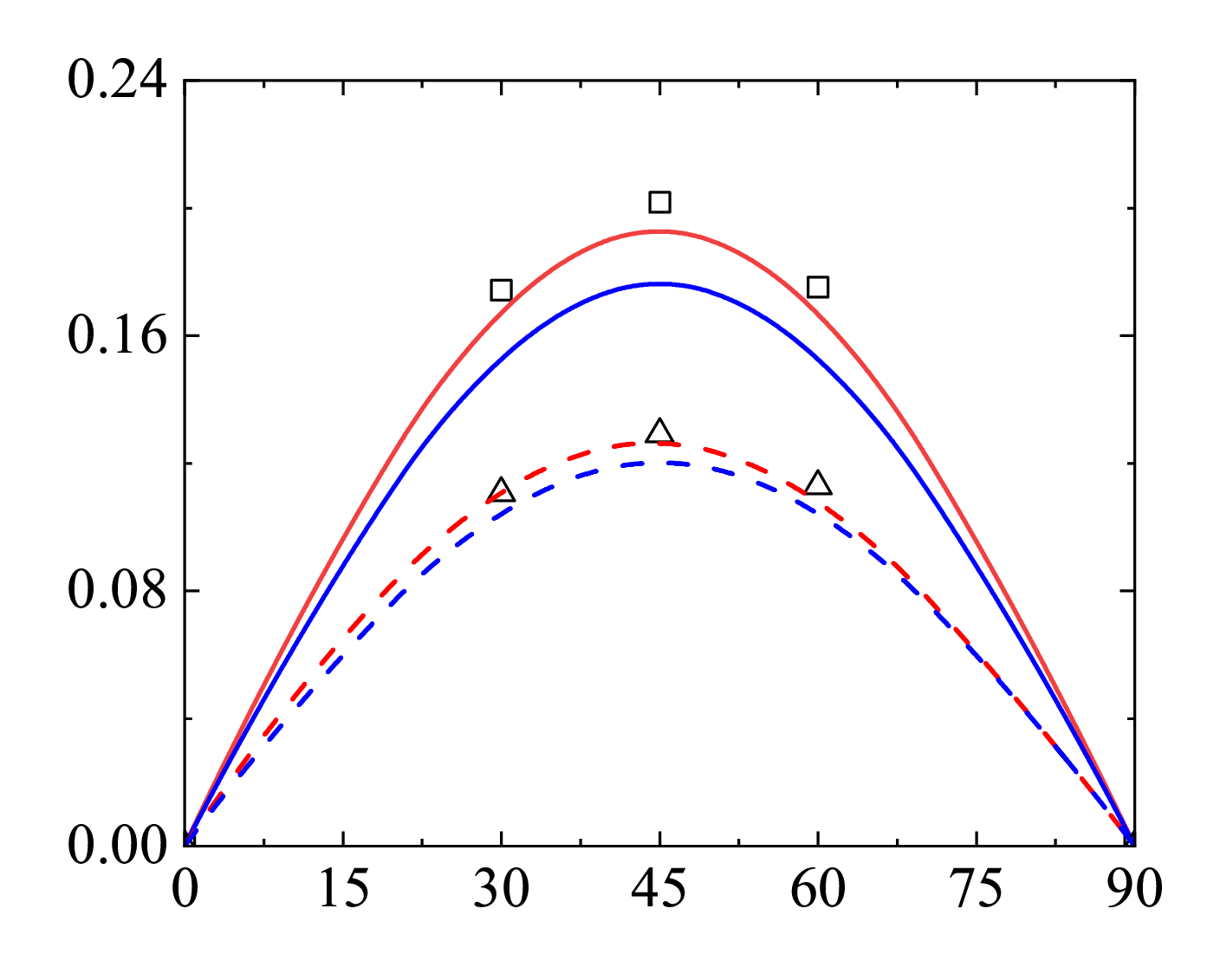}
		\put(-1,70){(f)}
		\put(3,40){\rotatebox{90}{$C_L$}}
		\put(52.5,0){$\alpha^\circ$}
\end{overpic}\\
	\caption{Variation of particle lift force coefficients $C_L$ against $\alpha$, at $M_p=0.1$.
	(a) $w$=1.25, (b) $w$=0.2, (c) $w$=2.5, (d) $w$=0.4, (e) $w$=5, (f) $w$=0.8. 
	Solid lines and square symbols: $Re_p=10$, dashed lines and triangle symbols: $Re_p=100$.
	Symbols: present simulations, 
	red lines: Eq.~\eqref{3.2.4-cl-incom}, 
	blue lines: (a,c,e) \citet{ouchene_new_2016}, (b,d,f) \citet{ouchene_numerical_2020}, 
	green lines: \citet{zastawny_derivation_2012},  
	orange lines: \citet{sanjeevi_drag_2018}.}
	\label{fig8:cl-a-incom}
\end{figure}

\begin{figure}[tp!]
	\centering
	\begin{overpic}[width=0.5\textwidth]{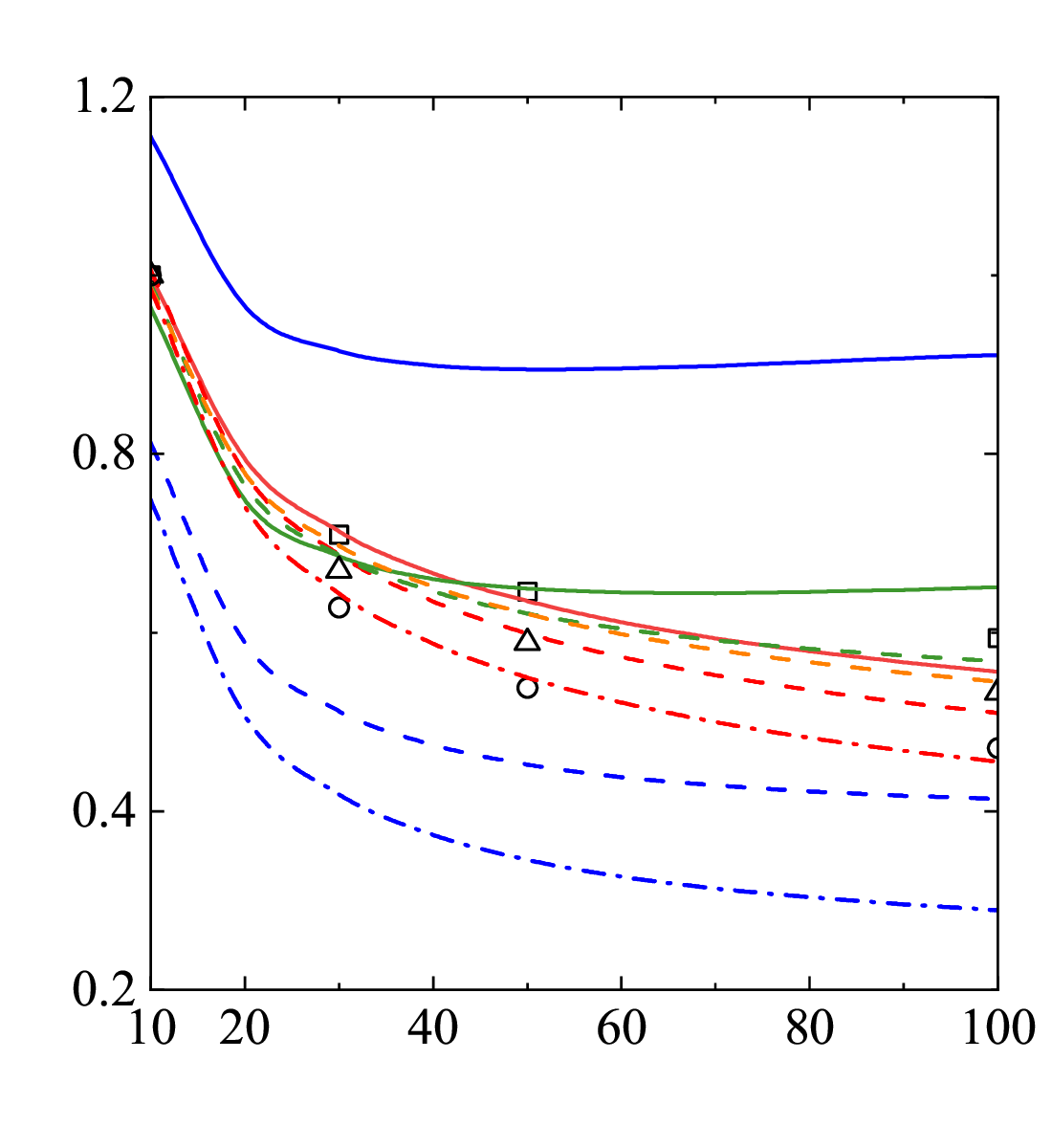}
		\put(0,90){(a)}
		\put(0,30){\rotatebox{90}{\textcolor{black}{$C_{L,\alpha=45^\circ}/C_{L,\alpha=45^\circ,Re_p=10}$}}}
		\put(52.5,2){$Re_p$}		
	\end{overpic}~
	\begin{overpic}[width=0.5\textwidth]{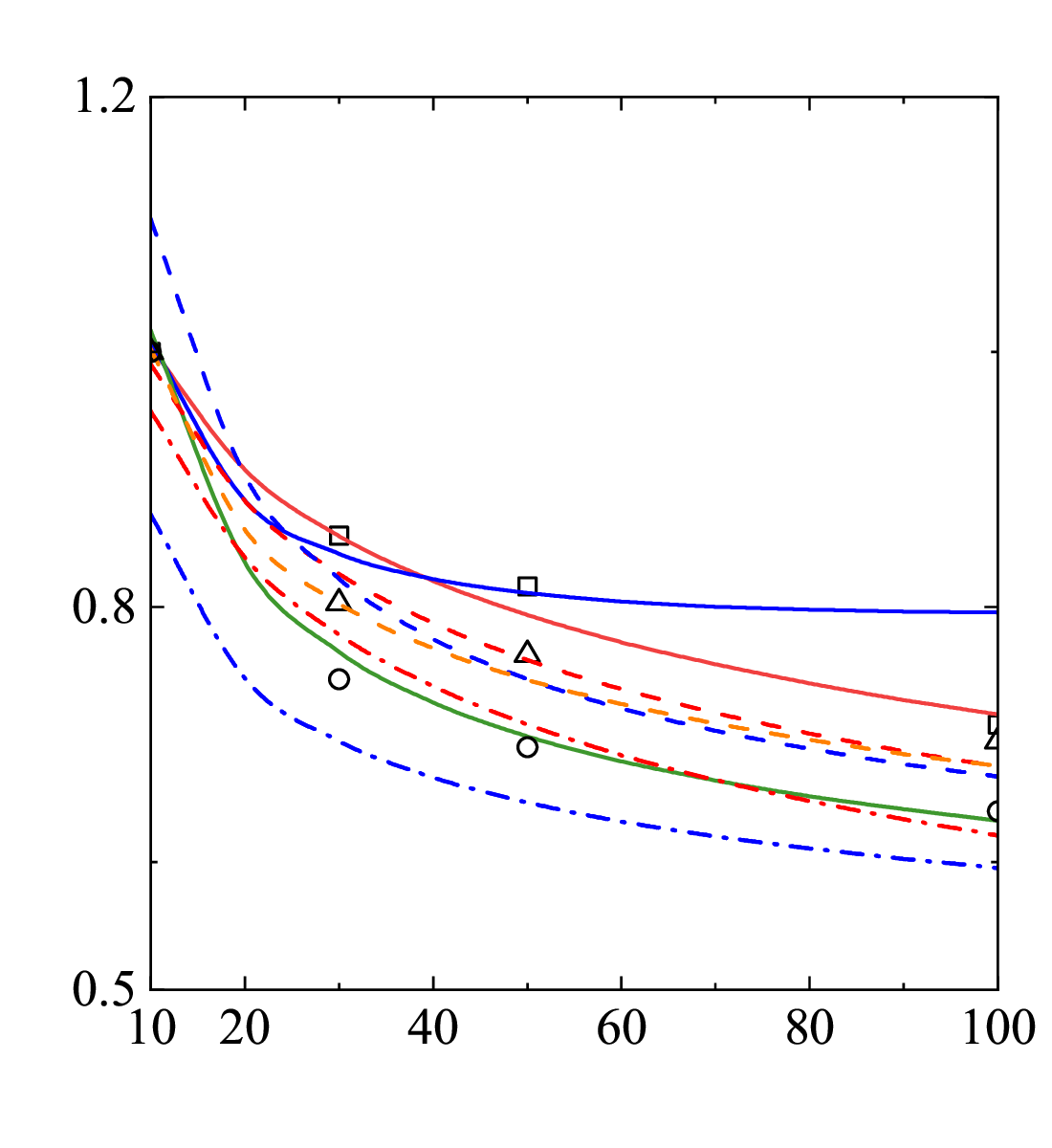}
		\put(0,90){(b)}
		\put(0,30){\rotatebox{90}{\textcolor{black}{$C_{L,\alpha=45^\circ}/C_{L,\alpha=45^\circ,Re_p=10}$}}}
		\put(52.5,2){$Re_p$}	
	\end{overpic}\\
	\caption{Variation of 
		\textcolor{black}{$C_{L,\alpha=45^\circ}/C_{L,\alpha=45^\circ,Re_p=10}$} against $Re_p$, at $M_p=0.1$,
		(a) $w=1.25$ (squares and solid lines), $w=2.5$ (triangles and dashed lines) and $w=5$ (circles and dash-dotted lines),
		(b) $w=0.2$ (squares and solid lines), $w=0.4$ (triangles and dashed lines) and $w=0.8$ (circles and dash-dotted lines).
		Symbols: present simulation, red lines: Eq.~\eqref{3.2.2-clmax-incom}, 
		blue lines: (a) \citet{ouchene_new_2016}, (b) \citet{ouchene_numerical_2020}, 
		green lines: \citet{zastawny_derivation_2012}, 
		orange lines: \citet{sanjeevi_drag_2018}.}
	\label{fig9:cl-re-incom}
\end{figure}

In Figure \ref{fig8:cl-a-incom}, we present a comparison of lift coefficients obtained from numerical simulations 
with empirical formulas proposed by \citet{zastawny_derivation_2012,ouchene_new_2016,ouchene_numerical_2020} and \citet{sanjeevi_drag_2018},
along with the revised formula given herein. 
At low Reynolds numbers, the lift coefficient exhibits symmetry around $\alpha=45^\circ$. 
However, for prolate \textcolor{black}{spheroidal} particles, increasing aspect ratios $w$ and Reynolds numbers $Re_p$ disrupt this symmetry, 
particularly pronounced at $w=5$ and $Re_p=100$. 
The variations of $C_L$ against $Re_p$ at $\alpha=45^\circ$ are depicted in Figure \ref{fig9:cl-re-incom}. 
Additionally, lift coefficients decrease with \textcolor{black}{the} Reynolds number.
The prolate particles, especially those with $w=5$, are more significantly affected. 
Both prolate and oblate particles experience enhanced lift coefficients as $w$ deviates from unity,
\textcolor{black}{and $C_{L,\alpha=45^\circ}/C_{L,\alpha=45^\circ,Re_p=10}$ increases with $w$.}
The proposed empirical formula accurately predicts lift coefficients with an average error of less than 3\%, 
outperforming previous studies that exhibit larger discrepancies across different flow scenarios.

We further investigate the impact of Mach numbers on the lift force coefficient.
Following our analysis of \textcolor{black}{the} drag coefficient modifications, 
we initially \textcolor{black}{conducted} Spearman correlation analysis for lift force coefficient.
The results, detailed in Table \ref{tab8:relat-cl}, indicate a strong dependence of lift force coefficients on 
$Re_p$ and $M_p$, but a much weaker correlation with $w$.

\begin{table}[tp!]
	\centering
	\caption{Spearman correlation analysis for lift coefficient $C_{L,\alpha=45^\circ}$ }
	\label{tab8:relat-cl}
	\begin{tabular}{cccc} 
		\hline
		~ & $Re_p$   & $M_p$     & $w$      \\ 
		\hline
	$w<1,M_p\leq1$	& 0.090 & 0.893  & 0.052  \\ 
		\hline
	$w<1,M_p>1$	& 0.864 & -0.465 & 0.074  \\ 
		\hline
	$w>1,M_p\leq1$	& 0.532 & 0.566  & 0.062  \\ 
		\hline
	$w>1,M_p>1$	& 0.864 & -0.427 & 0.112  \\
		\hline
	\end{tabular}
\end{table}

Similar to drag coefficients, lift coefficients exhibit significant variation at $M_p=1$.
Considering the \textcolor{black}{dependancy} of lift coefficients on $Re_p$ and $M_p$, the compressible impact factor is structured as
\begin{equation}
	\label{3.2.6-gcl}
	g_l(Re_p,M_p)=l_1Re_p^{l_2M_p^{l_3}}
\end{equation}
Henceforth, the lift coefficients of the spheroid particles in compressible flows can be empirically expressed as
\begin{equation}
	\label{3.2.7-cl-com}
	C_L(Re_p,w,M_p,\alpha)=2(\sin \alpha)^{c_{l,1}^{Re_pw}} (\cos \alpha)^{c_{l,2}^{Re_pw}} C_{L,\alpha=45^\circ}(Re_p,w)g_l(Re_p,M_p)
\end{equation}
The parameters in the aforementioned formulas, as given in Table \ref{tab8:relat-cl}, 
are obtained by data fitting utilizing the numerical simulation results.
The average relative error of the proposed formula is less than $4\%$. 

\begin{table}[tp!]
	\centering
	\small
	\caption{Parameters $l_i$ in Eq.\eqref{3.2.6-gcl} for lift coefficients and \textcolor{black}{average} relative error}
	\label{3.2.8-cl-com}
	\begin{tblr}{
			cell{2}{1} = {r=2}{},
			cell{2}{6} = {r=2}{},
			cell{4}{1} = {r=2}{},
			cell{4}{6} = {r=2}{},
			hline{1-2,4,6} = {-}{},
			colspec={Q[c] Q[c] Q[c] Q[c] Q[c] Q[c]}
		}
		~  & ~                        & $l_1$     & $l_2$       & $l_3$        & \textcolor{black}{Average} relative error(\%)  \\
		$w>1$ & $M_p\leq1$ & 1       & 0.1031 & 2.546  & 4.10                 \\
		& $M_p>1$                    & 0.5982         & 0.237 & -0.3432                       \\
		$w<1$ & $M_p\leq1$ & 1              & 0.0896 & 3.1416  & 3.77                 \\
		& $M_p>1$                    & 0.6129         & 0.2148 & -0.4256                      
	\end{tblr}
\end{table}

\begin{figure}[tp!]
	\centering
	\begin{overpic}[width=0.5\textwidth]{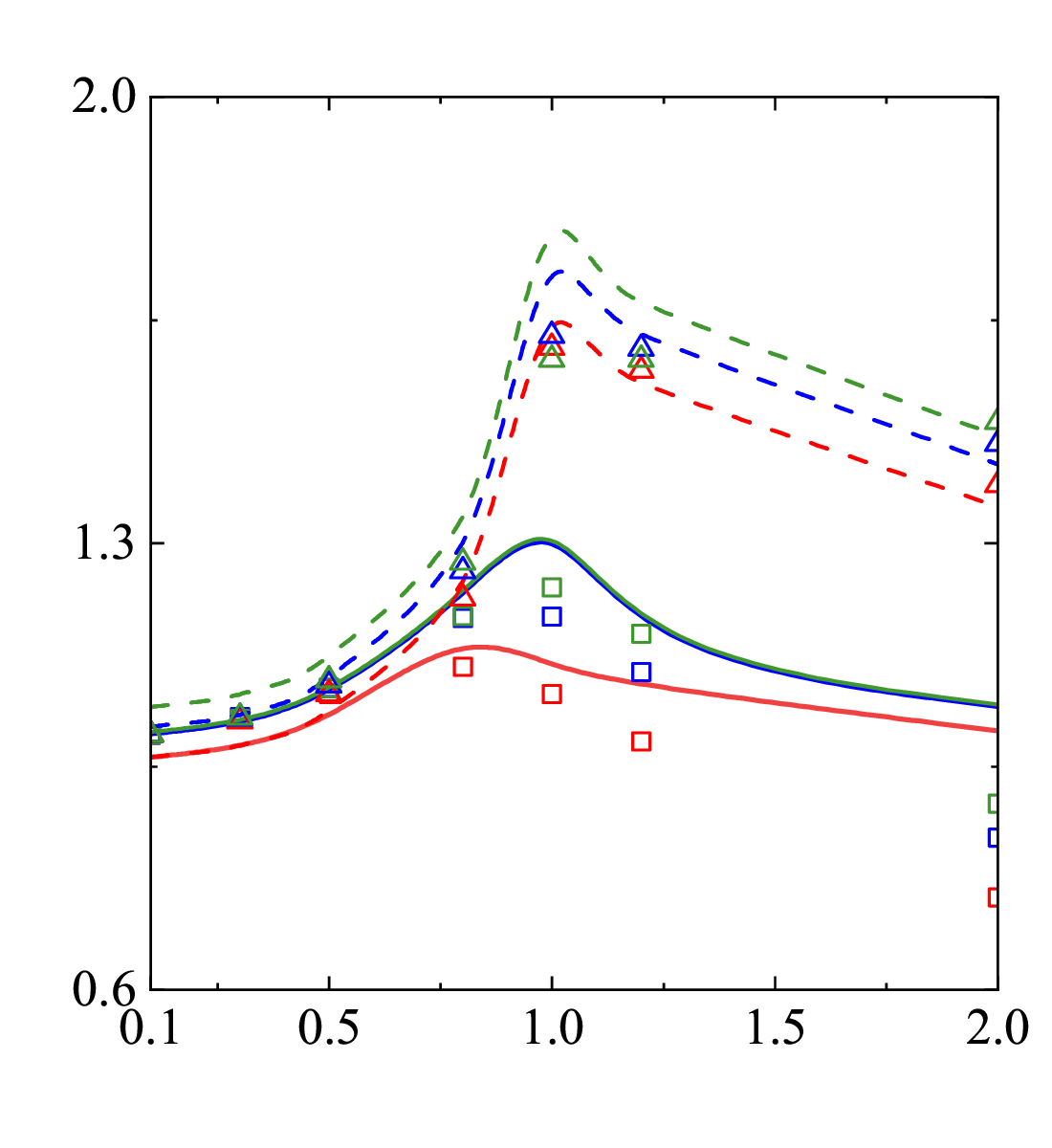}
		\put(0,90){(a)}
		\put(0,30){\rotatebox{90}{\textcolor{black}{$C_{L,\alpha=45^\circ}/C_{L,\alpha=45^\circ,M_p=0.1}$}}}
		\put(45,2){$M_p$}		
	\end{overpic}~
	\begin{overpic}[width=0.5\textwidth]{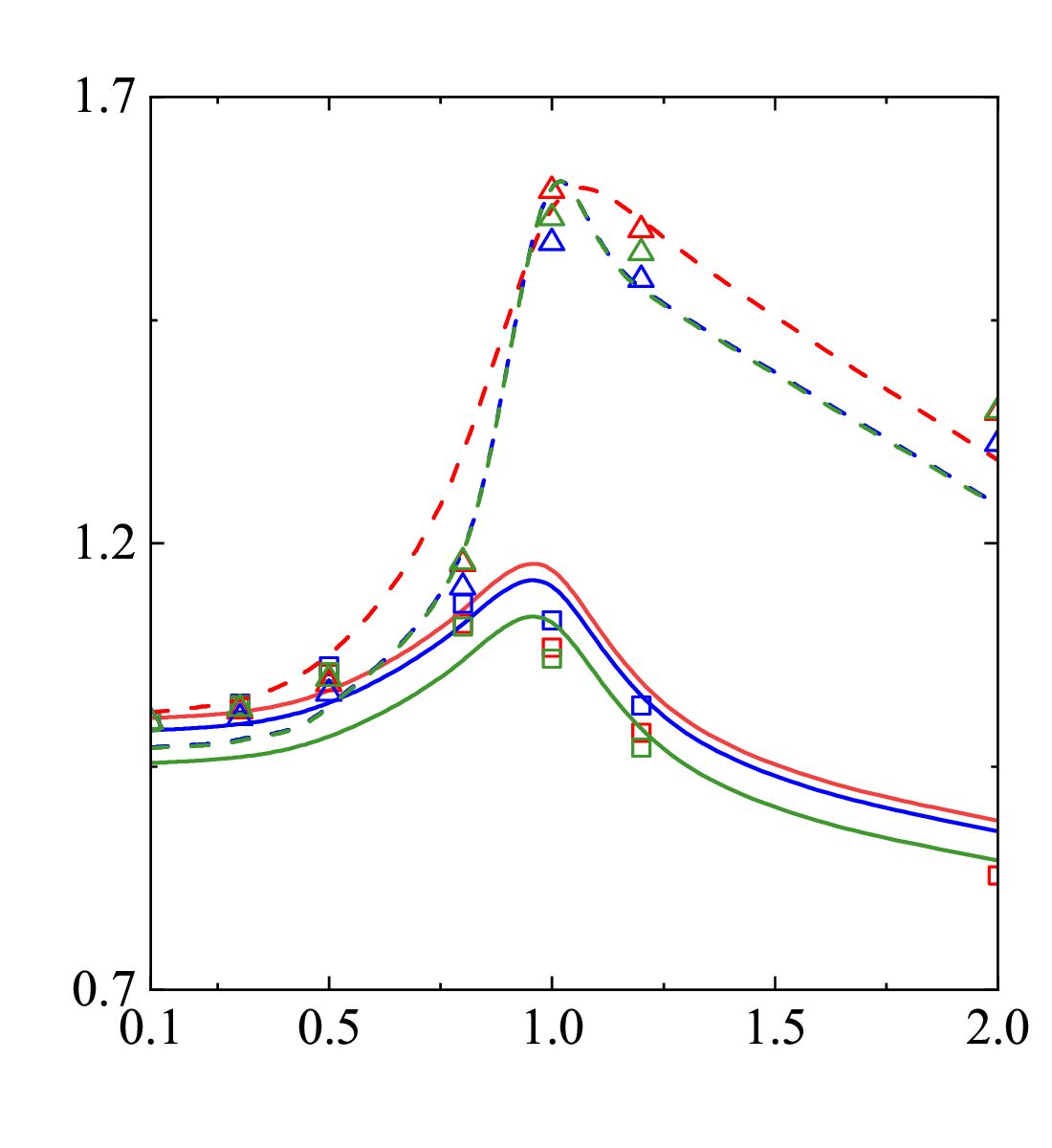}
		\put(0,90){(b)}
		\put(0,30){\rotatebox{90}{\textcolor{black}{$C_{L,\alpha=45^\circ}/C_{L,\alpha=45^\circ,M_p=0.1}$}}}
		\put(45,2){$M_p$}		
	\end{overpic}\\
	\caption{Variation of lift coefficients \textcolor{black}{$C_{L,\alpha=45^\circ}/C_{L,\alpha=45^\circ,M_p=0.1}$} against $M_p$, 
	(a) prolate particles: $w=1.25$ (red), $w=2.5$ (blue),  $w=5$ (green)
	and (b) oblate particles: $w=0.2$ (red), $w=0.4$ (blue),  $w=0.8$ (green). 
	Solid lines and squares: $Re_p=10$, dashed lines and triangles: $Re_p=100$. 
	Lines: Eq.~\eqref{3.2.7-cl-com}, symbols: present simulations.}
	\label{fig10:cl-ma}
\end{figure}

\begin{figure}[tp!]
	\centering
	\begin{overpic}[width=0.5\textwidth]{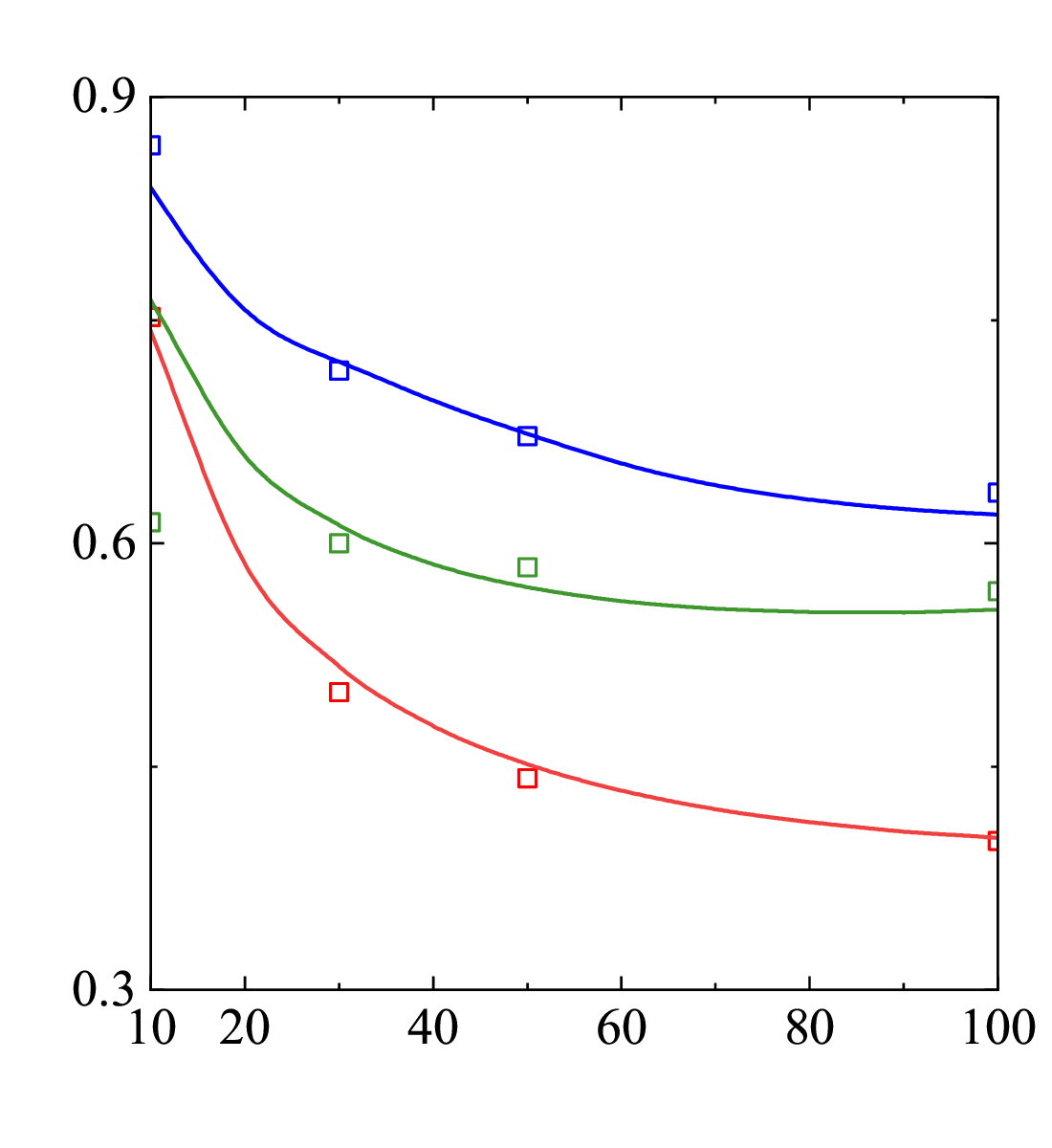}
		\put(0,90){(a)}
		\put(0,50){\rotatebox{90}{$C_{L,\alpha=45^\circ}$}}
		\put(45,2){$Re_p$}				
	\end{overpic}~
	\begin{overpic}[width=0.5\textwidth]{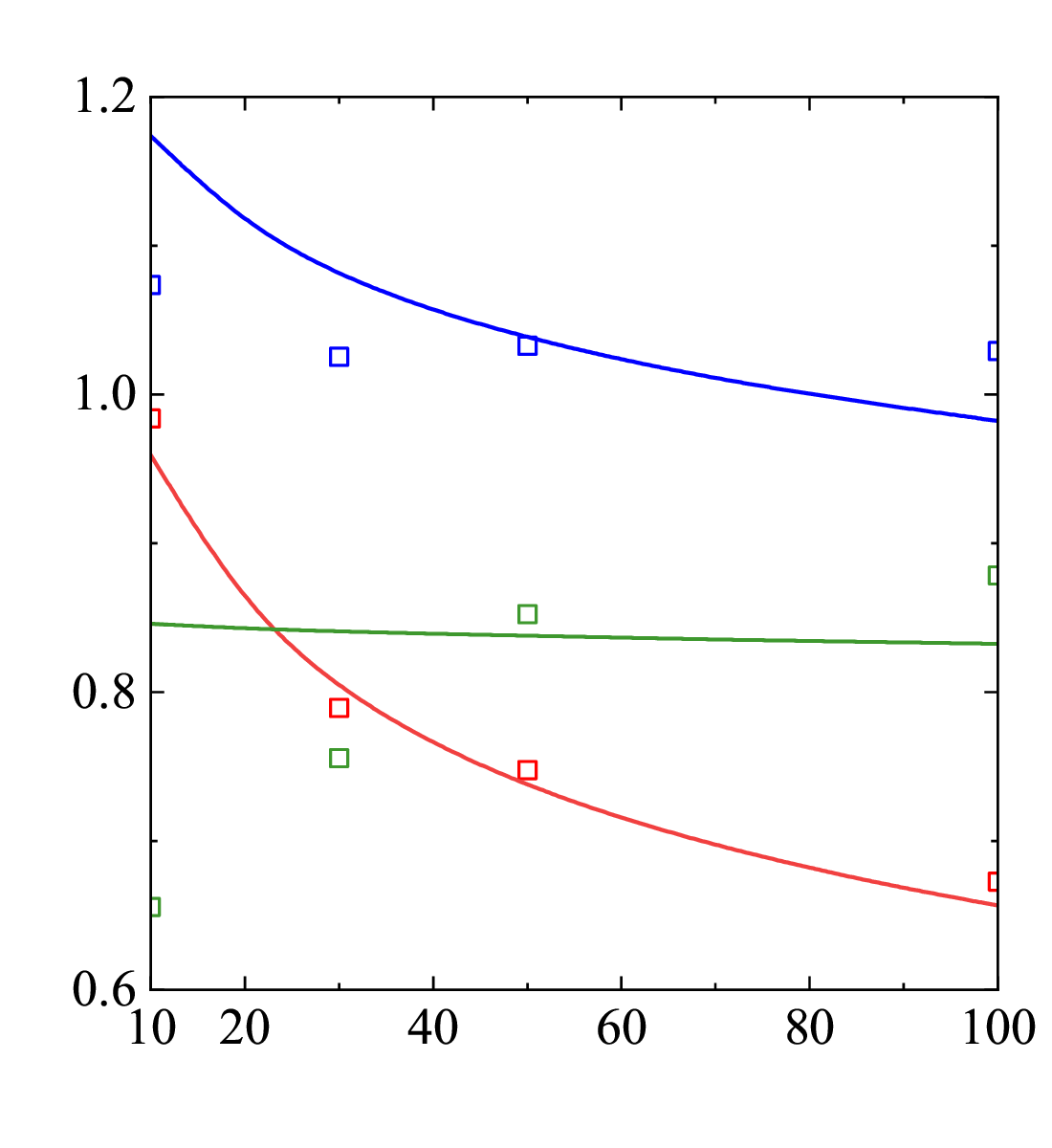}
		\put(0,90){(b)}
		\put(0,50){\rotatebox{90}{$C_{L,\alpha=45^\circ}$}}
		\put(45,2){$Re_p$}
	\end{overpic}\\
	\caption{Variation of lift coefficients $C_{L,\alpha=45^\circ}$ against $Re_p$, (a) $w=2.5$ and (b) $w=0.4$. 
		Red: $M_p=0.3$, 
		blue: $M_p=1$, 
		green: $M_p=2$.
		Lines: Eq.~\eqref{3.2.7-cl-com}, symbols: present simulations.}
	\label{fig11:cl-re-com}
\end{figure}

Figures \ref{fig10:cl-ma} and \ref{fig11:cl-re-com} illustrate the variation of the maximum lift coefficients 
$C_{L,\alpha=45^\circ}$ with Mach number $M_p$ and Reynolds number $Re_p$. 
When $Re_p$ is held constant, the maximum lift coefficients initially rise at $M_p<1$, peak around $M_p \approx 1$, 
and then decline beyond $M_p=1$, partially mirroring the behavior of drag coefficients discussed earlier.
At a fixed Mach number $M_p$, the maximum lift coefficients $C_{L,\alpha=45^\circ}$ exhibit a monotonic decrease with $Re_p$ 
for prolate particles.
This trend is prominent in subsonic scenarios but diminishes in supersonic flows. 
Conversely, for oblate particles, the maximum lift coefficient $C_{L,\alpha=45^\circ}$ decreases solely with $Re_p$ 
in subsonic conditions at $M_p=0.3$, while this decreasing trend disappears in transonic and supersonic cases. 
The proposed formula provides reasonably accurate predictions in terms of $Re_p$ and $M_p$, except for extreme cases 
with low $Re_p$ but high $M_p$, as depicted in Figure \ref{fig11:cl-re-com}. 
In such instances, rarefaction effects cannot be overlooked, rendering both simulations and empirical formulas unsuitable. 
However, in point-particle simulations within turbulence, such scenarios are infrequent, 
as $Re_p$ and $M_p$ typically increase concurrently. This observation holds for drag forces as well. 
The exploration of rarefaction effects falls outside the scope of this paper and will be considered in our future investigation.

\begin{figure}[tp!]
	\centering
	\begin{overpic}[width=0.5\textwidth]{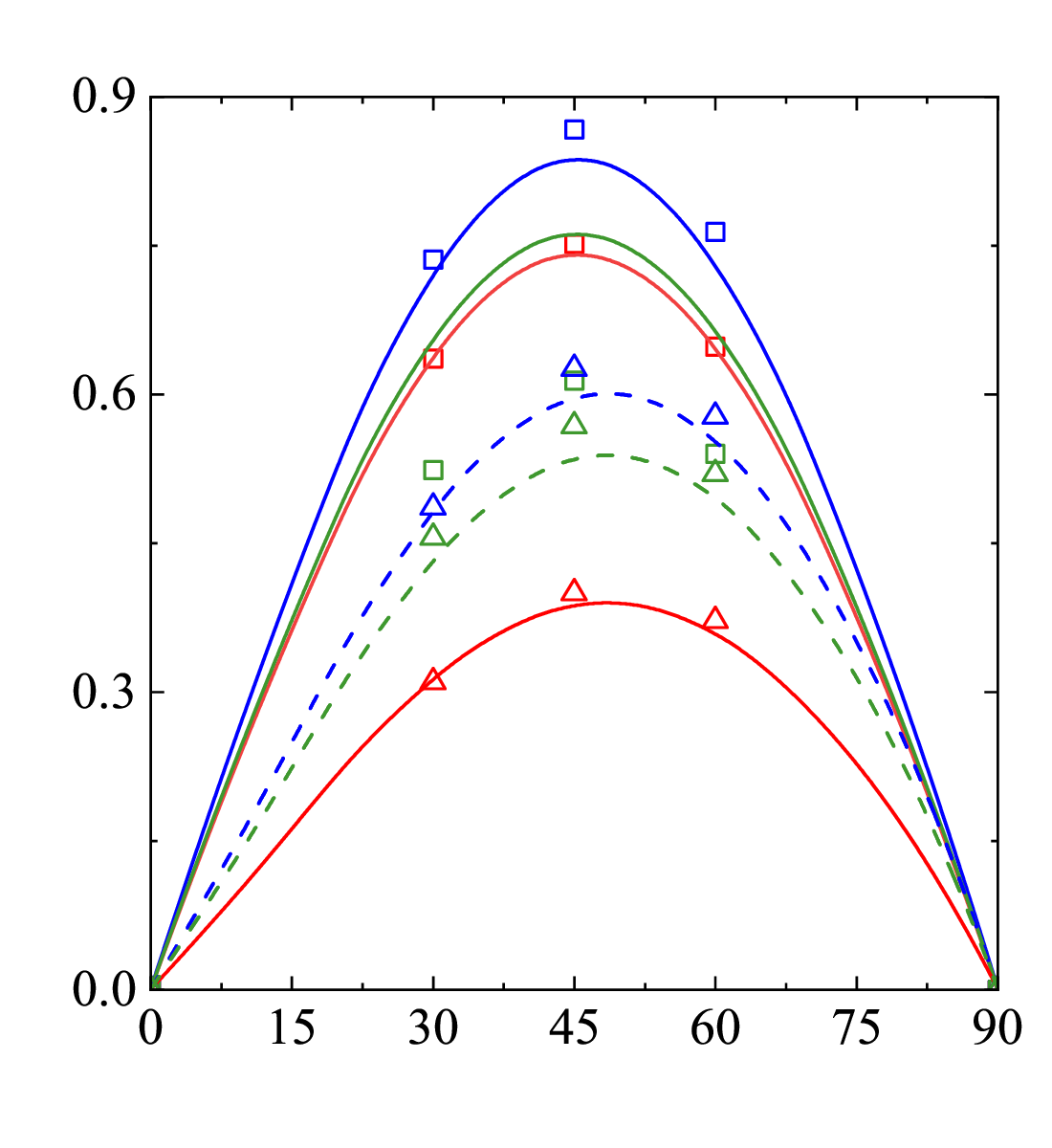}
		\put(0,90){(a)}
		\put(0,50){\rotatebox{90}{$C_L$}}
		\put(45,2){$\alpha^\circ$}		
	\end{overpic}~
	\begin{overpic}[width=0.5\textwidth]{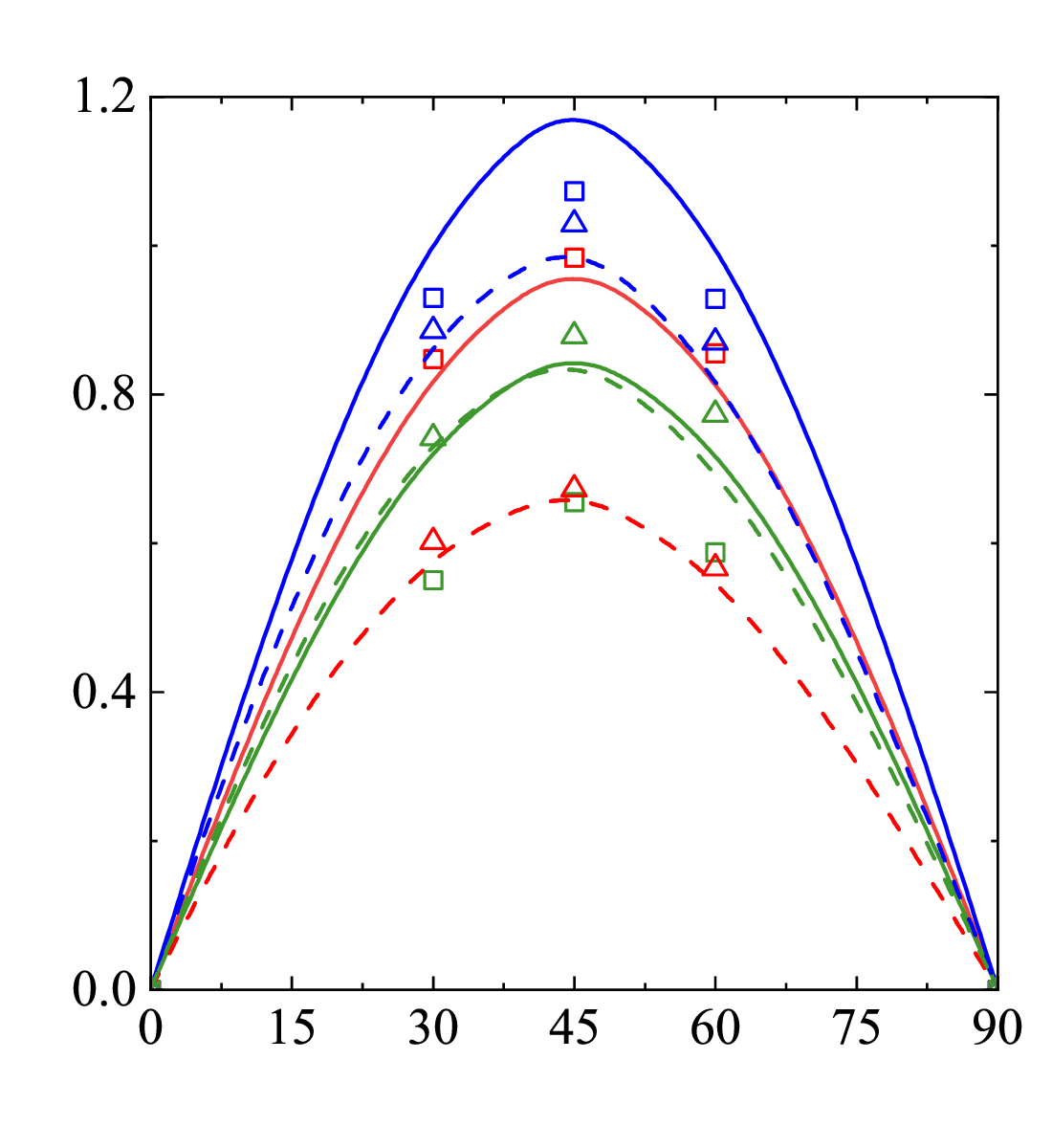}
		\put(0,90){(b)}
		\put(0,50){\rotatebox{90}{$C_L$}}
		\put(45,2){$\alpha^\circ$}
	\end{overpic}\\
	\caption{Variation of lift coefficients $C_{L}$ against $\alpha$, (a) $w=2.5$, (b) $w=0.4$.
		Solid lines and squares: $Re_p=10$, dashed lines and triangles: $Re_p=100$.
		Red: $M_p=0.3$, 
		blue: $M_p=1$, 
		green: $M_p=2$.
		Lines: Eq.~\eqref{3.2.7-cl-com}, symbols: present simulations.}
	\label{fig12:cl-a-com}
\end{figure}
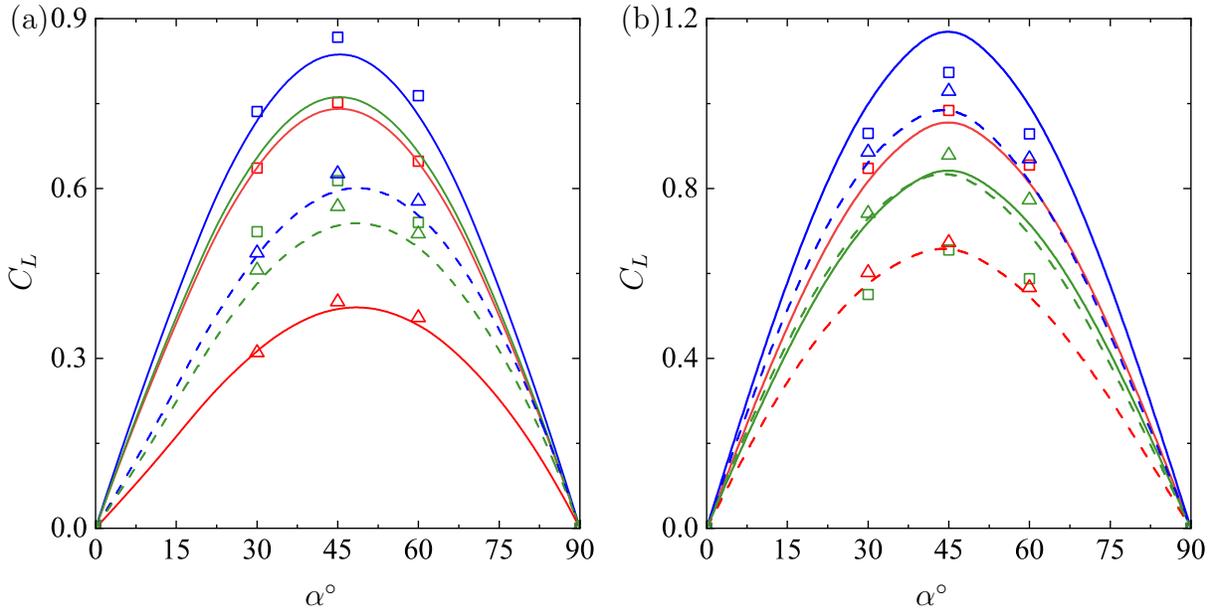

In Figure \ref{fig12:cl-a-com}, we present the variation of the lift coefficients with the angle of attack $\alpha$ 
at Mach numbers $M_p=0.3$, 1.0, and 2.0, and Reynolds numbers $Re_p=10$ and 100. 
The trend of variation resembles that of incompressible flow. 
The lift coefficient curves exhibit a slight shift to the right for prolate particles and \textcolor{black}{to} the left for oblate particles, 
a pattern well-captured by the proposed empirical formula. 
Particularly, the relative error for the lift coefficients of oblate particles at $Re_p=10$ and $M_p=2.0$ is comparatively high. 
These discrepancies can be neglected, for as we have stated in the previous subsection,
the high $M_p$ and low $Re_p$ correspond to the high particle Knudsen number in such cases, 
rendering the results meaningless.

This analysis demonstrates that the variation of lift coefficients with angle of attack at different Mach numbers is similar to 
that of incompressible flow. 
The lift force coefficients for prolate spheroidal particles exhibits a slight rightward shift, 
whereas that for oblate spheroidal particles displays a slight leftward shift.
When considering Reynolds numbers \textcolor{black}{at} 10 and 100 for analysis and comparison, at higher Reynolds numbers, 
the lift coefficients are notably lower than the lift coefficients at lower $Re_p$. 
This discrepancy diminishes gradually with an increase in Mach number.

In this section, we initially refined the lift coefficient for incompressible flows by analyzing cases 
at low Mach numbers and incorporating asymmetrical distributions with respect to the attack angle. 
Subsequently, we considered the effects of compressibility for higher Mach number cases
and elucidated the formulation of the compressibility factor as a function of both $Re_p$ and $M_p$.
Through a comparison of numerical simulation databases, we have demonstrated that the current proposed formula for 
the lift coefficient provides accurate results, except in scenarios of high $M_p$ and 
low $Re_p$ where rarefaction effects become significant. 

\subsection{Pitching torque} \label{subsec:torque}

In this subsection, we consider the pitching torque coefficients. 
Unlike drag and lift forces, pitching torques in creeping flows lack exact solutions, 
with low Reynolds numbers suggesting zero pitching torque. 
Empirical formulas are scarce at higher Reynolds numbers. 
\citet{zastawny_derivation_2012} highlighted the similarity between pitching torque coefficients and lift coefficients, 
leading to the modeling of pitching torque coefficients based on a formula akin to lift coefficients. 
\citet{ouchene_new_2016} refined the formula to accommodate higher Reynolds numbers and broader aspect ratios, 
focusing solely on the prolate \textcolor{black}{spheroid} formulation in this study. 
Subsequently, \citet{ouchene_numerical_2020} extended this analysis to oblate \textcolor{black}{spheroidal} particles. 
Building on prior research, the expression for pitching torque coefficients will be determined in conjunction 
with the lift coefficient expression in the current study, with that of prolate particles as
\begin{equation}
	\label{3.3.1-ctmax}
	C_{T,\alpha=45^\circ}(Re_p,w)=(C_{D,Stokes, \alpha=90^\circ}-C_{D,Stokes, \alpha = 0^\circ})f_{t,\alpha=45^\circ}(Re_p,w)
\end{equation}
with the factor $f_{t,\alpha=45^\circ}$ cast as
\begin{equation}
	\label{3.3.2-fct}
	f_{t,\alpha=45^\circ}(Re_p,w)=c_{t,3}Re_p^{c_{t,4}+c_{t,5}\ln w}w^{c_{t,6}+c_{t,7}Re_p}.
\end{equation}
\textcolor{black}{The pitching torque coefficient can be obtained at varying attack angles by employing the following formula}
\begin{equation}
	\label{3.3.3-ct-incom}
	C_T(Re_p,w,\alpha)=2(\sin \alpha)^{c_{t,1}^{Re_pw}} (\cos \alpha)^{c_{t,2}^{Re_pw}}C_{T,\alpha=45^\circ}(Re_p,w).
\end{equation}
The $w$ should be replaced with $1/w$ for oblate particles. 
The variation with the attack angle is expressed using a combination of trigonometric functions. 
\textcolor{black}{The pitching torque is zero for spherical particles in a uniform flow at relatively low Reynolds numbers, which results in a zero value for $C_T$ as $Re_p$ approaches zero.} 
The parameters included in the equations above are listed in Table \ref{tab10:ct-incom}, 
calculated using the results of cases at $M_p=0.1$.

\begin{table}[tp!]
	\centering
	\small
	\caption{Parameters $c_{t,i}$ in Eq.~\eqref{3.3.2-fct} and Eq.~\eqref{3.3.3-ct-incom}.}
	\label{tab10:ct-incom}
	\begin{tabular}{cccccccc} 
		\hline
		~ & $i=1$      & $i=2$     & $i=3$    & $i=4$     & $i=5$     & $i=6$    & $i=7$          \\ 
		\hline
		$w>1$ & 1.0001 & 0.9999   & 0.7095 & 0.7609 & -0.0087 & 0.0  & 0.0   \\ 
		$w<1$ & 0.9995    & 1.0005 & 0.1963 & 0.9024  & -0.0056  & 0.3346 & -0.0014    \\
		\hline
	\end{tabular}
\end{table}
	
\begin{figure}[tp!]
	\centering
	\begin{overpic}[width=0.5\textwidth]{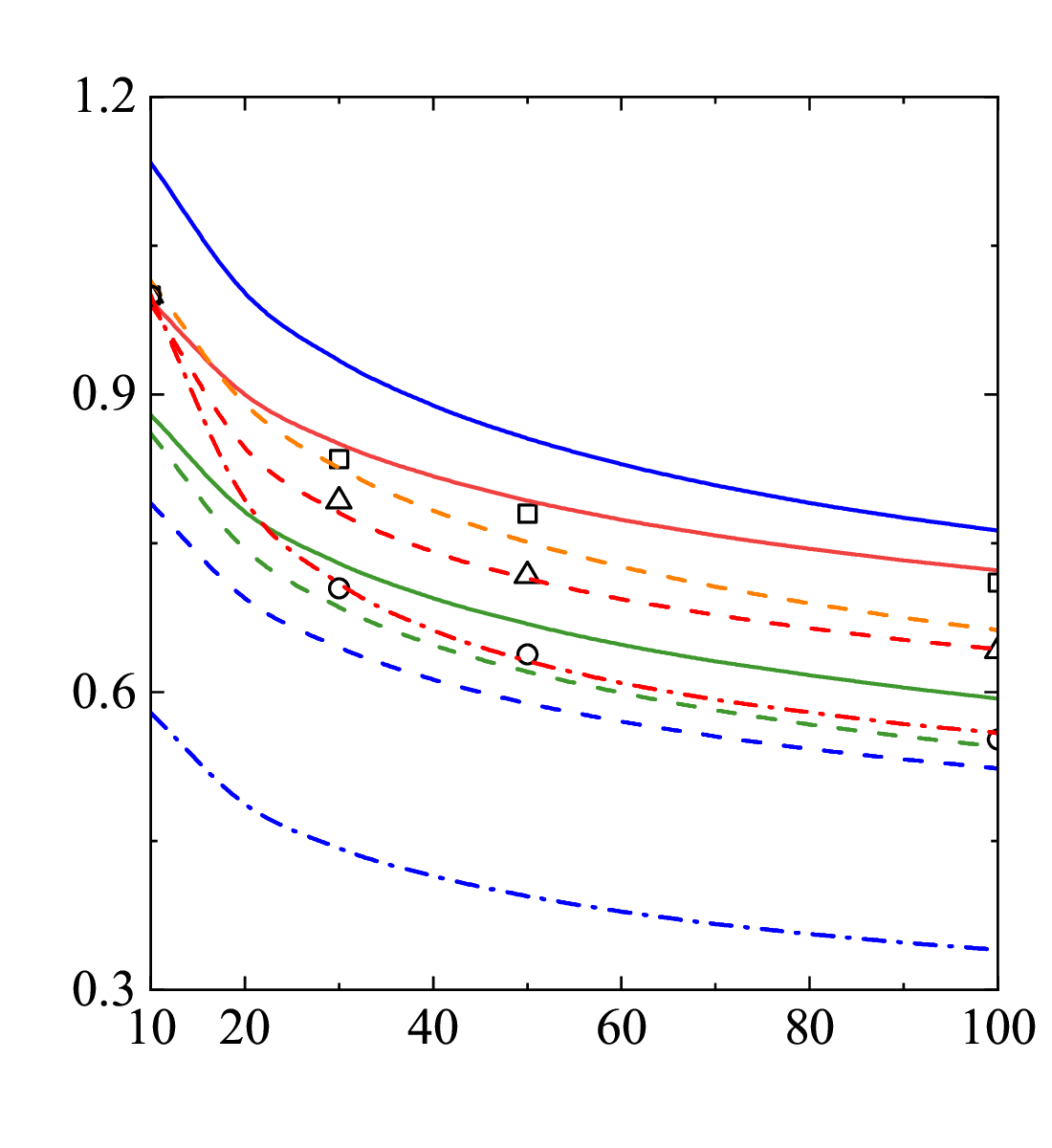}
		\put(0,90){(a)}
		\put(0,30){\rotatebox{90}{\textcolor{black}{$C_{T,\alpha=45^\circ}/C_{T,\alpha=45^\circ,Re_p=10}$}}}
		\put(45,2){$Re_p$}
	\end{overpic}~
	\begin{overpic}[width=0.5\textwidth]{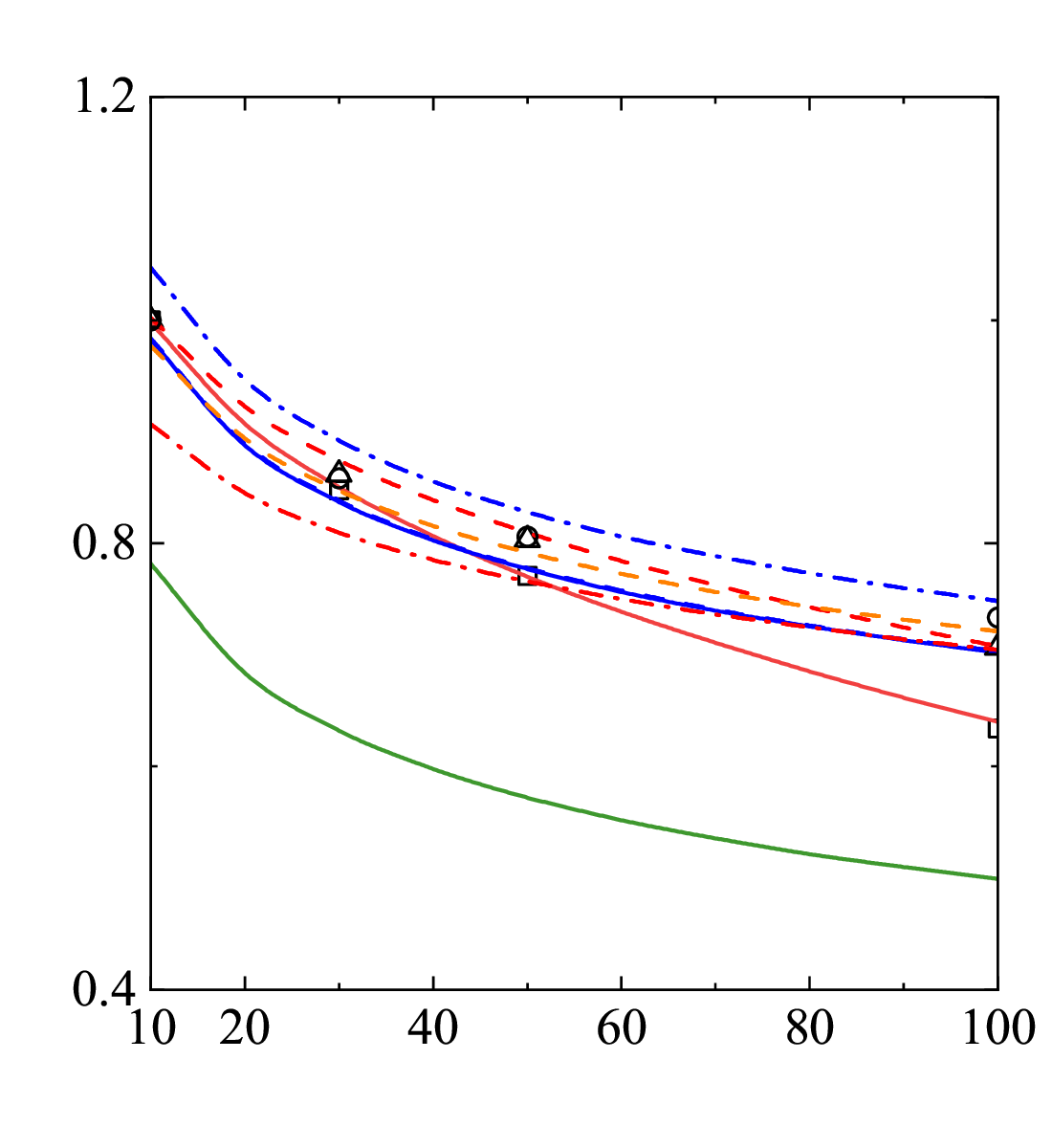}
		\put(0,90){(b)}
		\put(0,30){\rotatebox{90}{\textcolor{black}{$C_{T,\alpha=45^\circ}/C_{T,\alpha=45^\circ,Re_p=10}$}}}
		\put(45,2){$Re_p$}
	\end{overpic}\\
	\caption{Variation of 
		\textcolor{black}{$C_{T,\alpha=45^\circ}/C_{T,\alpha=45^\circ,Re_p=10}$} against $Re_p$, at $M_p=0.1$.
		(a) $w=1.25$ (squares and solid lines), $w=2.5$ (triangles and dashed lines) and $w=5$ (circles and dash-dotted lines),
		(b) $w=0.2$ (squares and solid lines), $w=0.4$ (triangles and dashed lines) and $w=0.8$ (circles and dash-dotted lines).
		Symbols: present simulation, red lines: Eq.~\eqref{3.3.1-ctmax}, blue lines: (a) \citet{ouchene_new_2016}, (b) \citet{ouchene_numerical_2020}, 
		green lines: \citet{zastawny_derivation_2012}, 
		orange lines: \citet{sanjeevi_drag_2018}.}
	\label{fig13:ct-re-incom}
\end{figure}

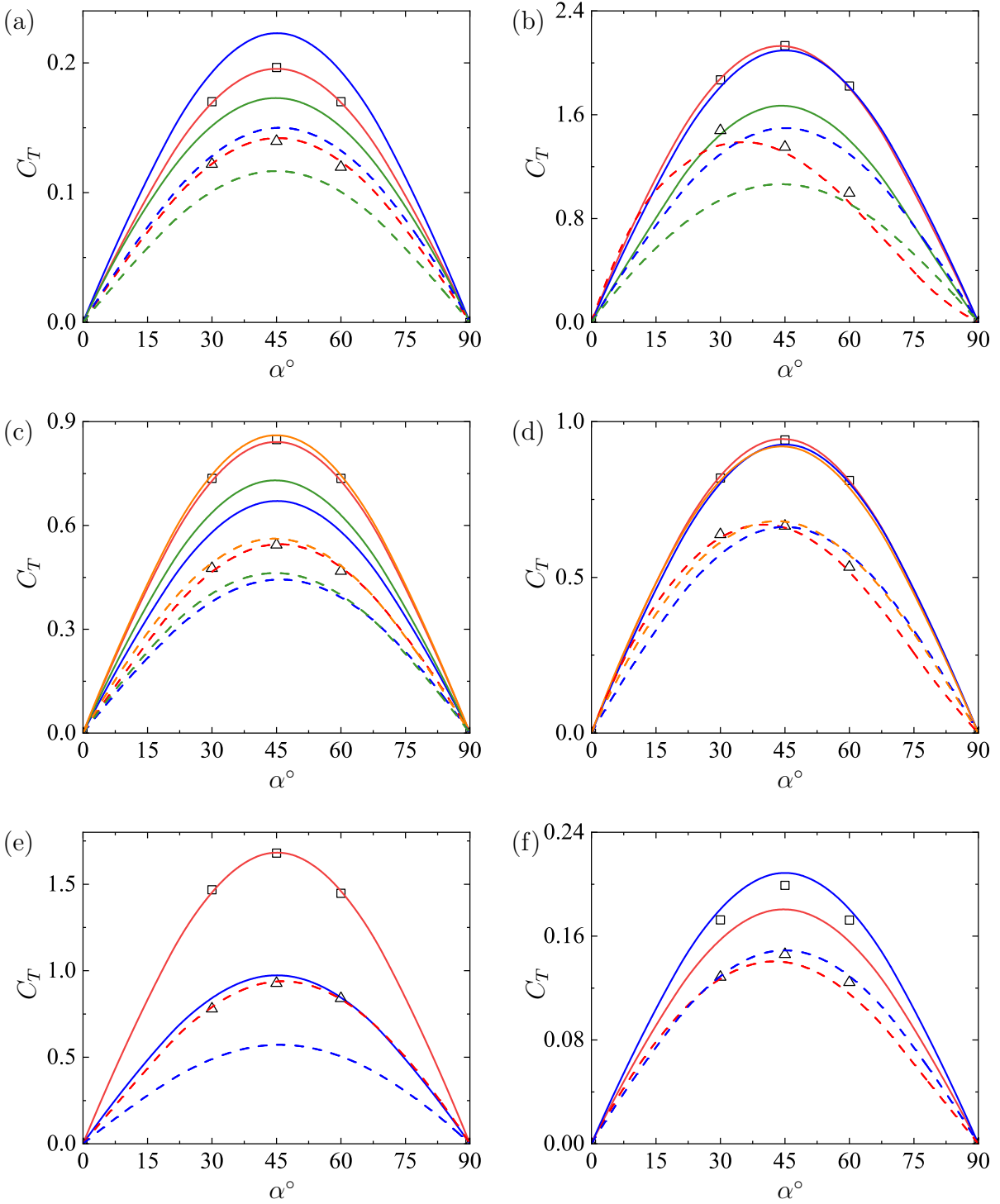
\begin{figure}[tp!]
	\centering
	\begin{overpic}[width=0.5\textwidth]{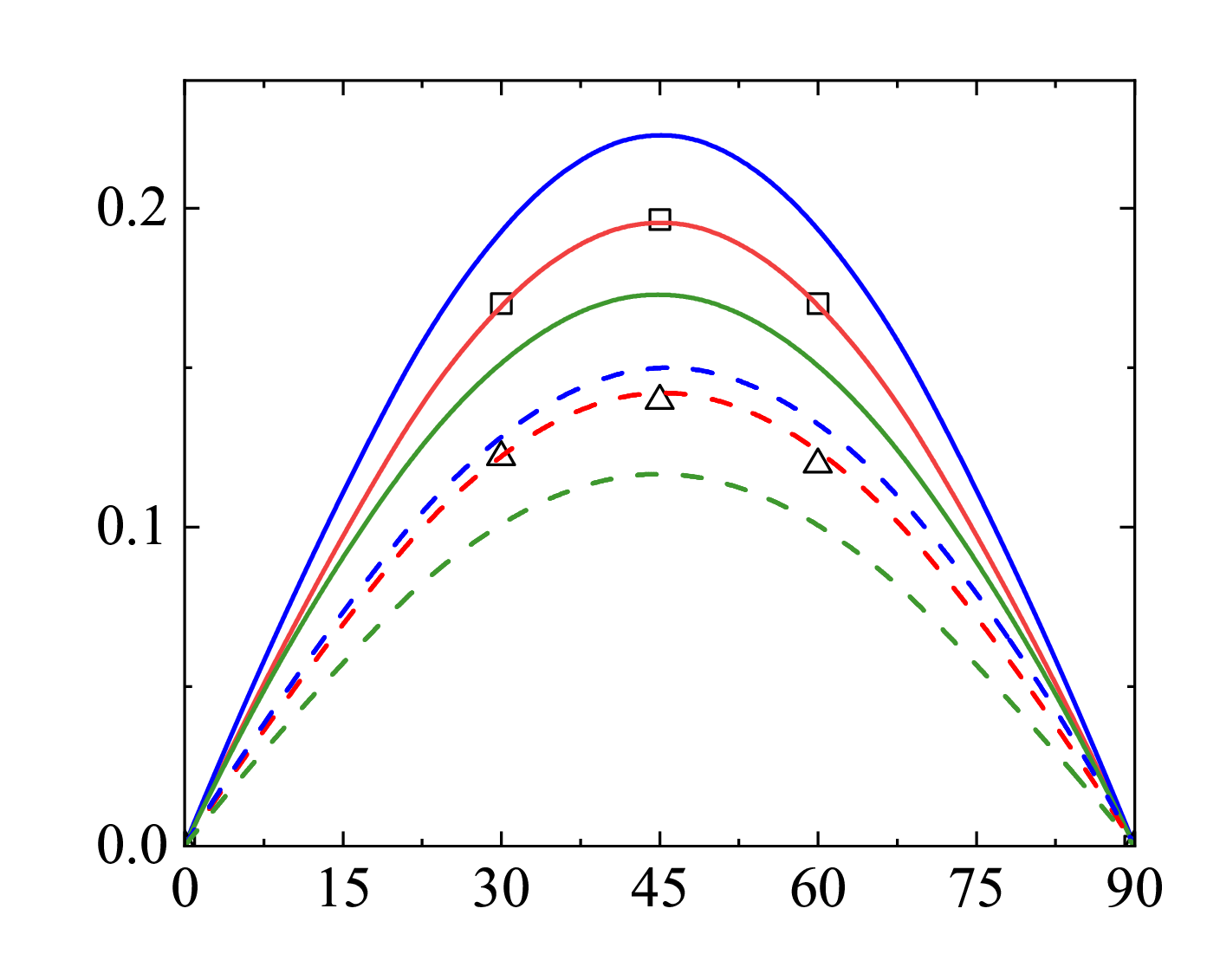}
		\put(-1,70){(a)}
		\put(2,40){\rotatebox{90}{$C_T$}}
		\put(52.5,0){$\alpha^\circ$}
	\end{overpic}~
	\begin{overpic}[width=0.5\textwidth]{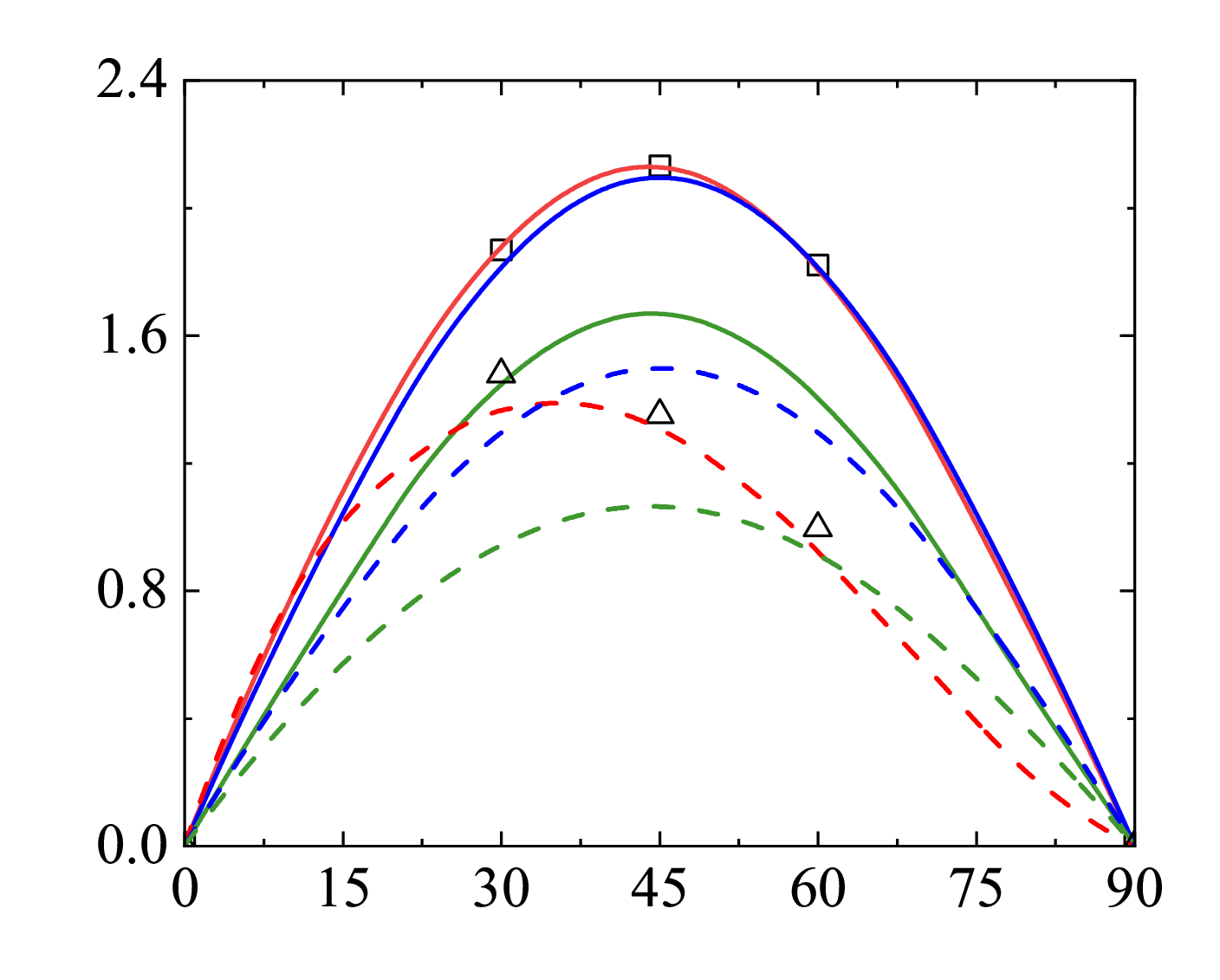}
		\put(-1,70){(b)}
		\put(2,40){\rotatebox{90}{$C_T$}}
		\put(52.5,0){$\alpha^\circ$}
	\end{overpic}\\[1.0ex]
	\begin{overpic}[width=0.5\textwidth]{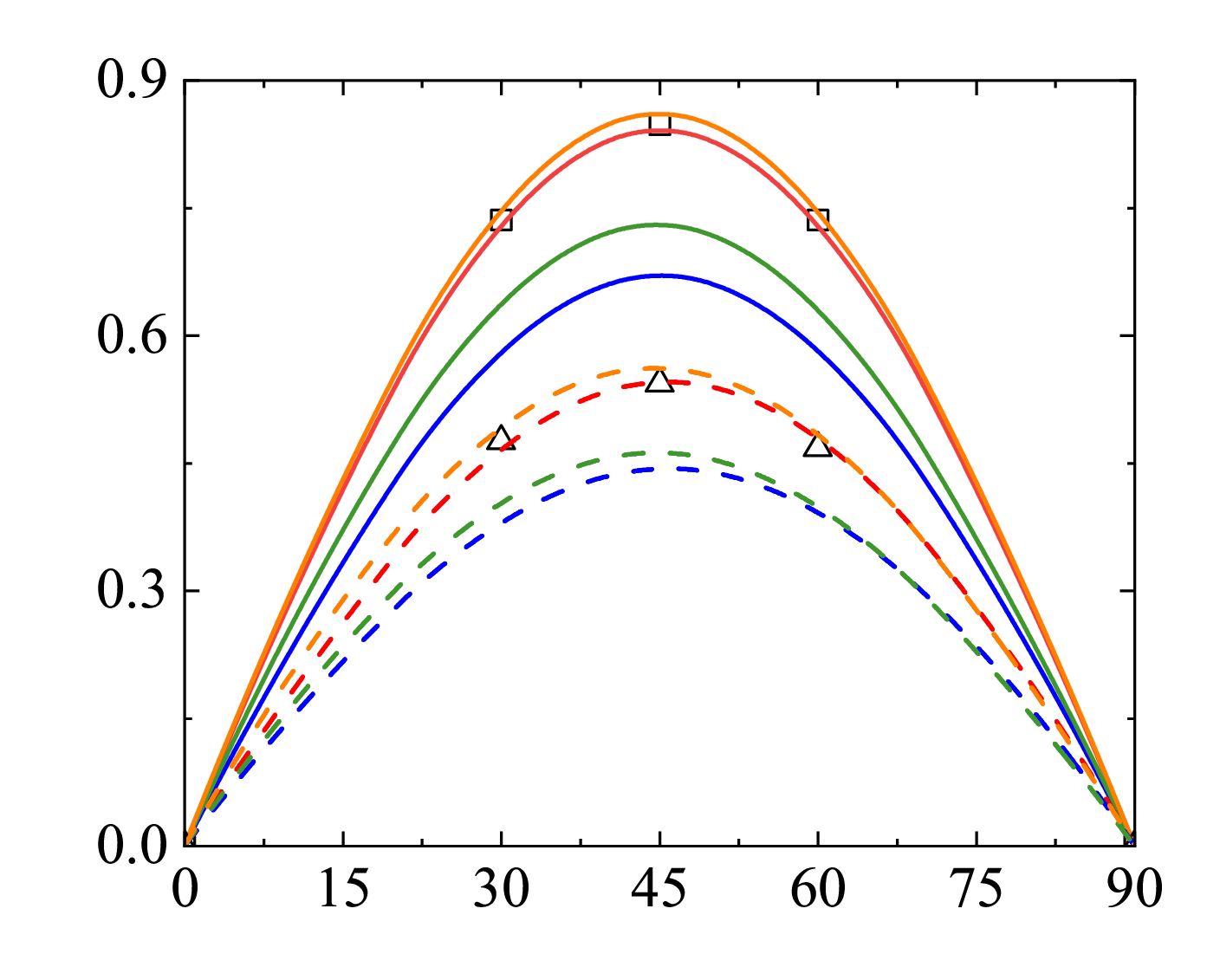}
		\put(-1,70){(c)}
		\put(2,40){\rotatebox{90}{$C_T$}}
		\put(52.5,0){$\alpha^\circ$}
	\end{overpic}~
	\begin{overpic}[width=0.5\textwidth]{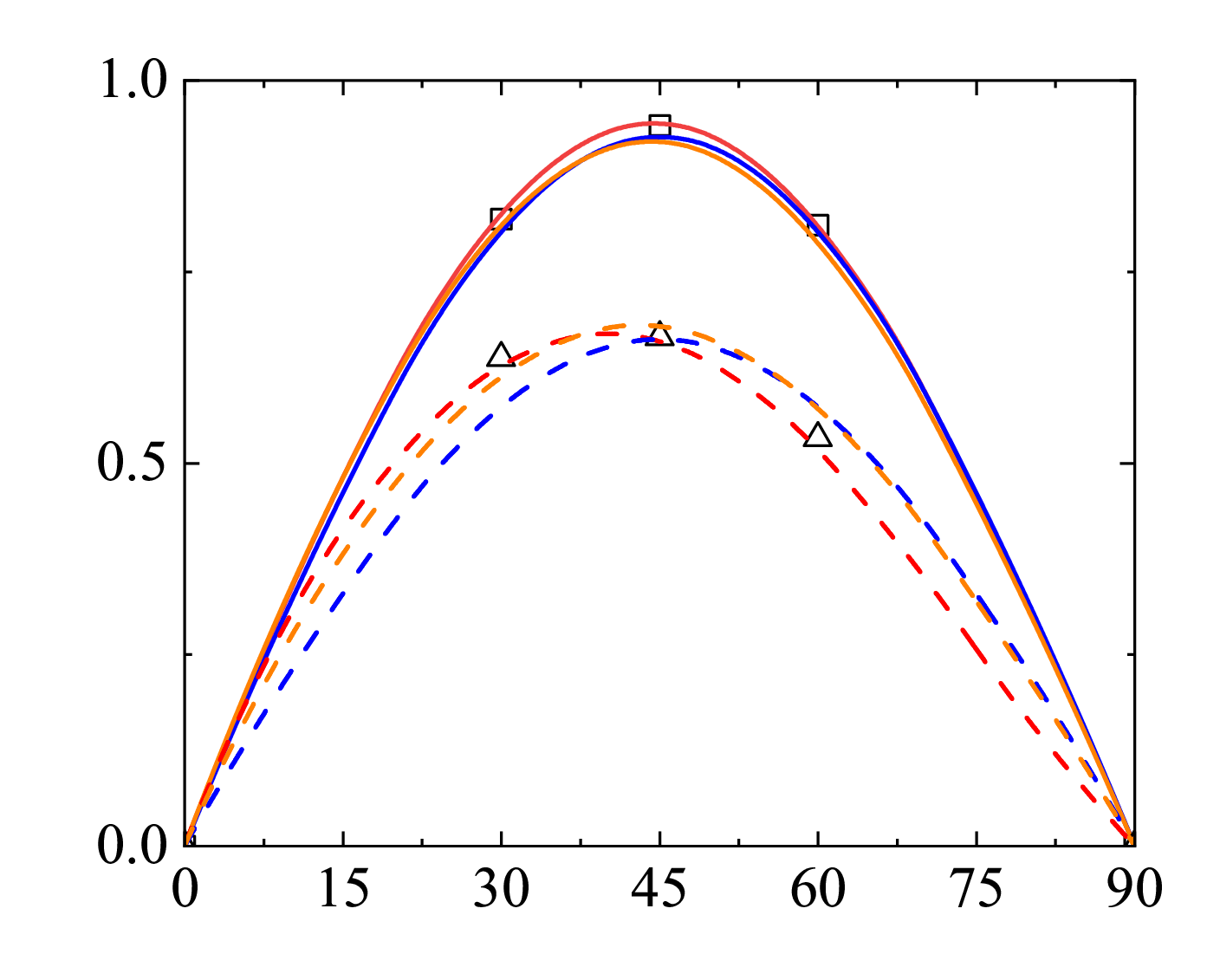}
		\put(-1,70){(d)}
		\put(2,40){\rotatebox{90}{$C_T$}}
		\put(52.5,0){$\alpha^\circ$}
	\end{overpic}\\[1.0ex]
	\begin{overpic}[width=0.5\textwidth]{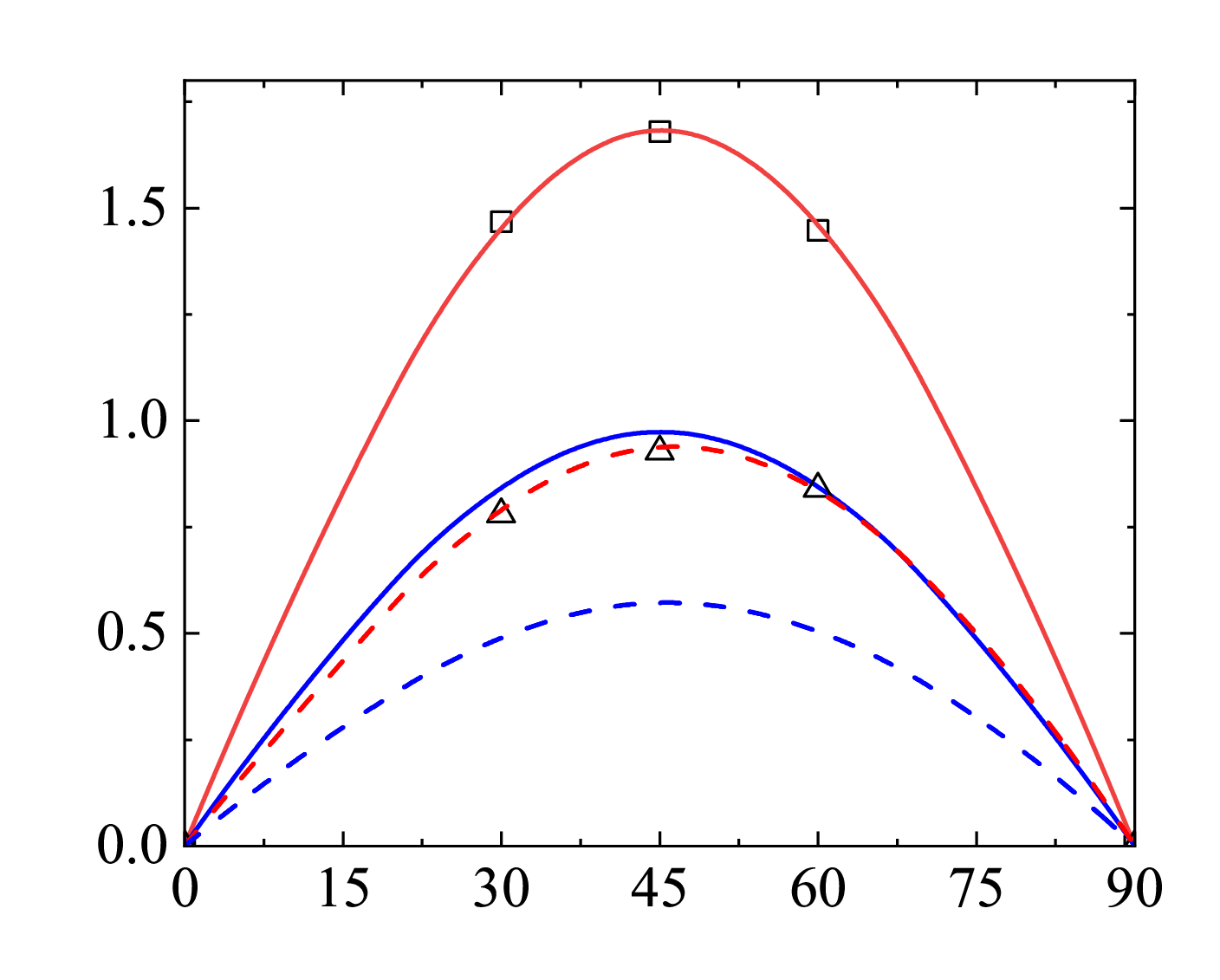}
		\put(-1,70){(e)}
		\put(2,40){\rotatebox{90}{$C_T$}}
		\put(52.5,0){$\alpha^\circ$}
	\end{overpic}~
	\begin{overpic}[width=0.5\textwidth]{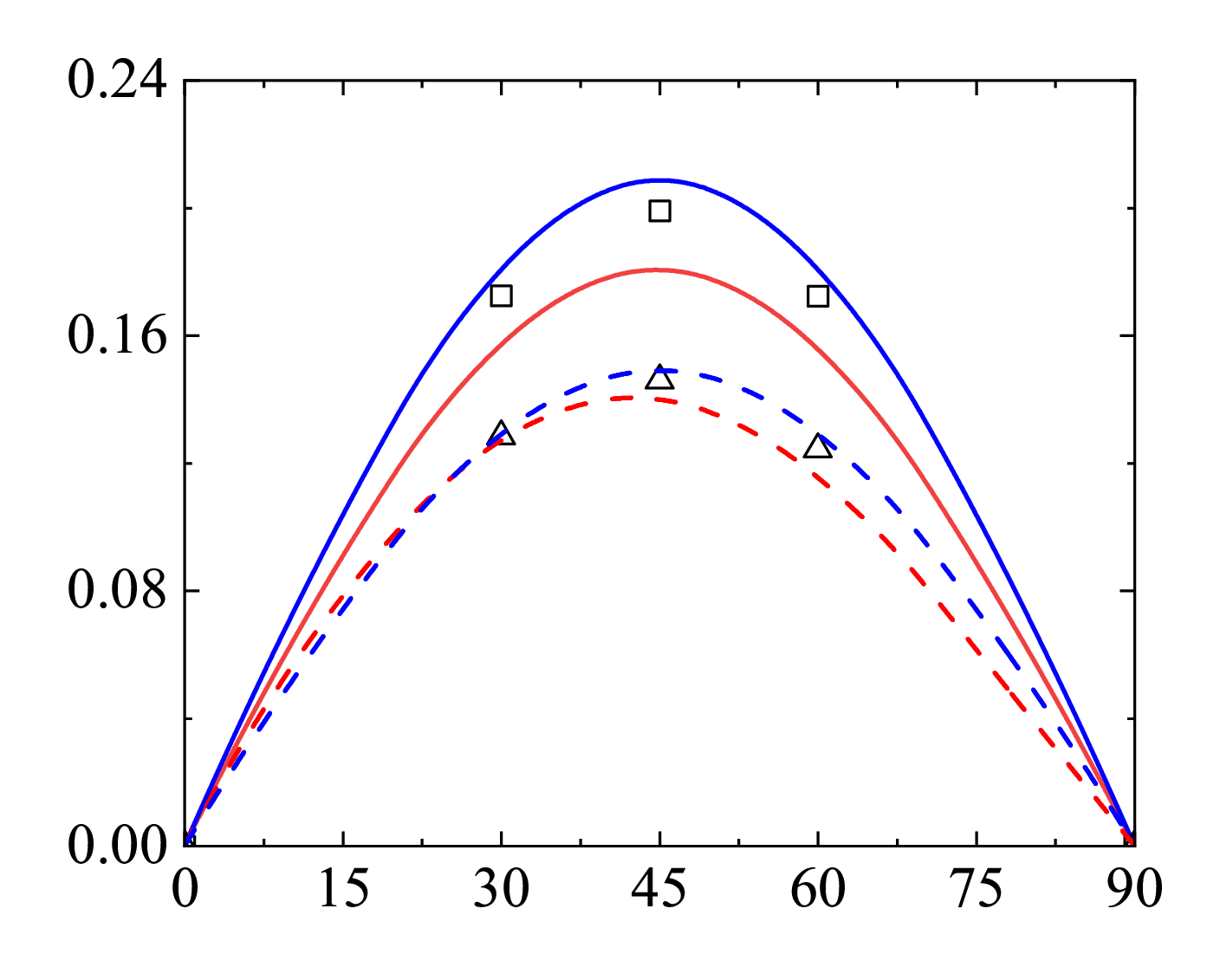}
		\put(-1,70){(f)}
		\put(2,40){\rotatebox{90}{$C_T$}}
		\put(52.5,0){$\alpha^\circ$}
	\end{overpic}\\
	\caption{Variation of particle pitching torque $C_T$ coefficients against $\alpha$, at $M_p=0.1$.
		\textcolor{black}{(a) $w = 1.25$, (b) $w = 0.2$, (c) $w = 2.5$, (d) $w = 0.4$, (e) $w = 5.0$, (f) $w = 0.8$.} 
		Solid lines and squares: $Re_p=10$, dashed lines and triangles: $Re_p=100$.
		Symbols :present simulation,
		red lines: Eq.~\eqref{3.3.3-ct-incom}, 
		blue lines: (a,c,e) \citet{ouchene_new_2016}, (b,d,f) \citet{ouchene_numerical_2020}, 
		green lines: \citet{zastawny_derivation_2012}, 
		orange lines: \citet{sanjeevi_drag_2018}.}
	\label{fig14:ct-a-incom}
\end{figure}	

Figure \ref{fig13:ct-re-incom} demonstrates that pitching torque coefficients $C_T$ decrease monotonically with Reynolds number 
at an attack angle of $\alpha=45^\circ$, similar to the trend observed for lift coefficients.
\textcolor{black}{At a given $Re_p$, $C_T$ is higher as the aspect ratio deviates from 1.0.} 
Comparative analysis with empirical formulas from previous studies \citep{ouchene_new_2016,zastawny_derivation_2012,sanjeevi_drag_2018} 
reveals that the currently proposed formula exhibits the lowest relative errors. 
In Figure \ref{fig14:ct-a-incom}, we further explore the variation of $C_T$ with the attack angle $\alpha$, 
comparing predictions from the proposed formula in this study with those of
\citet{ouchene_new_2016,ouchene_numerical_2020,zastawny_derivation_2012} and \citet{sanjeevi_drag_2018}. 
Overall, \textcolor{black}{the variation of $C_T$ follows a similar pattern to that of lift coefficients}, 
with asymmetrical distributions, particularly notable for oblate particles. 
The proposed formula provides the most precise predictions, with an average relative error of approximately 1.39\%.

The compressibility impact factor for the $C_T$ is similar to the drag and lift coefficients in compressible flow, 
in that the Spearman correlation analysis also show that the $C_T$ is monotonically correlated with $Re_p$ and $M_p$ (not shown here for brevity). 
However, it is noteworthy that the $C_T$ peaks around $M_p\approx0.8$, \textcolor{black}{and thus the segmentation was determined to be at $M_p = 0.8$}.
\begin{equation}
	\label{3.3.4-gt}
	g_t(Re_p,M_p)=t_1(t_2Re_p)^{t_3M_p^{t_4}}.
\end{equation}
With such refinement, the pitching torque coefficient for \textcolor{black}{spheroidal} particles in compressible flows can be written as
\begin{equation}
	\label{3.3.5-ct-com}
	C_T(Re_p,w,M_p,\alpha)=2(\sin \alpha)^{c_{t,1}^{Re_pw}} (\cos \alpha)^{c_{t,2}^{Re_pw}}C_{T,\alpha=45^\circ}(Re_p,w)g_t(Re_p,M_p)
\end{equation}
The parameters in the equation above is reported in Table \ref{tab11:ct-com}.
Compared with the numerical simulation databases, the presently proposed formula predicts $C_T$ with high accuracy, 
showing a relative error of approximately 4.6\%.

\begin{table}[tp!]
	\centering
	\small
	\caption{Parameters in Eq.\eqref{3.3.4-gt} for pitching torque coefficients $C_T$ and \textcolor{black}{average} relative error}
	\label{tab11:ct-com}
	\begin{tblr}{
			cell{2}{1} = {r=2}{c},
			cell{2}{7} = {r=2}{c},
			cell{4}{1} = {r=2}{c},
			cell{4}{7} = {r=2}{c},
			hline{1-2,4,6} = {-}{},
			colspec = {Q[c] Q[c] Q[c] Q[c] Q[c] Q[c] Q[c]}
		}
		~  & ~                        & $t_1$     & $t_2$       & $t_3$     & $t_4$      & \textcolor{black}{Average} relative
		error(\%) \\
		$w>1$ & $M_p\leq0.8$ & 1      & 0.0491 & 0.1714 & 2.2645  & 4.18                 \\
		& $M_p>0.8$                    & 0.3409 & 1        & 0.2368 & -1.2401 &                      \\
		$w<1$ & $M_p\leq0.8$ & 1      & 0.0331        & 0.1548 & 2.0034  & 5.03                 \\
		& $M_p>0.8$                    & 0.3398 & 1        & 0.2343 & -1.1002 &                      
	\end{tblr}
\end{table}

\begin{figure}[tp!]
	\centering
	\begin{overpic}[width=0.5\textwidth]{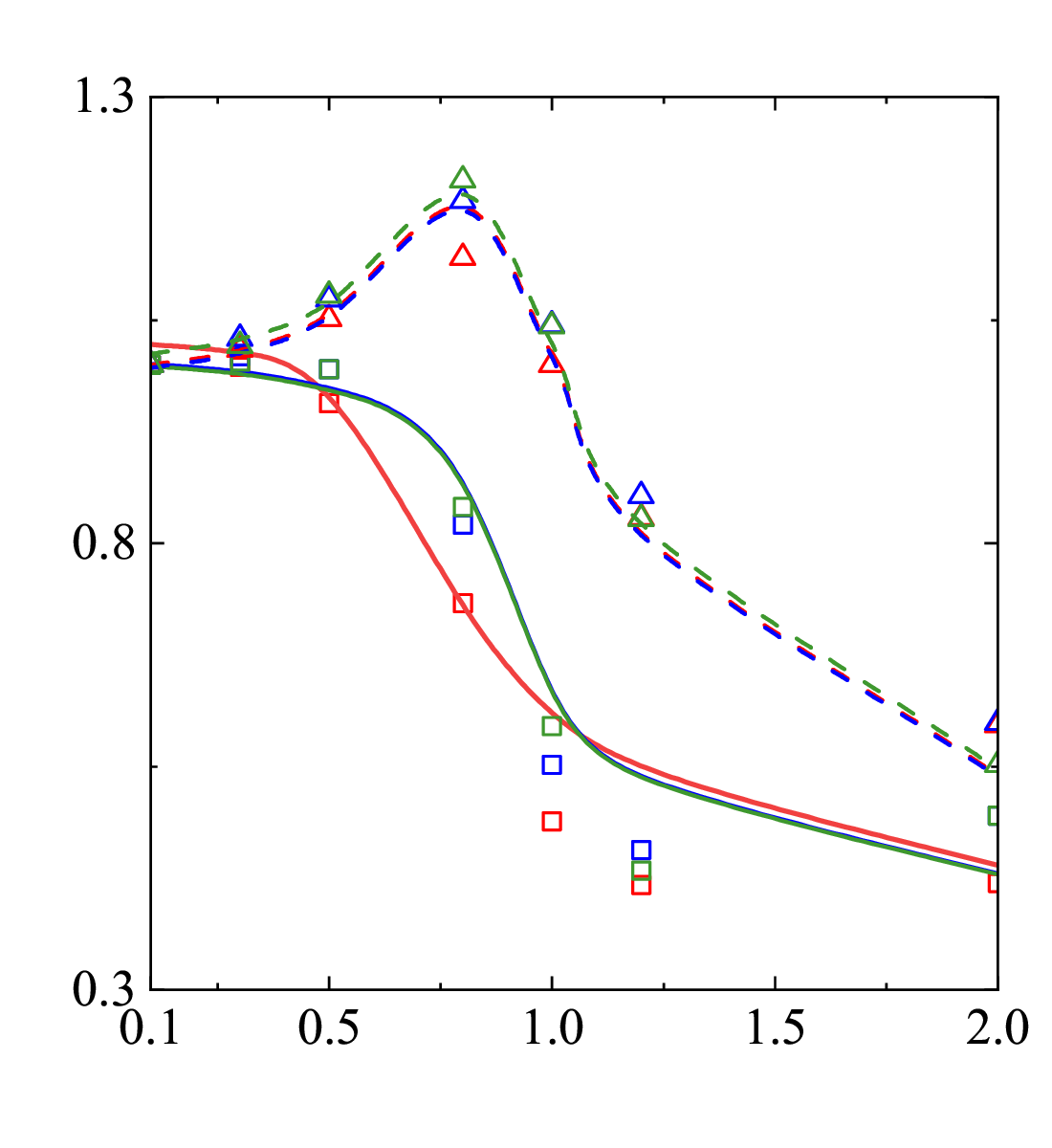}
		\put(0,90){(a)}
		\put(0,30){\rotatebox{90}{\textcolor{black}{$C_{T,\alpha=45^\circ}/C_{T,\alpha=45^\circ,M_p=0.1}$}}}
		\put(52.5,2){$M_p$}	
	\end{overpic}~
	\begin{overpic}[width=0.5\textwidth]{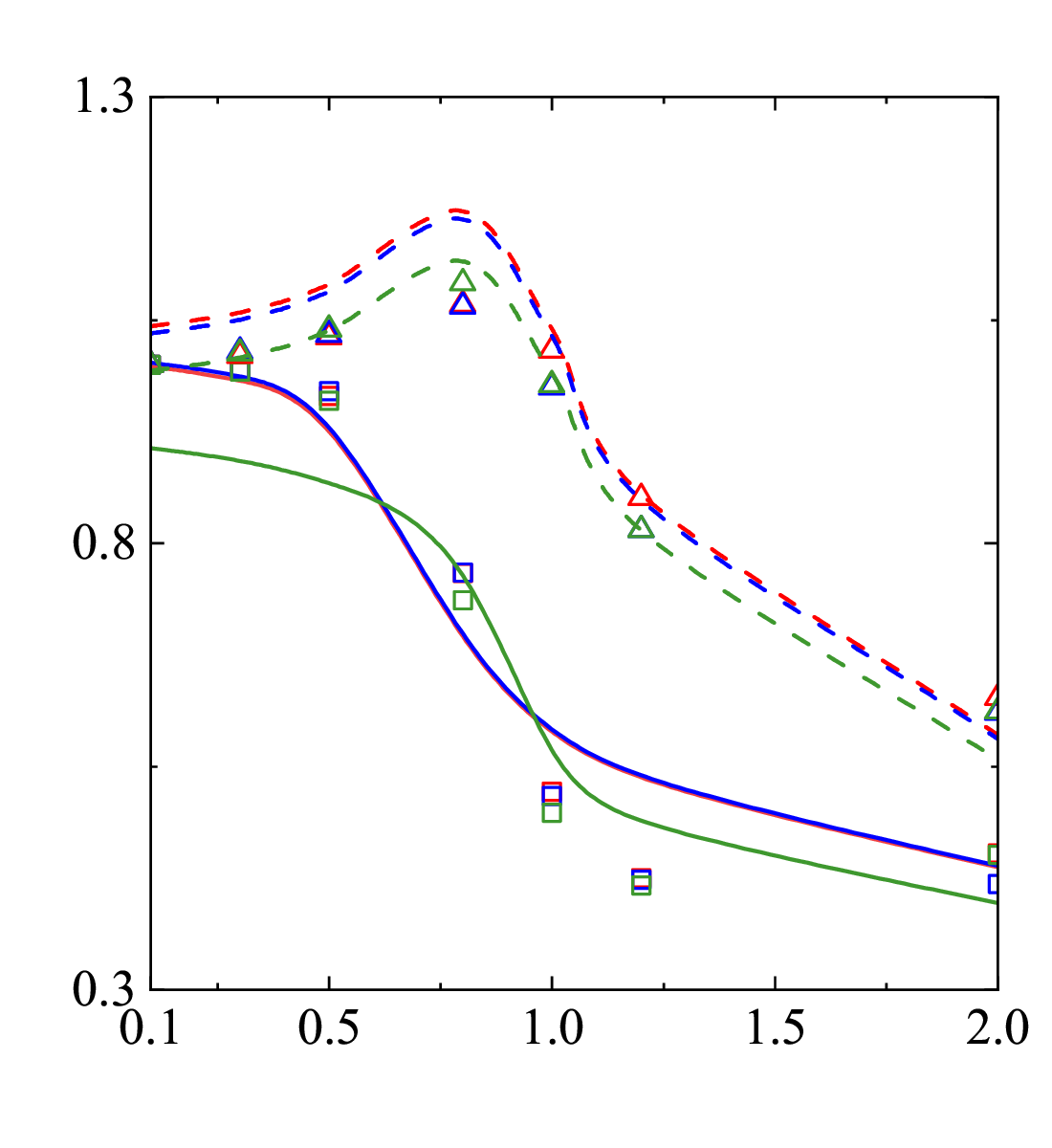}
		\put(0,90){(b)}
		\put(0,30){\rotatebox{90}{\textcolor{black}{$C_{T,\alpha=45^\circ}/C_{T,\alpha=45^\circ,M_p=0.1}$}}}	
		\put(52.5,2){$M_p$}
	\end{overpic}\\
	\caption{Variation of pitching torque coefficients \textcolor{black}{$C_{T,\alpha=45^\circ}/C_{T,\alpha=45^\circ,M_p=0.1}$} 
		against $M_p$, 
		(a) prolate particles: $w=1.25$ (red), $w=2.5$ (blue),  $w=5$ (green)
		and (b) oblate particles: $w=0.2$ (red), $w=0.4$ (blue),  $w=0.8$ (green). 
		Solid lines and squares: $Re_p=10$, dashed lines and triangles: $Re_p=100$. 
		Lines: Eq.~\eqref{3.3.5-ct-com}, symbols: present simulations.
		}
	\label{fig15:ct-ma}
\end{figure}

\begin{figure}[tp!]
	\centering
	\begin{overpic}[width=0.5\textwidth]{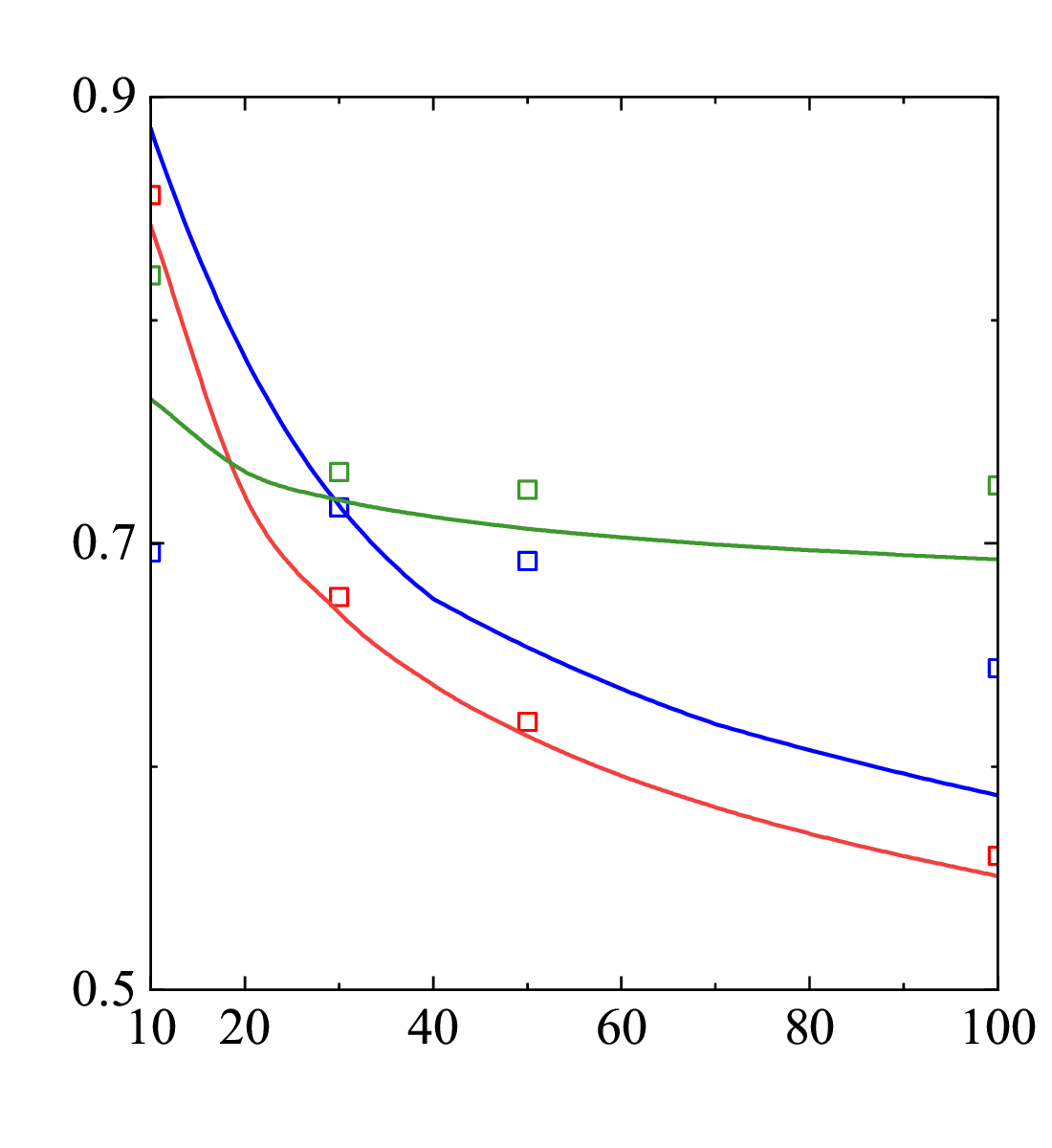}
		\put(0,90){(a)}
		\put(0,50){\rotatebox{90}{$C_{T,\alpha=45^\circ}$}}
		\put(52.5,2){$Re_p$}
	\end{overpic}~
	\begin{overpic}[width=0.5\textwidth]{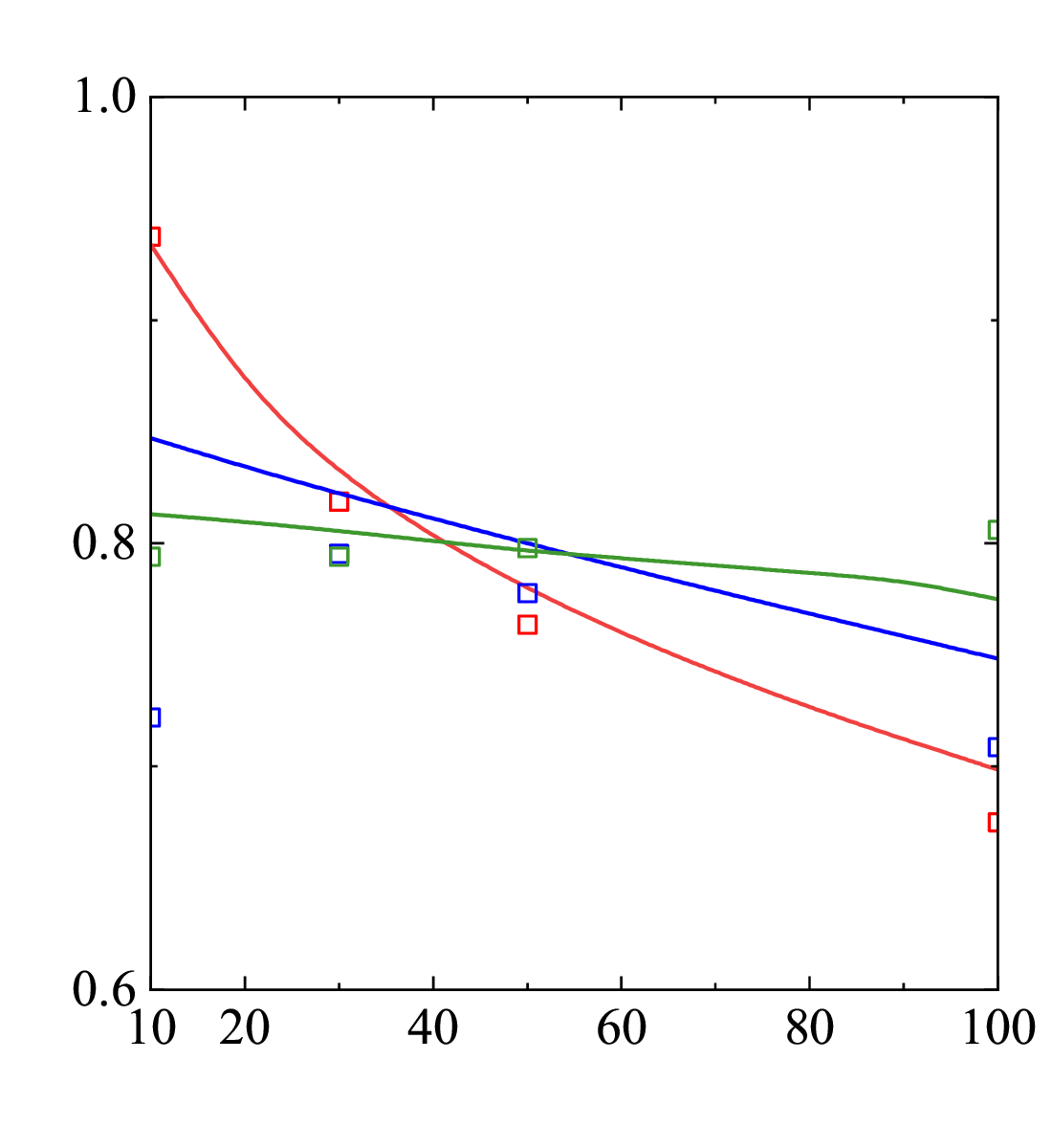}
		\put(0,90){(b)}
		\put(0,50){\rotatebox{90}{$C_{T,\alpha=45^\circ}$}}
		\put(52.5,2){$Re_p$}
	\end{overpic}\\
	\caption{Variation of pitching torque coefficients  $C_{T,\alpha=45^\circ}$ against $Re_p$
	\textcolor{black}{(When $M_p=2$, we add 0.4 to $C_{T,\alpha=45^\circ}$)}, (a) $w=2.5$ and (b) $w=0.4$. 
	Red: $M_p=0.3$, 
	blue: $M_p=0.8$, 
	green: $M_p=2$.
	Lines: Eq.~\eqref{3.3.5-ct-com}, symbols: present simulations.}
	\label{fig16:ct-re-com}	
\end{figure}

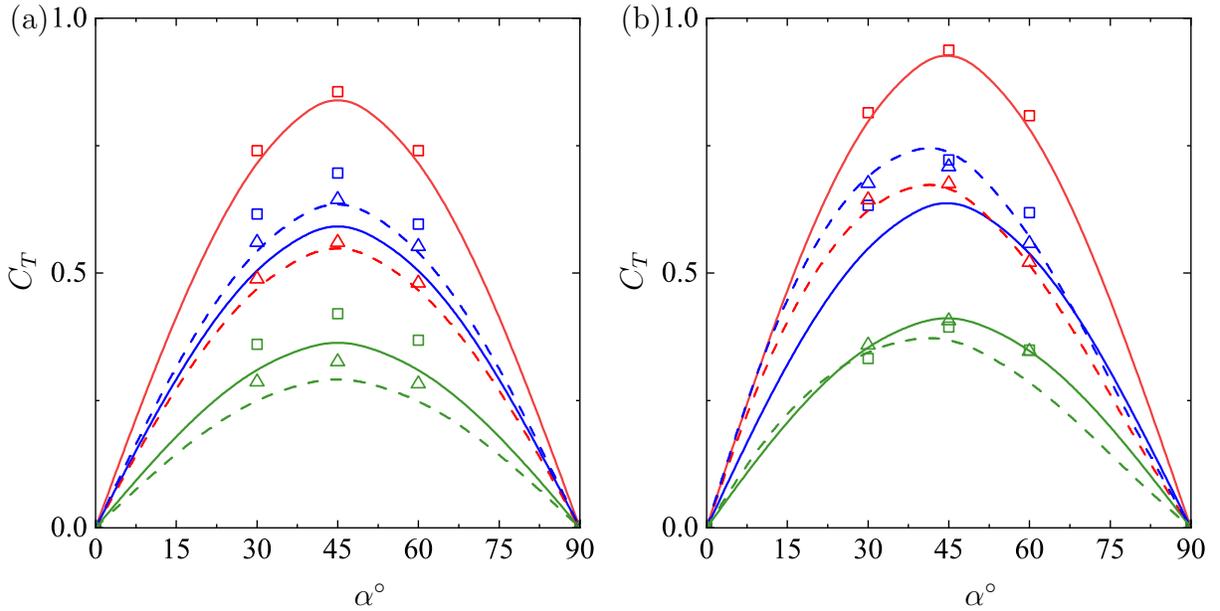
\begin{figure}[tp!]
	\centering
	\begin{overpic}[width=0.5\textwidth]{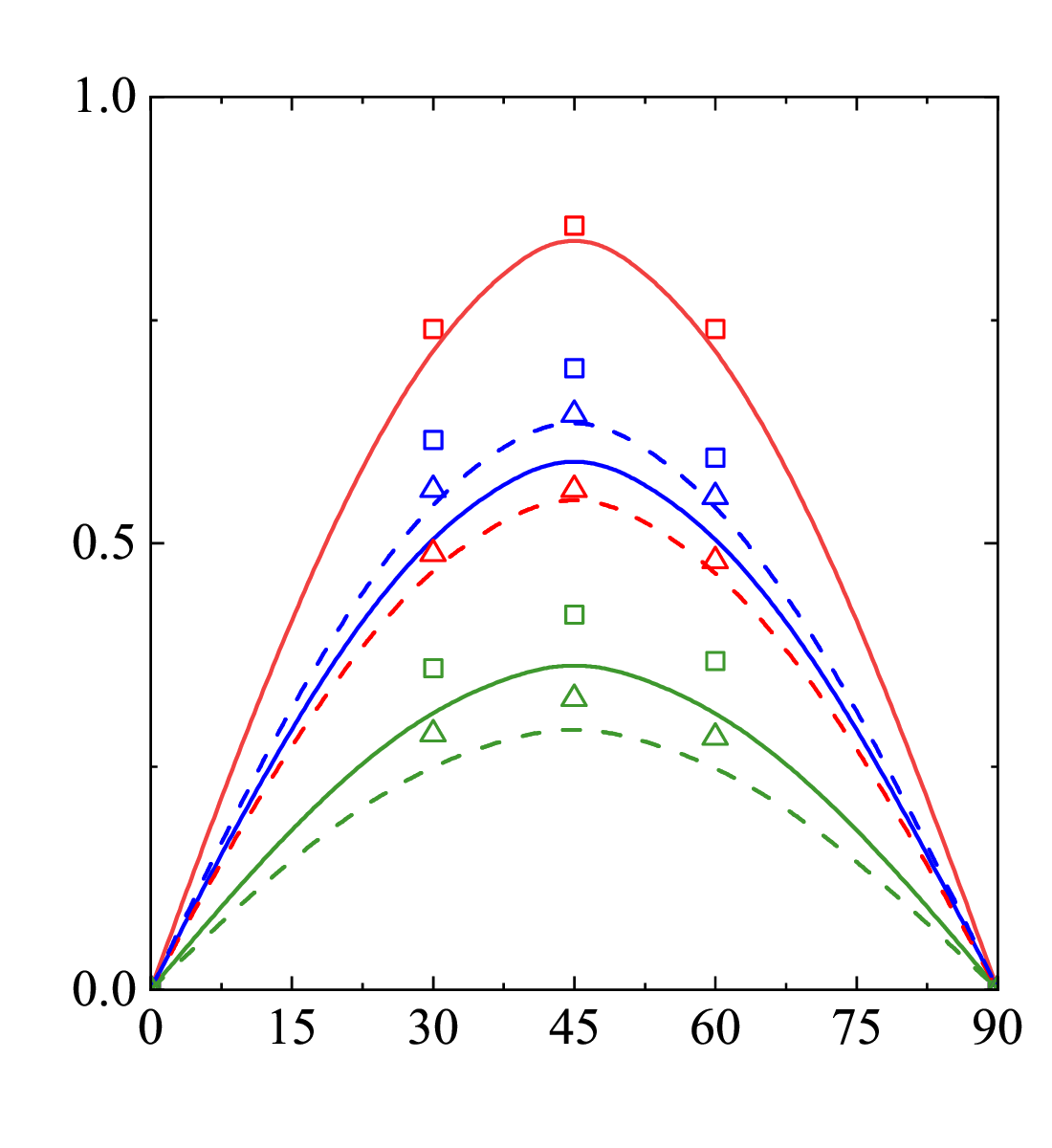}
		\put(0,90){(a)}		
		\put(0,50){\rotatebox{90}{$C_T$}}
		\put(52.5,2){$\alpha^\circ$}
	\end{overpic}~
	\begin{overpic}[width=0.5\textwidth]{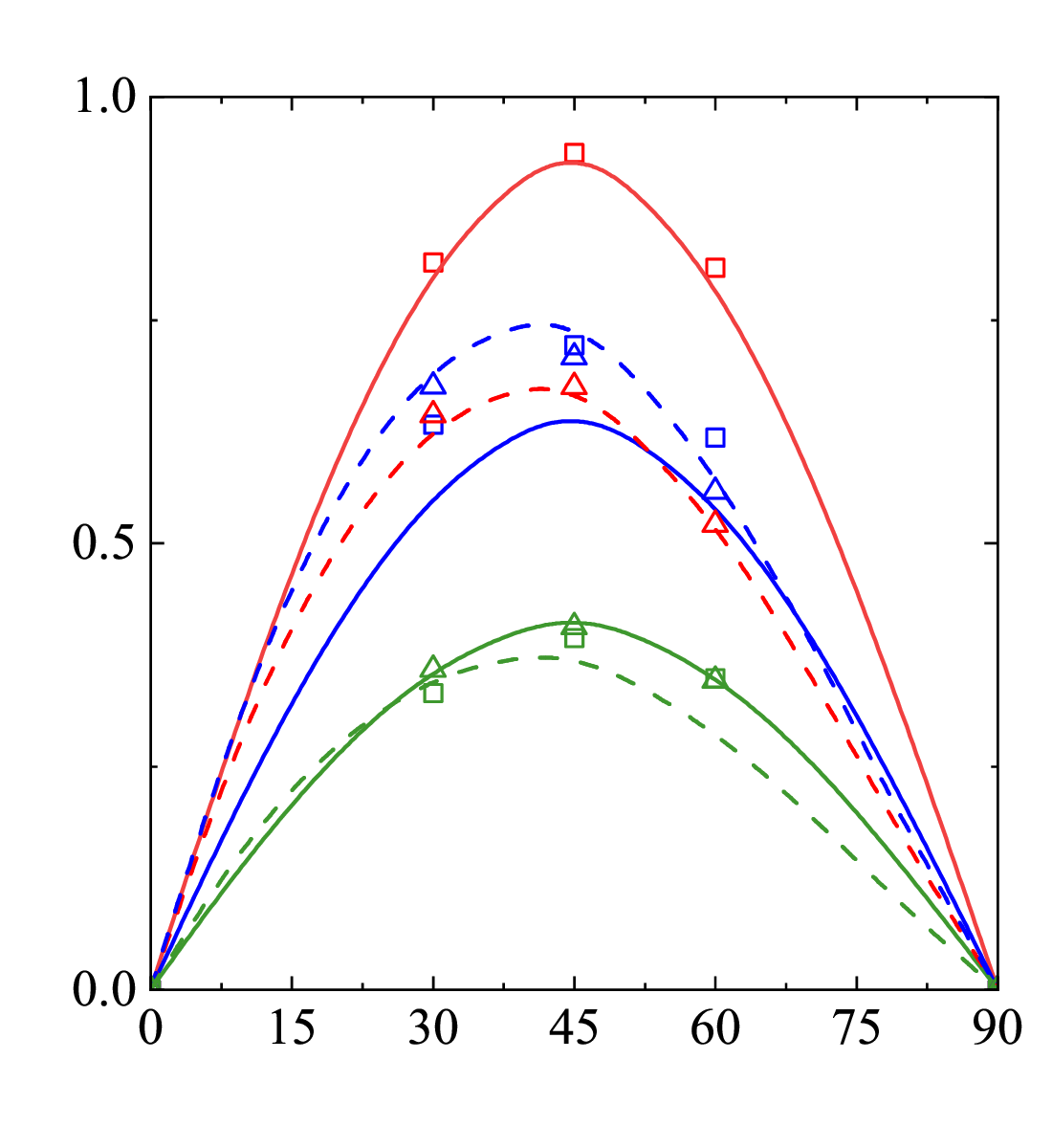}
		\put(0,90){(b)}
		\put(0,50){\rotatebox{90}{$C_T$}}
		\put(52.5,2){$\alpha^\circ$}
	\end{overpic}\\
	\caption{Variation of pitching torque coefficients $C_{T}$ against $\alpha$, (a) $w=2.5$, (b) $w=0.4$.
	Solid lines and squares: $Re_p=10$, dashed lines and triangles: $Re_p=100$.
	Red: $M_p=0.3$, 
	blue: $M_p=0.8$, 
	green: $M_p=2$.
	Lines: Eq.~\eqref{3.3.5-ct-com}, symbols: present simulations.}
	\label{fig17:ct-a-com}
\end{figure}

\textcolor{black}{The details of the comparison} are given as follows.
Figures \ref{fig15:ct-ma} and \ref{fig16:ct-re-com} illustrate the variation of pitching torque coefficients $C_T$ with Mach number 
$M_p$ and Reynolds number $Re_p$, respectively. 
For a constant Reynolds number $Re_p=10$, the $C_T$ of prolate \textcolor{black}{and oblate} particles \textcolor{black}{decreases} monotonically until 
$M_p=1.2$, beyond which the variation becomes insignificant across all six groups with different aspect ratios. However, 
at $Re_p=100$, the $C_T$ increases, peaks at $M_p=0.8$, and then decreases at higher $M_p$. 
Conversely, for a fixed $M_p=0.3$, the $C_T$ \textcolor{black}{decreases} monotonically with Reynolds number $Re_p$. 
In transonic cases at $M_p=0.8$, the $C_T$ initially \textcolor{black}{increases} and then \textcolor{black}{decreases}, peaking at $Re_p=30$. 
In supersonic cases at $M_p=2.0$, the $C_T$ are minimally affected by Reynolds number, 
indicating the significant influence of detached shock waves in these scenarios. 
Figure \ref{fig17:ct-a-com} showcases the variation of the pitching torque coefficient with \textcolor{black}{angle of attack} at different Mach numbers, 
demonstrating a consistent trend across higher Mach numbers, with a peak at $\alpha=45^\circ$ and slight asymmetrical distributions. 
The empirical formula proposed herein effectively captures the trend of variation and provides relatively accurate results 
for all cases.

\section{Conclusion} \label{sec:con}

In this research, we investigate the drag and lift forces and pitching torque of \textcolor{black}{spheroidal} particles in compressible uniform flows 
at various Reynolds and Mach numbers. 
The aim is to propose empirical formulas for point-particle simulations in high-speed flow applications. 
Both prolate and oblate particles are examined, encompassing aspect ratios ranging from 0.2 to 5.0, 
at Reynolds numbers from 10 to 100 and Mach numbers from 0.1 to 2.0, covering subsonic to supersonic flows without shedding vortices. 
Through an analysis of approximately \textcolor{black}{one} thousand numerical simulation cases, we refine empirical formulas for drag and lift forces 
and pitching torque coefficients, particularly focusing on improvements at low Mach number limits compared to previous studies. 
Additionally, Spearman correlation analysis highlights the significance of Reynolds and Mach numbers in compressible flow cases, 
leading to the proposal of compressibility impact factors to enhance incompressible flow formulas. Comprehensive comparisons 
demonstrate that the proposed empirical formulas yield accurate predictions, with average relative errors below $5\%$ 
compared to numerical simulation results.

This study addresses the gap in force and torque modeling of \textcolor{black}{spheroidal} particles in compressible flow
\textcolor{black}{and extends} 
the applicability of the point-particle approach to a broader range of flow conditions. 
Future research will concentrate on refining the formulas at higher Reynolds numbers to explore unsteady vortex shedding and 
at low Reynolds numbers but high Mach numbers where rarefaction effects become significant.



\setlength{\bibsep}{-0.1ex}
\bibliographystyle{elsarticle-harv}
\bibliography{bibfile}   

\begin{thebibliography}{34}
\expandafter\ifx\csname natexlab\endcsname\relax\def\natexlab#1{#1}\fi
\providecommand{\url}[1]{\texttt{#1}}
\providecommand{\href}[2]{#2}
\providecommand{\path}[1]{#1}
\providecommand{\DOIprefix}{doi:}
\providecommand{\ArXivprefix}{arXiv:}
\providecommand{\URLprefix}{URL: }
\providecommand{\Pubmedprefix}{pmid:}
\providecommand{\doi}[1]{\href{http://dx.doi.org/#1}{\path{#1}}}
\providecommand{\Pubmed}[1]{\href{pmid:#1}{\path{#1}}}
\providecommand{\bibinfo}[2]{#2}
\ifx\xfnm\relax \def\xfnm[#1]{\unskip,\space#1}\fi
\bibitem[{Arcen et~al.(2017)Arcen, Ouchene, Khalij and
  Tanière}]{arcen_prolate_2017}
\bibinfo{author}{Arcen, B.}, \bibinfo{author}{Ouchene, R.},
  \bibinfo{author}{Khalij, M.}, \bibinfo{author}{Tanière, A.},
  \bibinfo{year}{2017}.
\newblock \bibinfo{title}{Prolate spheroidal particles’ behavior in a
  vertical wall-bounded turbulent flow}.
\newblock \bibinfo{journal}{Phys. Fluids} \bibinfo{volume}{29},
  \bibinfo{pages}{093301}.
\bibitem[{Bailey and Starr(1976)}]{bailey_sphere_1976}
\bibinfo{author}{Bailey, A.B.}, \bibinfo{author}{Starr, R.F.},
  \bibinfo{year}{1976}.
\newblock \bibinfo{title}{Sphere drag at transonic speeds and high {Reynolds}
  numbers}.
\newblock \bibinfo{journal}{AIAA J.} \bibinfo{volume}{14},
  \bibinfo{pages}{1631--1631}.
\bibitem[{Batchelor(1970)}]{batchelor_stress_1970}
\bibinfo{author}{Batchelor, G.K.}, \bibinfo{year}{1970}.
\newblock \bibinfo{title}{The stress system in a suspension of force-free
  particles}.
\newblock \bibinfo{journal}{J. Fluid Mech.} \bibinfo{volume}{41},
  \bibinfo{pages}{545--570}.
\bibitem[{Bounoua et~al.(2018)Bounoua, Bouchet and
  Verhille}]{bounoua_tumbling_2018}
\bibinfo{author}{Bounoua, S.}, \bibinfo{author}{Bouchet, G.},
  \bibinfo{author}{Verhille, G.}, \bibinfo{year}{2018}.
\newblock \bibinfo{title}{Tumbling of {Inertial} {Fibers} in {Turbulence}}.
\newblock \bibinfo{journal}{Phys. Rev. Lett.} \bibinfo{volume}{121},
  \bibinfo{pages}{124502}.
\bibitem[{Brenner(1961)}]{brenner_oseen_1961}
\bibinfo{author}{Brenner, H.}, \bibinfo{year}{1961}.
\newblock \bibinfo{title}{The {Oseen} resistance of a particle of arbitrary
  shape}.
\newblock \bibinfo{journal}{J. Fluid Mech.} \bibinfo{volume}{11},
  \bibinfo{pages}{604}.
\bibitem[{Brenner(1964)}]{brenner_stokes_1964}
\bibinfo{author}{Brenner, H.}, \bibinfo{year}{1964}.
\newblock \bibinfo{title}{The {Stokes} resistance of an arbitrary
  particle—{III}}.
\newblock \bibinfo{journal}{Chem. Eng. Sci.} \bibinfo{volume}{19},
  \bibinfo{pages}{631--651}.
\bibitem[{Capecelatro and Wagner(2024)}]{capecelatro_gas-particle_2023}
\bibinfo{author}{Capecelatro, J.}, \bibinfo{author}{Wagner, J.},
  \bibinfo{year}{2024}.
\newblock \bibinfo{title}{Gas--particle dynamics in high-speed flows}.
\newblock \bibinfo{journal}{Annu. Rev. Fluid Mech.} \bibinfo{volume}{56},
  \bibinfo{pages}{379--403}.
\bibitem[{Clift and Gauvin(1971)}]{clift_motion_1971}
\bibinfo{author}{Clift, R.}, \bibinfo{author}{Gauvin, W.H.},
  \bibinfo{year}{1971}.
\newblock \bibinfo{title}{Motion of entrained particles in gas streams}.
\newblock \bibinfo{journal}{Can. J. Chem. Eng.} \bibinfo{volume}{49},
  \bibinfo{pages}{439--448}.
\bibitem[{Cui et~al.(2023)Cui, Qiu, Jiang and Zhao}]{cui_eect_nodate}
\bibinfo{author}{Cui, Z.}, \bibinfo{author}{Qiu, J.}, \bibinfo{author}{Jiang,
  X.}, \bibinfo{author}{Zhao, L.}, \bibinfo{year}{2023}.
\newblock \bibinfo{title}{Effect of fluid inertial torque on the rotational and
  orientational dynamics of tiny spheroidal particles in turbulent channel
  flow}.
\newblock \bibinfo{journal}{J. Fluid Mech.} \bibinfo{volume}{977},
  \bibinfo{pages}{A20}.
\bibitem[{Fröhlich et~al.(2020)Fröhlich, Meinke and
  Schröder}]{frohlich_correlations_2020}
\bibinfo{author}{Fröhlich, K.}, \bibinfo{author}{Meinke, M.},
  \bibinfo{author}{Schröder, W.}, \bibinfo{year}{2020}.
\newblock \bibinfo{title}{Correlations for inclined prolates based on highly
  resolved simulations}.
\newblock \bibinfo{journal}{J. Fluid Mech.} \bibinfo{volume}{901},
  \bibinfo{pages}{A5}.
\bibitem[{Gustavsson et~al.(2014)Gustavsson, Einarsson and
  Mehlig}]{gustavsson_tumbling_2014}
\bibinfo{author}{Gustavsson, K.}, \bibinfo{author}{Einarsson, J.},
  \bibinfo{author}{Mehlig, B.}, \bibinfo{year}{2014}.
\newblock \bibinfo{title}{Tumbling of {Small} {Axisymmetric} {Particles} in
  {Random} and {Turbulent} {Flows}}.
\newblock \bibinfo{journal}{Phys. Rev. Lett.} \bibinfo{volume}{112},
  \bibinfo{pages}{014501}.
\bibitem[{Happel and Brenner(1983)}]{happel_low_1983}
\bibinfo{author}{Happel, J.}, \bibinfo{author}{Brenner, H.},
  \bibinfo{year}{1983}.
\newblock \bibinfo{title}{Low {Reynolds} number hydrodynamics: with special
  applications to particulate media}. volume~\bibinfo{volume}{1}.
\newblock \bibinfo{publisher}{Springer Netherlands}.
\bibitem[{He et~al.(2016)He, Zhao, Ma, Wang and Zhang}]{he2016validation}
\bibinfo{author}{He, X.}, \bibinfo{author}{Zhao, Z.}, \bibinfo{author}{Ma, R.},
  \bibinfo{author}{Wang, N.}, \bibinfo{author}{Zhang, L.},
  \bibinfo{year}{2016}.
\newblock \bibinfo{title}{Validation of hyperflow in subsonic and transonic
  flow}.
\newblock \bibinfo{journal}{Acta Aerodyn.Sin.} \bibinfo{volume}{34},
  \bibinfo{pages}{267--275}.
\bibitem[{Huang et~al.(2012)Huang, Yang, Krafczyk and Lu}]{huang_rotation_2012}
\bibinfo{author}{Huang, H.}, \bibinfo{author}{Yang, X.},
  \bibinfo{author}{Krafczyk, M.}, \bibinfo{author}{Lu, X.},
  \bibinfo{year}{2012}.
\newblock \bibinfo{title}{Rotation of spheroidal particles in {Couette} flows}.
\newblock \bibinfo{journal}{J. Fluid Mech.} \bibinfo{volume}{692},
  \bibinfo{pages}{369--394}.
\bibitem[{Hölzer and Sommerfeld(2008)}]{holzer_new_2008}
\bibinfo{author}{Hölzer, A.}, \bibinfo{author}{Sommerfeld, M.},
  \bibinfo{year}{2008}.
\newblock \bibinfo{title}{New simple correlation formula for the drag
  coefficient of non-spherical particles}.
\newblock \bibinfo{journal}{Powder Technol.} \bibinfo{volume}{184},
  \bibinfo{pages}{361--365}.
\bibitem[{Loth(2008)}]{loth_drag_2008}
\bibinfo{author}{Loth, E.}, \bibinfo{year}{2008}.
\newblock \bibinfo{title}{Drag of non-spherical solid particles of regular and
  irregular shape}.
\newblock \bibinfo{journal}{Powder Technol.} \bibinfo{volume}{182},
  \bibinfo{pages}{342--353}.
\bibitem[{Loth et~al.(2021)Loth, Tyler~Daspit, Jeong, Nagata and
  Nonomura}]{loth_supersonic_2021}
\bibinfo{author}{Loth, E.}, \bibinfo{author}{Tyler~Daspit, J.},
  \bibinfo{author}{Jeong, M.}, \bibinfo{author}{Nagata, T.},
  \bibinfo{author}{Nonomura, T.}, \bibinfo{year}{2021}.
\newblock \bibinfo{title}{Supersonic and {Hypersonic} {Drag} {Coefficients} for
  a {Sphere}}.
\newblock \bibinfo{journal}{AIAA J.} , \bibinfo{pages}{3261--3274}.
\bibitem[{Marchioli et~al.(2016)Marchioli, Zhao and
  Andersson}]{marchioli_relative_2016}
\bibinfo{author}{Marchioli, C.}, \bibinfo{author}{Zhao, L.},
  \bibinfo{author}{Andersson, H.I.}, \bibinfo{year}{2016}.
\newblock \bibinfo{title}{On the relative rotational motion between rigid
  fibers and fluid in turbulent channel flow}.
\newblock \bibinfo{journal}{Phys. Fluids} \bibinfo{volume}{28},
  \bibinfo{pages}{013301}.
\bibitem[{Nagata et~al.(2016)Nagata, Nonomura, Takahashi, Mizuno and
  Fukuda}]{nagata_investigation_2016}
\bibinfo{author}{Nagata, T.}, \bibinfo{author}{Nonomura, T.},
  \bibinfo{author}{Takahashi, S.}, \bibinfo{author}{Mizuno, Y.},
  \bibinfo{author}{Fukuda, K.}, \bibinfo{year}{2016}.
\newblock \bibinfo{title}{Investigation on subsonic to supersonic flow around a
  sphere at low {Reynolds} number of between 50 and 300 by direct numerical
  simulation}.
\newblock \bibinfo{journal}{Phys. Fluids} \bibinfo{volume}{28},
  \bibinfo{pages}{056101}.
\bibitem[{Oberbeck(1876)}]{oberbeck_lieber_nodate}
\bibinfo{author}{Oberbeck, A.}, \bibinfo{year}{1876}.
\newblock \bibinfo{title}{lieber stationäre {Müssigkeitsbewegungen} mit
  {Berück}- sichtigung der inneren {Keibung}.}
\newblock \bibinfo{journal}{J. Reine Angew. Math.} \bibinfo{volume}{81},
  \bibinfo{pages}{62--90}.
\bibitem[{Ouchene(2020)}]{ouchene_numerical_2020}
\bibinfo{author}{Ouchene, R.}, \bibinfo{year}{2020}.
\newblock \bibinfo{title}{Numerical simulation and modeling of the hydrodynamic
  forces and torque acting on individual oblate spheroids}.
\newblock \bibinfo{journal}{Phys. Fluids} \bibinfo{volume}{32},
  \bibinfo{pages}{073303}.
\bibitem[{Ouchene et~al.(2016)Ouchene, Khalij, Arcen and
  Tanière}]{ouchene_new_2016}
\bibinfo{author}{Ouchene, R.}, \bibinfo{author}{Khalij, M.},
  \bibinfo{author}{Arcen, B.}, \bibinfo{author}{Tanière, A.},
  \bibinfo{year}{2016}.
\newblock \bibinfo{title}{A new set of correlations of drag, lift and torque
  coefficients for non-spherical particles and large {Reynolds} numbers}.
\newblock \bibinfo{journal}{Powder Technol.} \bibinfo{volume}{303},
  \bibinfo{pages}{33--43}.
\bibitem[{Ouchene et~al.(2015)Ouchene, Khalij, Tanière and
  Arcen}]{ouchene_drag_2015}
\bibinfo{author}{Ouchene, R.}, \bibinfo{author}{Khalij, M.},
  \bibinfo{author}{Tanière, A.}, \bibinfo{author}{Arcen, B.},
  \bibinfo{year}{2015}.
\newblock \bibinfo{title}{Drag, lift and torque coefficients for ellipsoidal
  particles: {From} low to moderate particle {Reynolds} numbers}.
\newblock \bibinfo{journal}{Comput. Fluids} \bibinfo{volume}{113},
  \bibinfo{pages}{53--64}.
\bibitem[{Parmar et~al.(2010)Parmar, Haselbacher and
  Balachandar}]{parmar_improved_2010}
\bibinfo{author}{Parmar, M.}, \bibinfo{author}{Haselbacher, A.},
  \bibinfo{author}{Balachandar, S.}, \bibinfo{year}{2010}.
\newblock \bibinfo{title}{Improved {Drag} {Correlation} for {Spheres} and
  {Application} to {Shock}-{Tube} {Experiments}}.
\newblock \bibinfo{journal}{AIAA J.} \bibinfo{volume}{48},
  \bibinfo{pages}{1273--1276}.
\bibitem[{Richter and Nikrityuk(2013)}]{richter_new_2013}
\bibinfo{author}{Richter, A.}, \bibinfo{author}{Nikrityuk, P.},
  \bibinfo{year}{2013}.
\newblock \bibinfo{title}{New correlations for heat and fluid flow past
  ellipsoidal and cubic particles at different angles of attack}.
\newblock \bibinfo{journal}{Powder Technol.} \bibinfo{volume}{249},
  \bibinfo{pages}{463--474}.
\bibitem[{Richter and Nikrityuk(2012)}]{richter_drag_2012}
\bibinfo{author}{Richter, A.}, \bibinfo{author}{Nikrityuk, P.A.},
  \bibinfo{year}{2012}.
\newblock \bibinfo{title}{Drag forces and heat transfer coefficients for
  spherical, cuboidal and ellipsoidal particles in cross flow at sub-critical
  {Reynolds} numbers}.
\newblock \bibinfo{journal}{Int. J. Heat Mass Transfer} \bibinfo{volume}{55},
  \bibinfo{pages}{1343--1354}.
\bibitem[{Rosén et~al.(2014)Rosén, Lundell and Aidun}]{rosen_effect_2014}
\bibinfo{author}{Rosén, T.}, \bibinfo{author}{Lundell, F.},
  \bibinfo{author}{Aidun, C.K.}, \bibinfo{year}{2014}.
\newblock \bibinfo{title}{Effect of fluid inertia on the dynamics and scaling
  of neutrally buoyant particles in shear flow}.
\newblock \bibinfo{journal}{J. Fluid Mech.} \bibinfo{volume}{738},
  \bibinfo{pages}{563--590}.
\bibitem[{Sanjeevi et~al.(2018)Sanjeevi, Kuipers and
  Padding}]{sanjeevi_drag_2018}
\bibinfo{author}{Sanjeevi, S.}, \bibinfo{author}{Kuipers, J.},
  \bibinfo{author}{Padding, J.T.}, \bibinfo{year}{2018}.
\newblock \bibinfo{title}{Drag, lift and torque correlations for non-spherical
  particles from {Stokes} limit to high {Reynolds} numbers}.
\newblock \bibinfo{journal}{Int. J. Multiphase Flow} \bibinfo{volume}{106},
  \bibinfo{pages}{325--337}.
\bibitem[{Sanjeevi and Padding(2017)}]{sanjeevi_2017_orientational}
\bibinfo{author}{Sanjeevi, S.}, \bibinfo{author}{Padding, J.},
  \bibinfo{year}{2017}.
\newblock \bibinfo{title}{On the orientational dependence of drag experienced
  by spheroids}.
\newblock \bibinfo{journal}{J. Fluid Mech.} \bibinfo{volume}{820},
  \bibinfo{pages}{R1}.
\bibitem[{Schiller and Naumann(1933)}]{schiller_uber_1933}
\bibinfo{author}{Schiller, L.}, \bibinfo{author}{Naumann, A.},
  \bibinfo{year}{1933}.
\newblock \bibinfo{title}{Über die {Grundlegenden} {Berechungen} bei der
  {Schwerkraftaufbereitung} {Zeitschrift} des {Vereines} {Deutscher}
  {Ingenieure}}.
\newblock \bibinfo{journal}{Ver. Deut. Ing.} \bibinfo{volume}{77},
  \bibinfo{pages}{No. 12}.
\bibitem[{Stokes(1901)}]{stokes_mathematical_2009}
\bibinfo{author}{Stokes, G.}, \bibinfo{year}{1901}.
\newblock \bibinfo{title}{Mathematical and physical papers}.
  volume~\bibinfo{volume}{3}.
\bibitem[{Zastawny et~al.(2012)Zastawny, Mallouppas, Zhao and
  Van~Wachem}]{zastawny_derivation_2012}
\bibinfo{author}{Zastawny, M.}, \bibinfo{author}{Mallouppas, G.},
  \bibinfo{author}{Zhao, F.}, \bibinfo{author}{Van~Wachem, B.},
  \bibinfo{year}{2012}.
\newblock \bibinfo{title}{Derivation of drag and lift force and torque
  coefficients for non-spherical particles in flows}.
\newblock \bibinfo{journal}{Int. J. Multiphase Flow} \bibinfo{volume}{39},
  \bibinfo{pages}{227--239}.
\bibitem[{Zhao et~al.(2015)Zhao, George and Van~Wachem}]{zhao_four-way_2015}
\bibinfo{author}{Zhao, F.}, \bibinfo{author}{George, W.K.},
  \bibinfo{author}{Van~Wachem, B.G.M.}, \bibinfo{year}{2015}.
\newblock \bibinfo{title}{Four-way coupled simulations of small particles in
  turbulent channel flow: {The} effects of particle shape and {Stokes} number}.
\newblock \bibinfo{journal}{Phys. Fluids} \bibinfo{volume}{27},
  \bibinfo{pages}{083301}.
\bibitem[{Zhao et~al.(2014)Zhao, Marchioli and Andersson}]{zhao_slip_2014}
\bibinfo{author}{Zhao, L.}, \bibinfo{author}{Marchioli, C.},
  \bibinfo{author}{Andersson, H.I.}, \bibinfo{year}{2014}.
\newblock \bibinfo{title}{Slip velocity of rigid fibers in turbulent channel
  flow}.
\newblock \bibinfo{journal}{Phys. Fluids} \bibinfo{volume}{26},
  \bibinfo{pages}{063302}.

\end{thebibliography}
\end{document}